\documentclass[aps,prb,twocolumn,nofootinbib,notitlepage,superscriptaddress,longbibliography,preprintnumbers]{revtex4-2}
\usepackage{times}
\usepackage{graphicx}
\usepackage[whole]{bxcjkjatype} 

\usepackage[top=30truemm,bottom=30truemm,left=25truemm,right=25truemm]{geometry}

\usepackage[colorlinks=true,linktoc=page,citecolor=red,linkcolor=blue]{hyperref}	
\graphicspath{{./figs/}}  
\usepackage[usenames]{xcolor}
\usepackage{amsmath,amssymb,amsfonts}	
\usepackage{bm,braket,array}
\usepackage{multirow}
\usepackage[all]{xy}
\usepackage{amsthm}
\usepackage{amscd}
\usepackage{slashed} 
\usepackage{calligra} 
\usepackage[normalem]{ulem} 

\newtheorem{thm}{Theorem}

\newtheorem{assumption}[thm]{Assumption}

\theoremstyle{definition}

\theoremstyle{remark}

\def\im{{\rm Im\,}}

\def\mod{{\rm\ mod\ }}

\def\tr{{\rm tr\,}}

\def\dim{{\rm dim\,}}
\def\p{\partial}

\def\wh{\widehat}
\def\ker{{\rm Ker\,}}
\def\coker{{\rm Coker\,}}

\newcommand{\s}{\sigma}
\newcommand{\g}{\gamma}
\newcommand{\la}{\lambda}
\newcommand{\G}{\Gamma}

\newcommand{\C}{\mathbb{C}}
\newcommand{\N}{\mathbb{N}}

\newcommand{\R}{\mathbb{R}}
\newcommand{\Z}{\mathbb{Z}}

\newcommand{\bx}{{\bm{x}}}
\newcommand{\bk}{{\bm{k}}}
\newcommand{\br}{{\bm{r}}}

\newcommand{\bR}{{\bm{R}}}

\def\widebar{\accentset{{\cc@style\underline{\mskip10mu}}}} 
\def\wideubar{\underaccent{{\cc@style\underline{\mskip10mu}}}} 

\begin{document}
\title{Atiyah-Hirzebruch spectral sequence for topological insulators and superconductors: \\
$E_2$ pages for 1651 magnetic space groups}
\author{Ken Shiozaki}
\affiliation{Center for Gravitational Physics and Quantum Information, Yukawa Institute for Theoretical Physics, Kyoto University, Kyoto 606-8502, Japan}
\author{Seishiro Ono}
\thanks{Present address:~Interdisciplinary Theoretical and Mathematical Sciences Program (iTHEMS), RIKEN, Wako 351-0198, Japan}
\affiliation{Department of Applied Physics, University of Tokyo, Tokyo 113-8656, Japan}
\date{\today}
\preprint{YITP-23-17}
\preprint{RIKEN-iTHEMS-Report-23}
\begin{abstract}
We compute the $E_2$ pages of the momentum-space and real-space Atiyah-Hirzebruch spectral sequence (AHSS) for topological crystalline insulators and superconductors up to three spatial dimensions, considering the cell decomposition in which if a group action fixes a cell setwise then its group action fixes the same cell pointwise.
We provide a detailed description of the implementation for computing the $E_2$ pages of AHSS.
Under a physically reasonable assumption, we enumerate all possible $K$-groups that are compatible with the $E_2$ pages for both momentum and real-space AHSS.
As a result, we determined the $K$-groups for approximately 59\% of symmetry settings in three spatial dimensions.
All the results can be found at this http \href{https://www2.yukawa.kyoto-u.ac.jp/~ken.shiozaki/ahss/e2.html}{URL}.
\end{abstract}
\maketitle
\tableofcontents
\parskip=\baselineskip

\section{Introduction}
\label{intro}
Topological crystalline insulators and superconductors are topological phases of electronic systems protected by crystalline symmetries~\cite{SchnyderRyu_Classification2008, Kitaev_periodic2009, RyuTenFold, ChiuYaoRyu_Reflection2013, Morimoto_Clifford2013, Slager:2013aa, ShiozakiSato2014, ShiozakiSatoGomi_nonsymmorphic2016, Chiu_RMP_2016, Kruthoff_2017, Po_SymmetryIndicator2017, Bradlyn2017topological, Watanabe_1651MSG_2018, ShiozakiSatoGomi_Atiyah-Hirzebruch_2022, CornfeldChapman2019, PhysRevResearch.3.013052, MTQC, Shiozaki_PTEP2022, ZhangRenQiFang_intrinsicTSC_2022, Wire_Fang, OnoShiozakiWatanabe_classification_SC_2022}.
Classification of topological crystalline phases involves listing possible higher-order topological phases that exhibit surface, hinge, and corner states~\cite{Benalcazar_multipole_insulators2016, FangFu_RotaionAnomaly2019, Song_Fang_Fang_(d-2)-dimensional_edge2017, Langbehn_Peng_Trifunovic_vonOppen_Brouwer_second_order_2017, Schindler_Higher_order2018, TrifunovicBrouwer_HigherOrderBBC_2019, Shiozaki_homology, FuHermele_prx_point_group_2017, PhysRevB.96.205106, SongHuangQiFangHermele_topological_crystals_2019, SongFangQi_Real-space_recipes2020, freed2019invertible}.
Nowadays, it is well-known that this task is achieved by enumerating configurations of lower-dimensional topological phases protected by internal symmetries in real space~\cite{Xiong_2018, SongHuangQiFangHermele_topological_crystals_2019, SongFangQi_Real-space_recipes2020,Shiozaki_homology,ElseThorngren_Crystalline_topological_phases_as_defect_networks2019}.
While topological phases are generally defined in quantum many-body systems, free fermionic systems with translation symmetry have a unique feature: their single-particle nature allows us to use an alternative approach based on the topology of band structures in momentum space, which is dual to the real-space description.

A practical framework as a computational method for exhaustive classification is provided by $K$-theory~\cite{Kitaev_periodic2009, FreedMoore, Thiang_Ktheory2016, ShiozakiSatoGomi_crystalline_2017, Gomi2017FreedMoore}.
More precisely, the classification in momentum space is described by $K$-cohomology~\cite{Atiyah_KTheory, Karoubi_K-theory}, while the classification in real space is described by $K$-homology~\cite{Kasparov_operator_k-functor_1981, Higson2000analytic, Kubota_twisted_equivariant_K-theory2016}, and these $K$-groups are isomorphic to each other for topological crystalline insulators and superconductors~\cite{GomiKubotaThiang_crystallographic_T-duality2021}.
In the $K$-theory classification of topological insulators and superconductors, two gapped Hamiltonians $H_0$ and $H_1$ defined on a common set of atomic orbitals are regarded as in the same topological phase if there is a gapped Hamiltonian $H'$ such that there is a path from $H_0 \oplus H'$ to $H_1\oplus H'$ without closing a gap. 
This equivalence condition is called the stable equivalence. 
In the $ K$-theory, the classification is given as a $\Z$-module called the $K$-group. 
A set of pair $(H_0, H_1)$ represents an element of $K$-group, and we denote the equivalence class by $[H_0,H_1]$. 
If $[H_0,H_1] \neq 0$ as an element of $K$-group, there is no adiabatic paths between $H_0$ and $H_1$. 
Note that the inverse is not true in general: 
Two Hamiltonians $H_0$ and $H_1$ can be stably equivalent even if there is no adiabatic path between them.  

Although the weakening of the identity condition by stable equivalence gives a slightly coarser classification, $K$-theory has the technical advantage of being computationally feasible.
The $ K$-theory is a generalized (co)homology theory, meaning that one can apply various tools of the generalized (co)homology theory to compute the $K$-group we are interested in. 
For example, the Mayer-Vietoris sequence gives us a long exact sequence for a decomposition of momentum/real space $X = U \cup V$, where the $K$-group over $X$ is constrained by the $K$-groups over more small spaces $U$ and $V$~\cite{ShiozakiSatoGomi_crystalline_2017}. 
A systematic framework of this kind of bottom-up approach is the Atiyah-Hirzebruch spectral sequence (AHSS)~\cite{AtiyahHirzebruch}, which was introduced in \cite{ShiozakiSatoGomi_Atiyah-Hirzebruch_2022} for the band theory and in \cite{Shiozaki_homology} for real-space classification.
See also \cite{ElseThorngren_Crystalline_topological_phases_as_defect_networks2019, SongHuangQiFangHermele_topological_crystals_2019, SongFangQi_Real-space_recipes2020} for real-space approaches. 
In the AHSS, $E_1$-page, $E_2$-page, $E_3$-page,... are computed sequentially, and the convergent $E_\infty$-page approximates the $K$-group. 
For three-dimensional systems, $E_4$-page is the $E_\infty$-page.
While it is currently unknown how to systematically compute higher-order pages, such as $E_3$ and $E_4$, except for particular symmetry classes~\cite{OnoShiozakiWatanabe_classification_SC_2022}, the computation of $E_2$-page is easier than that of higher-order pages.

In this work, based on the physical picture and mathematical structure of the AHSS discussed in \cite{ShiozakiSatoGomi_Atiyah-Hirzebruch_2022} and \cite{Shiozaki_homology}, we propose an efficient and systematic computation of $E_2$-pages for a certain class of decomposition of space, and we present computed $E_2$-page for free fermionic insulators and superconductors in one, two, and three dimensions. 
The symmetry settings we compute are 1651 magnetic space groups (MSGs), 528 magnetic layer groups, and 393 magnetic rod groups. 
For superconductors, all one-dimensional representations of pairing symmetry are considered.
All the results can be found at the following http \href{https://www2.yukawa.kyoto-u.ac.jp/~ken.shiozaki/ahss/e2.html}{URL}.

Furthermore, we discuss a technique to find candidate $K$-groups from the $E_2$-page in the momentum-space AHSS and the real-space AHSS.
Although it is generally difficult to obtain $E_3$- and $E_4$-pages, we can tabulate all the possible $E_3$- and $E_4$- pages by considering all possible higher differentials. 
For each candidate of $E_4$-page, we can also tabulate all the possible $K$-groups compatible with the $E_4$-page. 
Importantly, the two sets of candidate $K$-groups are obtained from the momentum-space and real-space AHSSs for a symmetry setting, and the true $K$-group lies in the intersection of these two sets.
These facts give us a strong constraint on possible $K$-groups. As a result, the number of candidate $K$-groups for each symmetry setting is limited and countable. 
Surprisingly, the $K$-groups are determined for about 59\% of symmetry settings we consider in three dimensions. 

The organization of this paper is as follows. 
In Sec.~\ref{sec:symmetry and factor system}, we summarize the crystal symmetries, factor systems for electronic fermions, pairing symmetries in superconductors, and algebraic relations of symmetry actions targeted in this paper.
In Sec.~\ref{Preliminary for AHSS in general}, we describe common preliminary matters in momentum-space and real-space AHSS, particularly a class of cell decomposition used in this paper and the implementation of winding numbers for each irrep.
In Sec.~\ref{sec:Momentum-space AHSS} and Sec.~\ref{sec:Real-space AHSS}, we elaborate on the calculation details of the $E_2$ pages in the momentum-space AHSS and real-space AHSS, respectively.
In Sec.~\ref{sec:Analysis of E2 pages}, we summarize the calculation technique for imposing constraints on the possible $K$-groups from the $E_2$ pages obtained by the momentum-space and real-space AHSS.
In Sec.~\ref{sec:Other symmetry settings}, we comment on the symmetry settings other than the MSGs calculated in this paper.
We provide the conclusion in Sec.~\ref{sec:Conclusion}. 
Several computational details are summarized in Appendices of the four sections.

\section{Symmetry and factor system
\label{sec:symmetry and factor system}
}
In this paper, we compute AHSSs for symmetry groups that are either the MSGs or the combination of MSGs and Particle-Hole Symmetry (PHS). 
In this section, we summarize these symmetry groups and the factor system in momentum space.

We introduce an abbreviation to specify electronic insulators or superconductors for the physical system under consideration.
The electronic insulating system is abbreviated to TI (topological insulator), and the electric superconducting system to SC (superconductor).

\subsection{MSG}
Let ${\cal G}$ be an MSG.
We denote the lattice translation group and magnetic point group by $\Pi$ and $G={\cal G}/\Pi$, respectively.
An element of ${\cal G}$ is specified by the Seitz symbol and the homomorphism $\phi:{\cal G} \to \Z_2=\{\pm 1\}$, which are introduced below.
The group ${\cal G}$ acts on the three-dimensional Euclidean space $\mathbb{E}^3$ as $g(\bm{x}) = p_g \bm{x} + \bm{t}_g$ for $\bm{x} \in \mathbb{E}^3$, where $p_g \in O(3)$ is a three-dimensional rotation matrix and $\bm{t}_g \in \R^3$ is a vector of (fractional) translation. 
We employ the Seitz symbol $g = \{p_g|\bm{t}_g\}$ to specify an element $g \in {\cal G}$.
(It should be noted that $\{p_g|\bm{t}_g\}$ is, in fact, a representation of the group ${\cal G}$.)
They satisfy $p_gp_h=p_{gh}$ and $\bm{t}_{gh} = p_g\bm{t}_h+\bm{t}_g$ due to the group structure of ${\cal G}$.
The homomorphism $\phi$ specifies whether $g \in {\cal G}$ acts unitarily or antiunitarily on the one-particle Hilbert space. 
The symmetry group ${\cal G}$ acts as a projective representation on the one-particle Hilbert space:
Let $u_g$ be representation matrices for $g \in {\cal G}$. 
These matrices satisfy the group law, except for a $U(1)$ phase, as in 
\begin{align}
    u_gu_h^{\phi_g}=z_{g,h}u_{gh},\quad g,h \in {\cal G}.
\end{align}
Here, we introduced a short-hand notation for matrices $A$ and the sign $\phi_g$ so that $A^{\phi_g} = A$ for $\phi_g=1$ and $A^{\phi_g}=A^*$ (complex conjugate of $A$) for $\phi_g=-1$. 
The $U(1)$ phases $z_{g,h}$ are called the factor system. 
We simply assume that the factor system is independent of lattice translations.
In other words, $z_{g,h}=z_{g \tau,h \tau'}$ holds for any lattice translations $\tau,\tau' \in \Pi$.
(This is not the case for the so-called magnetic translation symmetry.)

We introduce the Bravais lattice 
\begin{align}
L := \{\tau(\bm{0}) \in \mathbb{R}^3 | \tau \in \Pi\}    
\end{align}
as the orbit of the lattice translation group for a fixed center $\bm{0}$ of a unit cell, and we specify a unit cell by $\bm{R} \in L$.
To specify an MSG, it is useful to employ a complete set of left coset representatives of $\Pi$ in ${\cal G}$, which can be written using the Seitz notation as $\{p_g|\bm{a}_g\}$ for elements $g \in G$.
From the group structure of ${\cal G}$, the vector 
\begin{align}
p_g \bm{a}_h+\bm{a}_g-\bm{a}_{gh} \in L 
\end{align}
is in the Bravais lattice. 
All the data for $\phi, p_g$, and $\bm{a}_g$ of the MSGs are available at \cite{Stokes_ISO-MAG}, and we utilized them.

\subsection{Factor system
\label{sec:Factor system}
}
Although the set of inequivalent factor systems $z_{g,h}$ is classified by the group cohomology $H^2(G,U(1)_\phi)$, where $U(1)_\phi$ is the group $U(1)$ with the left $G$-action defined as $g.z=z^{\phi_{g}}$ for $z\in U(1)$, in this paper, we restrict our scope to factor systems that are realized in realistic electronic systems. 
For electrons with integer spin, $z_{g,h} = z^{\rm sp}_{g,h}\equiv 1$.
For electrons with half-integer spin, the factor system is constructed as follows. It is enough to derive the factor system for spin 1/2. 
The rotation matrix $p_g \in O(3)$ for $g\in \mathcal{G}$ is written as $p_g = R_{\hat{n}_g,\theta_g} I^{\frac{1-\det p_g}{2}}$, where $R_{\hat{n}_g,\theta_g} \in SO(3)$ is a rotation matrix along the $\hat n_g$-axis by the $\theta_g$ angle and $I$ is the space inversion.
The space inversion trivially acts on the spin internal degrees of freedom of electrons, meaning that the space inversion $I$ does not produce a nontrivial factor system. While the space rotation and the time-reversal symmetry (TRS) act on the electron wave function by elements of $Spin(3)=SU(2)$ group as 
\begin{align}
u_g=
\begin{cases}
    e^{-i \theta_g \hat n_g \cdot \bm{\sigma}/2} & \phi_g=1, \\
    e^{-i \theta_g \hat n_g \cdot \bm{\sigma}/2} (i\sigma_y) & \phi_g=-1. \\
\end{cases}
\end{align}
Here, $\bm{\sigma}=(\sigma_x,\sigma_y,\sigma_z)$ are the Pauli matrices. Using explicit representation matrices above, we have the factor system $z^{\rm sp}_{g,h}$ by the defining relation $u_g u_h^{\phi_g}=z^{\rm sp}_{g,h}u_{gh}$. As a result, the factor systems $z^{\rm sp}_{g,h}$ take values in the sign $\pm 1$.

\subsection{Symmetry action in momentum space}
Let $\hat c_\s(\bm{R}+\delta \bm{r})$ and $\hat c^\dag_\s(\bm{R}+\delta \bm{r})$ be annihilation and creation operators of a complex fermion localized at the position $\bm{R}+\delta \bm{r}$. 
Here, $\bm{R} \in L$ represents the center position of unit cells, $\delta \br$ is the displacement vector from the unit cell center for the fermions, and the subscript $\s$ is the index for the internal degrees of freedom like spin and orbital at $\bR+\delta \br$. 
Let ${\cal G}$ be an MSG. 
An element $g \in {\cal G}$ acts on the fermions as 
\begin{align}
    &\hat g \hat c^\dag_\s(\bm{R}+\delta \bm{r}) \hat g^{-1}\nonumber \\
    &= \hat c^\dag_{\s'}(p_g(\bm{R}+\delta \bm{r})+\bm{t}_g) [D_g]_{\s'\s}, 
    \label{eq:msg_def}
\end{align}
Here, $\hat g$ is unitary when $\phi_g=1$, and antiunitary when $\phi_g=-1$, and $D_g$ matrices are a set of unitary matrices satisfying 
\begin{align}
    D_g D_h^{\phi_g} = z_{g,h} D_{gh} 
\end{align}
with the factor system $z_{g,h}$ introduced before and does not depend on lattice translations. 

In this paper, we introduce the fermion annihilation and creation operators in momentum space so that they are periodic by reciprocal lattice vectors. 
Namely, 
\begin{align}
   \hat c^\dag_{\delta \br,\s}(\bk) 
   = \sum_{\bR\in\Pi} \hat c^\dag_\s(\bm{R}+\delta \br) e^{i\bk \cdot \bR}.
   \label{eq:ck_def}
\end{align}
(A reciprocal lattice vector is a vector in the dual lattice $\hat L:= \{{\bf G} \in \R^3 | e^{i \bm{{\bf G}} \cdot \bm{R}}=1 {\rm\ for\ all\ }\bm{R} \in L\}$.)
This definition does not reflect the spatial position of the degrees of freedom, which requires caution when calculating physical quantities, etc. However, since it does not affect the classification of topological phases, we adopt this definition in this paper.
Let us introduce a permutation matrix 
\begin{align}
    [P_g]_{\delta \br',\delta \br} = \left\{
    \begin{array}{ll}
       1  & \delta \br' \equiv p_g \delta \br + \bm{t}_g \mod L, \\
       0  & {\rm else}, 
    \end{array}
    \right.
\end{align}
for $g \in {\cal G}$, we have for $g \in {\cal G}$, 
\begin{align}
    \hat g \hat c^\dag_i(\bk) \hat g^{-1}
    = \sum_{i'} \hat c^\dag_{i'}(\phi_g p_g \bk) [u_g(\bk)]_{i'i}
\end{align}
with 
\begin{align}
    &[u_g(\bk)]_{i'i}\nonumber\\
    &=[P_g]_{\delta \br'\delta \br} [D_g]_{\s'\s} e^{-i\phi_g p_g \bk \cdot (p_g \delta \br +\bm{t}_g-\delta \br')}, 
\end{align}
where the indices $\delta \br$ and $\s$ were merged as a single index $i = (\delta \br,\s)$. 
In particular, $[u_{\tau}(\bk)]_{i'i} = \delta_{i'i} e^{-i \bk \cdot \bm{t}}$ for lattice translations $\tau = \{1|\bm{t}\} \in \Pi$.
Note that $u_g(\bk)$ is periodic $u_g(\bk+{\bf G})=u_g(\bk)$ for reciprocal lattice vectors ${\bf G}$. 
It is straightforward to show that 
\begin{align}
    u_g(\phi_g p_g \bk) u_h(\bk)^{\phi_g} = z_{g,h} u_{gh}(\bk) 
\end{align}
for $g,h \in {\cal G}$.

A free fermion Hamiltonian expressed in the momentum space is 
\begin{align}
    \hat H = \sum_{\bk,ij} c^\dag_i(\bk) [h(\bk)]_{ij} c_j(\bk).
\end{align}
The matrix $h(\bk)$ is also called a Hamiltonian. 
With the definition (\ref{eq:ck_def}), $h(\bk)$ is also periodic $h(\bk+\bm{G})=h(\bk)$. 
For free fermions, the MSG symmetry of the Hamiltonian $\hat H$ is that the matrix $h(\bk)$ satisfies 
\begin{align}
    u_g(\bk)h(\bk)^{\phi_g} u_g(\bk)^{-1} = h(\phi_g p_g \bk) 
\end{align}
for $g \in {\cal G}$. 
Note that the lattice translation symmetry is fulfilled as it is written in momentum space. 
Only the constraint conditions coming from the magnetic point group $G$ are meaningful.

\subsection{Symmetry of gap function
\label{sec:Symmetry of gap function}
}
In superconductors, the mean-field Hamiltonian in momentum space is written as 
\begin{align}
    \hat H_{\rm MF}
    &= \sum_{\bk,ij} c^\dag_i(\bk) [h(\bk)]_{ij} c_j(\bk) \nonumber \\
    &+ \frac{1}{2}\sum_{\bk,ij} (c^\dag_i(\bk) [\Delta(\bk)]_{ij} c^\dag_j(-\bk)+{\rm h.c.}).
\end{align}
The matrix $\Delta(\bk)$ is the gap function and satisfies 
\begin{align}
\Delta(\bk)^T = - \Delta(-\bk) \label{eq:gap_tr}    
\end{align}
due to the anticommutation relation of fermion operators.
The gap function $\Delta(\bk)$ is supposed to be a vector $\Delta(\bk)= \sum_{a=1}^{\dim \rho} \eta_a \Delta_a(\bk), \eta_a \in \C,$ of some basis functions $\{\Delta_a(\bk)\}_{a=1}^{\dim \rho}$ satisfying 
\begin{align}
    &u_g(\bk) \Delta_a(\bk)^{\phi_g} u_g(-\bk)^T \nonumber\\
    &= \sum_{b=1}^{\dim \rho} \Delta_b(\phi_g p_g \bk) [D^\rho_g]_{ba}
    \label{eq:gap_func_rep}
\end{align}
for $g \in {\cal G}$. 
We assume $D^\rho_g$ is independent of $\bk$ and lattice translations, meaning that $\rho$ is a representation of the magnetic point group $G$. 
From (\ref{eq:gap_func_rep}), the factor system for $D^\rho_g$ is $(z_{g,h})^2$ and thus trivial for spinless and spinful electrons. 
Since different irreps do not coexist as a solution of the gap equation in general, $\rho$ is an irrep of the magnetic point group $G$. 
When $\rho$ is not the trivial irrep of $G$, meaning that $D^\rho_g \neq 1$ for some $g \in {\cal G}$, the gap function $\Delta(\bk)$ breaks the original MSG symmetry defined by (\ref{eq:msg_def}).
However, when $\rho$ is a one-dimensional irrep of $G$, one can recover the MSG symmetry using $U(1)$ phase rotation $\hat {\sf u}(e^{i\theta}) c^\dag_i \hat {\sf u}(e^{i\theta})^{-1}=c^\dag_i e^{i\theta}$ of complex fermions.
Let us write $D^\rho_g = \xi_g\in U(1)$ for a one-dimensional irrep.
For each $\xi_g$, we pick a sign of the square root of $\xi_g$ and denote it by $\xi_g^{\frac{1}{2}}$. 
Then, the combined transformation $\hat g \hat {\sf u}((\xi_g^{\frac{1}{2}})^*)$ becomes symmetry of the Hamiltonian $\hat H_{\rm MF}$.

The mean-field Hamiltonian $\hat H_{\rm MF}$ can be written as 
\begin{align}
    \hat H_{\rm MF}
    &=\frac{1}{2}\sum_{\bk,ij} (c^\dag_i(\bk),c_i(-\bk)) \nonumber \\
    &[H_{\rm BdG}(\bk)]_{ij}
    \begin{pmatrix}
        c_j(\bk) \\
        c^\dag_j(-\bk)\\
    \end{pmatrix}, \nonumber
\end{align}
\begin{align}
    H_{\rm BdG}(\bk)
    =\begin{pmatrix}
        h(\bk)&\Delta(\bk)\\
        \Delta(\bk)^\dag&-h(-\bk)^T\\
    \end{pmatrix}.
\end{align}
The two-component spinor $\Psi_i(\bk) = (c_i(\bk),c^\dag_i(-\bk))^T$ and the matrix $H_{\rm BdG}(\bk)$ are called the Nambu spinor and the Bogoliubov-de Gennes (BdG) Hamiltonian, respectively. 
The relation (\ref{eq:gap_tr}) implies that the BdG Hamiltonian $H_{\rm BdG}(\bk)$ satisfies the following particle-hole ``symmetry" 
\begin{align}
    &U_c H_{\rm BdG}(\bk)^* U_c^{-1} = - H_{\rm BdG}(-\bk), \nonumber\\
    &U_c = \begin{pmatrix}
        &1\\
        1\\
    \end{pmatrix}. \label{eq:D_phs}
\end{align}
Therefore, the total symmetry group for the BdG Hamiltonian becomes ${\cal G} \times \Z_2^C$ with $\Z_2^C$ generated by PHS. 
Because $U_C U_C^*=1$ we call the PHS in the form (\ref{eq:D_phs}) the class D PHS in the Altland-Zirnbauer (AZ symmetry class~\cite{AltlandZirnbauer}. 
The one-dimensional irrep $\xi_g$ is encoded in the factor system of ${\cal G} \times \Z_2^C$. 
On the Nambu spinor, the combined symmetry is 
\begin{align}
    &\hat g \hat {\sf u}((\xi_g^{\frac{1}{2}})^*) \Psi^\dag(\bk) (\hat g \hat {\sf u}((\xi_g^{\frac{1}{2}})^*))^{-1} \nonumber\\
    &=\Psi^\dag(\phi_g p_g \bk) U_g(\bk) 
\end{align}    
with
\begin{align}
    U_g(\bk)=\begin{pmatrix}
        u_g(\bk) (\xi_g^{\frac{1}{2}})^* \\
        &u_g(-\bk)^* \xi_g^{\frac{1}{2}} \\
    \end{pmatrix}
\end{align}
for $g \in {\cal G}$.
We find that 
\begin{align}
    &U_g(\phi_h p_h \bk)U_h(\bk)^{\phi_g}
    =
    z_{g,h} z_{g,h}^{\xi} U_{gh}(\bk), \label{eq:UgUh} \\
    &U_c U_g(\bk)^* = U_g(-\bk) U_C^{\phi_g}, 
\end{align}
with 
\begin{align}
    z^\xi_{g,h}:= \xi_g^{\frac{1}{2}}(\xi_h^{\frac{1}{2}})^{\phi_g} (\xi_{gh}^{\frac{1}{2}})^{-1} \in \{\pm 1\}.
    \label{eq:facsys_PHS}
\end{align}
In (\ref{eq:UgUh}) we have used that $z_{g,h}$ is a sign so that $z_{g,h}^*=z_{g,h}$. 

Alternatively, the following phase choice, which is meaningful only for BdG Hamiltonian, is useful. 
\begin{align}
    U'_g(\bk)=\begin{pmatrix}
        u_g(\bk)\\
        &u_g(-\bk)^* \xi_g \\
    \end{pmatrix}.
\end{align}
With this, 
\begin{align}
    &U'_g(\phi_h p_h \bk)U'_h(\bk)^{\phi_g} = z_{g,h} U'_{gh}(\bk), \\
    &U'_g(-\bk) U_C^{\phi_g}=\xi_g U_C U'_g(\bk)^*.
\end{align}

\subsection{$SU(2)$ symmetry and class C}
Consider the cases in the presence of full $SU(2)$ internal symmetry of normal state $h(\bk)$ for either the spin-1/2 or a pseudo-spin-1/2 internal degree of freedom.
We denote the Pauli matrices for the (pseudo) spin-1/2 degrees of freedom by $\bm{\s}$. 
The normal part is written as $h(\bk) = \tilde h(\bk) \otimes \s_0$. 
When the gap function also preservers $SU(2)$ symmetry, the gap function is in the form 
\begin{align}
    \Delta(\bk)
    = \tilde \Delta(\bk) \otimes (i\s_y). 
\end{align}
The relation (\ref{eq:gap_tr}) means that 
\begin{align}
    \tilde \Delta(\bk)^T = \tilde \Delta(-\bk). \label{eq:gap_tr_su2}
\end{align}
On the basis of Nambu spinor $\Psi'_i(\bk) = (c_i(\bk),(i\s_y) c_i^\dag(-\bk))^T$, where $i$ denotes the internal degrees of freedom excluding the (pseudo) spin-1/2, the BdG Hamiltonian is
\begin{align}
    &H_{\rm BdG}(\bk)
    =\tilde H_{\rm BdG}(\bk) \otimes \s_0, \\
    &\tilde H_{\rm BdG}(\bk)=
    \begin{pmatrix}
        \tilde h(\bk) & \tilde \Delta(\bk)\\
        \tilde \Delta(\bk)^\dag & -\tilde h(-\bk)^T.
    \end{pmatrix}
\end{align}
The relation (\ref{eq:gap_tr_su2}) implies that the BdG Hamiltonian $\tilde H(\bk)$ satisfies the following PHS 
\begin{align}
&\tilde U_C \tilde H_{\rm BdG}(\bk)^* \tilde U_C^{-1}
=-\tilde H_{\rm BdG}(-\bk), \\
&\tilde U_C = \begin{pmatrix}
        &1\\
        -1\\
    \end{pmatrix}, \quad \tilde U_C \tilde U_C^*=-1.
\end{align}
This is the class C PHS~\cite{SchnyderRyu_Classification2008}. 

If the $SU(2)$ symmetry in the electron system comes from the electron spin, the BdG Hamiltonian $\tilde H_{\rm BdG}(\bk)$ satisfies the factor system for spinless electron systems.
On the other hand, if the $SU(2)$ symmetry in the electron system comes from the pseudo-spin originating from an orbital degree of freedom, the BdG Hamiltonian $H_{\rm BdG}(\bk)$ satisfies the factor system for spinful electron systems.
We call the former cases the class C spinful SC and the latter class C spinless SC, respectively.

\section{Preliminary for AHSS in general
\label{Preliminary for AHSS in general}
}

We describe common preliminaries in momentum-space and real-space AHSS, including a class of cell decomposition used in this paper and the definition of the symmetry-resolved winding number. 

\subsection{Symmetry in one-particle Hilbert space}
Although this paper focuses only on symmetries and factor systems in electron systems, we summarize here the more general symmetry classes~\cite{FreedMoore}.

Let ${\cal G}$ be a discrete group that fits into the short exact sequence 
\begin{align}
    \Pi \to {\cal G} \to G 
\end{align}
with $\Pi \cong \Z^d$ being the translational group in $d$-space dimensions.
A symmetry class is characterized by ${\cal G}$ with the quintet $(p_g,\bm{t}_g,\phi_g,c_g,z^{\rm int}_{g,h})$ for $g,h \in {\cal G}$, explained below. 
The matrix $p_g$ is an $O(d)$ matrix, and $\bm{t}_g \in \R^d$ is a translation vector, meaning that $g \in {\cal G}$ acts on the real space as $\bm{x} \mapsto g(\bm{x}) = p_g \bm{x}+\bm{t}_g$. 
We denote the one-particle Hilbert space on a $d$-dimensional lattice by ${\cal H}$ and the symmetry action on ${\cal H}$ by $\hat g$.
The homomorphism $\phi:{\cal G} \rightarrow \{\pm 1\}$ specifies whether $\hat g$ is unitary or antiunitary on ${\cal H}$. 
We assume that $\phi(\Pi)=\{1\}$, i.e., the translation group $\Pi$ is composed only of unitary elements. 
The factor system $z^{\rm int}_{g,h}$ specifies how ${\cal G}$ is represented projectively on the Hilbert space ${\cal H}$, such that $\hat g \hat h = z^{\rm int}_{g,h} \wh{gh}$. 
We assume that $z^{\rm int}_{g,h}$ does not depend on translations in the sense that $z^{\rm int}_{gt,gt'}=z^{\rm int}_{g,h}$ for $\tau,\tau' \in \Pi$, meaning that $z$ is a two-cocycle $z^{\rm int} \in Z^2(G,U(1)_\phi)$, where $U(1)_\phi$ means that $g \in G$ acts on $U(1)$ as $g.z=z^{\phi_g}$ for $z \in U(1)$. 
Let $\hat H$ be a Hamiltonian on ${\cal H}$. 
The homomorphism $c:{\cal G}\to \{\pm 1\}$ specifies whether $\hat g$ commutes or anticommutes with the Hamiltonian $\hat H$, i.e., 
\begin{align}
    \hat g \hat H \hat g^{-1} = c_g \hat H,\quad g \in {\cal G}. 
\end{align}
We also assume that $c(\Pi)=\{1\}$, i.e., lattice translations commute with the Hamiltonian $\hat H$. 

Now we summarize the symmetry classes discussed in this paper.

\subsubsection{TIs}
For TIs, a symmetry class is specified by an MSG ${\cal G}_{\rm MSG}$ and whether it is spinless or spinful. 
The group ${\cal G}$ is an MSG ${\cal G}_{\rm MSG}$ equipped with the data $p_g,\bm{t}_g,\phi_g$.
For spinless TIs, the factor system $z^{\rm int}_{g,h}$ is a trivial one, $z^{\rm int}_{g,h} \equiv 1$.
For spinful TIs, the factor system is that for spin-1/2 electrons, $z^{\rm int}_{g,h}=z^{\rm sp}_{g,h} \in \{\pm 1\}$, defined in Sec.~\ref{sec:Factor system}. 

\subsubsection{SCs}
For SCs, a symmetry class is specified by an MSG ${\cal G}_{\rm MSG}$, a one-dimensional irrep $\xi$ of the magnetic point group $G_{\rm MPG}$ of ${\cal G}_{\rm MSG}$, whether it is spinless or spinful, and the type of PHS, either class D or class C. 
The total symmetry group ${\cal G}$ is the product ${\cal G} = {\cal G}_{\rm MSG} \times \Z_2^C$ with $\Z_2^C = \{e,C\}$ being the group of PHS. 
PHS is an internal symmetry, meaning that the PHS $C$ does not change the spatial position $C(\bm{x})=\bm{x}$, i.e., $p_C={\bf 1}_d$ and $\bm{t}_C=\bm{0}$. 
As discussed in Sec.~\ref{sec:Symmetry of gap function}, PHS behaves as an antiunitary symmetry and anticommutes with the BdG Hamiltonian, meaning that $\phi_C=-1$ and $c_C=-1$. 
For generic elements $g \in {\cal G}$, $p_g,\bm{t}_g,\phi_g,$ and $c_g$ are extended with the group structure. 

For a given irrep $\xi$, let $z_{g,h}^{\xi}$ for $g,h \in {\cal G}_{\rm MSG}$ be the factor system introduced in (\ref{eq:facsys_PHS}). 
Let $\pi_1:{\cal G} \to {\cal G}_{\rm MSG}$ and $\pi_2:{\cal G} \to \Z_2^C$ be the projections onto ${\cal G}_{\rm MSG}$ and $\Z_2^C$, respectively. 
The factor system can be summarized in the form 
\begin{align}
    &z^{\rm int}_{g,h} = z^{\rm sp}_{\pi_1(g),\pi_1(h)} z^{\xi}_{\pi_1(g),\pi_1(h)} z^{\rm PHS}_{\pi_2(g),\pi_2(h)},\nonumber \\
    &g,h \in {\cal G}. 
\end{align}
Here, $z^{\rm PHS}_{g,h} \equiv 1$ for class D, and 
\begin{align}
    z^{\rm PHS}_{g,h} = \left\{\begin{array}{ll}
        -1 & g=h=C, \\
        1 & {\rm else},  
    \end{array} \right.
\end{align}
for class C.

\subsection{Cell decomposition
\label{sec:Cell decomposition}
}
In this section, we introduce a class of cell decompositions used for the AHSS in this paper. 
Let $X$ be a space over which we want to compute the $K$-group. 
The space $X$ is either the Brillouin zone (BZ) torus $T^d$ or the infinite real space $\R^d$, where $d$ is the space dimension. 
Let $G$ be a symmetry group acting on $X$. 
For $d=3$, the group $G$ is the MSG for the real space, whereas $G$ is the magnetic point group for the momentum space.
We introduce a sequence of spaces
\begin{align}
\emptyset = X_{-1} \subset X_0 \subset X_1 \subset \cdots \subset X_d =X
\end{align}
such that each $X_p$ is obtained from $X_{p-1}$ by gluing $p$-cells $D^p_i$, which are each homeomorphic to a $p$-dimensional disk $D^p$, to $X_{p-1}$ along their boundary $(p-1)$-dimensional spheres $\p D^p_i \cong S^{p-1}$. Additionally, there is a symmetry constraint on the $p$-cells:
For each $g \in G$, $p$-cells $D^p_i$ are mapped to other $p$-cells by $g$.
In other words, for each $g\in G$, $g(D^p_i) = D^p_j$ holds with some $j$. 
Each $X_p$ is called the $p$-skeleton. 
We refer to such a decomposition of the space $X$ as a cell decomposition. 

While the AHSS is defined for the above cell decomposition, we impose the following additional condition on the cell decomposition in this paper: 

-- If $g \in G$ fixes the $p$-cell $D^p_i$ setwise, then $g$ fixes $D^p_i$ pointwise. 
Namely, if $g(D^p_i) = D^p_i$, then $g(x) = x$ for all $x \in D^p_i$. 

Note that even with the additional constraint, the cell decomposition is not unique. 
Nevertheless, the $E_2$-page is known to be unique.

In addition, we assign an orientation to each $p$-cell in such a way as to satisfy the symmetry.
All orientations of 0-cells are fixed to be positive. 
Orientation is not necessary for the AHSS in general, but it is required to construct the first differential $d_1$, which will be developed later.

In Appendix \ref{app:An algorithm computing cell decomposition}, we present an algorithm for computing cell decomposition such that the condition above is satisfied and the cell of maximum dimension ($d$-cell) is a convex fundamental domain.
For instance, FIGs.~\ref{fig:kcell_P2_11'} and \ref{fig:rcell_P2_11'} show fundamental domains in the momentum and real spaces for MSG $P2_11'$.

\subsection{Chiral symmetry and winding number}
In odd spatial dimensions, one can define the winding number in the presence of chiral symmetry. 
Let $G = G_0 \coprod \gamma G_0$ be an internal symmetry group composed of unitary and chiral type symmetry such that 
\begin{align}
    u_g H(\bk) u_g^{-1}=\left\{\begin{array}{ll}
        H(\bk) & g \in G_0,  \\
        -H(\bk) & g \in \gamma G_0,
    \end{array}\right. 
\end{align}
where $\gamma$ is a representative element of $\{g \in \mathcal{G} \vert \phi_g = -c_g = +1 \}$. 
Let $\alpha$ be an irrep of $G_0$. 
If the mapped irrep $\gamma[\alpha]$ is unitarily equivalent to $\alpha$, one can define the winding number $W^\alpha_{2n-1}$ for the irrep $\alpha$ as follows. 
The construction of the winding number in this section is based on \cite{Shiozaki_PTEP2022}.

Let $z_{g,h} \in Z^2(G,U(1))$ be a factor system of projective representation appearing as $u_g u_h = z_{g,h} u_{gh}$ for $g,h \in G$. 
Let $\chi^\alpha_{g\in G_0}$ be the character of the irrep $\alpha$ of $G_0$. 
The character of the mapped irrep $\gamma [\alpha]$ is given by $\chi^{\gamma [\alpha]}_{g\in G_0} = \frac{z_{g,\gamma}}{z_{\gamma,\gamma^{-1}g\gamma}} \chi^\alpha_{\gamma^{-1}g\gamma}$.
If $\sum_{g\in G_0} (\chi^\alpha_g)^*\chi^{\gamma [\alpha]}_g = 1$, the two irreps $\alpha$ and $\gamma[\alpha]$ are unitarily equivalent to each other. 
If this is the case, there are exactly two irreps $\alpha+$ and $\alpha-$ of $G_0 \coprod \gamma G_0$ such that the character $\chi^{\alpha\pm}_g$ of the irrep $\alpha \pm$ satisfies 
\begin{align}
    \chi^{\alpha \pm}_{g \in G_0} =\chi^\alpha_g,\quad
    \chi^{\alpha -}_{g \in \gamma G_0} =-\chi^{\alpha +}_g.
\end{align}
The orthogonality $\frac{1}{|G_0 \coprod \gamma G_0|} \sum_{g \in G_0 \coprod \gamma G_0} (\chi^{\alpha+}_g)^* \chi^{\alpha-}_g=0$ leads to 
\begin{align}
    \frac{1}{|G_0|} \sum_{g \in \gamma G_0} (\chi^{\alpha+}_g) \chi^{\alpha \pm}_g = \pm 1.
    \label{eq:chiral_sign}
\end{align}
With these characters, we introduce the projection $P_{\alpha\pm}$ onto the $\alpha\pm$ irrep as 
\begin{align}
    P_{\alpha\pm} = \frac{{\rm dim}(\alpha\pm)}{|G|} \sum_{g \in G} (\chi^{\alpha\pm}_g)^* u_g.
\end{align}
here, ${\rm dim}(\alpha\pm)$ is the dimension of the representation $\alpha \pm$. 
The chiral matrix of the irrep $\alpha$ is defined as
\begin{align}
\Gamma_\alpha = P_{\alpha+}-P_{\alpha-},
\end{align}
and the winding number $W^\alpha_{2n-1}$ of the irrep $\alpha$ is given by
\begin{align}
W^\alpha_{2n-1}=\frac{n!}{(2\pi i)^n(2n)!} \int \tr[(H^{-1}dH)^{2n-1}\Gamma_\alpha].
\end{align}
For convenience, here, we have written $\alpha+$ and $\alpha-$, but there is no way to choose one or the other as $\alpha+$ or $\alpha-$.
Interchanging $\alpha+$ and $\alpha-$ flips the sign of the chiral matrix $\Gamma_\alpha$ and the winding number $W^\alpha_{2n-1}$.
Therefore, the sign of the winding number $W^\alpha_{2n-1}$ depends on the choice of the sign of the chiral matrix $\Gamma_\alpha$, which must be properly incorporated in the AHSS formulated in later sections.
The following expression of $W^\alpha_{2n-1}$ is also useful.
\begin{align}
W^\alpha_{2n-1}
&=\frac{n!}{(2\pi i)^n(2n)!} \frac{1}{|G_0|} \sum_{g \in \gamma G_0} (\chi^{\alpha+}_g)^* \nonumber \\
&\times \int \tr[(H^{-1}dH)^{2n-1}u_g].
\end{align}

\section{Momentum-space AHSS
\label{sec:Momentum-space AHSS}
}
We provide a method for calculating the first differential $d_1$ of the momentum-space AHSS~\cite{ShiozakiSatoGomi_Atiyah-Hirzebruch_2022}. 
In this section, we refer to $g, h, \dots$ as elements of the group $G = \mathcal{G}/\Pi$.

In momentum space, we can consider the symmetry group of the Hamiltonian $H(\bk)$ as the group $G$.
Let $\{\{p_g | \bm{a}_g\}\}_{g \in G}$ be a set of left coset representatives of $\Pi$ in $\mathcal{G}$. 
The symmetry constraints on the Hamiltonian $H(\bk)$ are written as 
\begin{align}
    u_g(\bk) H(\bk)^{\phi_g} u_g(\bk)^{-1} = c_g H(\phi_g p_g \bk),\quad g \in G, 
\end{align}
together with the $\bk$-dependent factor system 
\begin{align}
    &u_g(\phi_h p_h\bk)u_h(\bk)^{\phi_g}=z_{g,h}(\bk) u_{gh}(\bk),\\
    &z_{g,h}(\bk):=z^{\rm int}_{g,h} e^{-i\phi_{gh}p_{gh}\bk \cdot (p_g\bm{a}_h+\bm{a}_g-\bm{a}_{gh})} 
\end{align}
for $g, h \in G$. 

What we aim to compute is the twisted equivariant $K$-group ${}^\phi K_G^{(z,c)-n}(X)$ over the BZ torus $X$ by Freed and Moore~\cite{FreedMoore}. 
Here, $n$ takes values in integers, and the $K$-group enjoys the Bott periodicity ${}^\phi K_G^{(z,c)-(n+8)}(X) \cong {}^\phi K_G^{(z,c)-n}(X)$. 
Roughly speaking, the integers $n$ have the following physical meanings: 
$n=1, n=0$, and $n=-1$ correspond to gauge transformations in momentum space, gapped Hamiltonians, and gapless Hamiltonians realized only as a boundary of gapped Hamiltonians, respectively.
In particular, the $0$th K-group ${}^\phi K_G^{(z,c)-0}(X)$ represents the classification of gapped Hamiltonians. 

Hereafter, we set the spatial dimension to three.

\subsection{Preliminary}
For a cell decomposition $X_0 \subset X_1 \subset X_2 \subset X_3 = X$ introduced in Sec.~\ref{sec:Cell decomposition}, the $E_1$-page $E_1^{p,-n}$ is defined as the relative $K$-group 
\begin{align}
    E_1^{p,-n}
    := {}^\phi K^{(z,c)+p-n}_G(X_p,X_{p-1}). 
\end{align}
We denote the label set of orbits of $p$-cells by $I^p_{\rm orb}$. 
Since the $p$-skeleton is obtained by gluing $p$-cells to the $(p-1)$-skeleton $X_{p-1}$, this is the direct sum of the $K$-groups over each orbit 
\begin{align}
    &E_1^{p,-n} \cong {}^\phi K_G^{(z,c)+p-n} (\coprod_{{\sf a} \in I^p_{\rm orb}} G/G_{D^p_{\sf a}} \times D^p_{\sf a},\nonumber \\
    &\hspace{50pt} \coprod_{{\sf a} \in I^p_{\rm orb}} G/G_{D^p_{\sf a}} \times \partial D^p_{\sf a}) \nonumber\\
    &\cong 
    \bigoplus_{{\sf a} \in I^p_{\rm orb}} {}^{\phi|_{D^p_{\sf a}}} K^{(z|_{D^p_{\sf a}},c|_{D^p_{\sf a}})+p-n}_{G_{D^p_{\sf a}}}(D^p_{\sf a},\partial D^p_{\sf a}). 
\end{align}
Here, $D^p_{\sf a}$ is a $p$-cell corresponding to a representative orbit ${\sf a} \in I^p_{\rm orb}$, $G_{D^p_{\sf a}} = \{g \in G|\phi_g p_g \bk \equiv \bk {\rm\ for\ } \bk \in D^p_{\sf a}\}$ is the little group of $G$ for the $p$-cell $D^p_{\sf a}$, and $\phi|_{D^p_{\sf a}},c|_{D^p_{\sf a}},z|_{D^p_{\sf a}}$ are symmetry data restricted in the $p$-cell $D^p_{\sf a}$. 
Since the little group $G_{D^p_{\sf a}}$ fixes the $p$-cell $D^p_{\sf a}$ pointwise, $E_1^{p,-n}$ is further simplified as 
\begin{align}
    E_1^{p,-n} 
    &\cong \bigoplus_{{\sf a} \in I^p_{\rm orb}} {}^{\phi|_{D^p_{\sf a}}} \tilde K^{(z|_{D^p_{\sf a}},c|_{D^p_{\sf a}})+p-n}_{G_{D^p_{\sf a}}}(D^p_{\sf a}/\p D^p_{\sf a}) \nonumber \\
    &\cong \bigoplus_{{\sf a} \in I^p_{\rm orb}} {}^{\phi|_{\bk^p_{\sf a}}} K^{(z|_{\bk^p_{\sf a}},c|_{\bk^p_{\sf a}})-n}_{G_{\bk^p_{\sf a}}}(\{\bk^p_{\sf a}\}).
\end{align}
Thus, the group $E_1^{p,-n}$ is the direct sum of $K$-groups over a point $\bk^p_{\sf a} \in D^p_{\sf a}$ for each with the degree shift by $-n$. 

There are several ways to represent the $K$-group ${}^{\phi_{\bk^p_{\sf a}}}K^{(z|_{\bk^p_{\sf a}},c|_{\bk^p_{\sf a}})-n}_{G_{\bk^p_{\sf a}}}(\{\bk^p_{\sf a}\})$. 
One is adding $n$ chiral symmetries to the symmetry group $G_{\bk^p_{\sf a}}$ for the gapped Hamiltonians $H(\bk^p_{\sf a})$ over the point $\bk^p_{\sf a}$~\cite{ShiozakiSatoGomi_crystalline_2017}. 
In the following, instead of adding $n$ chiral symmetries, we use suspension isomorphism
\begin{align}
    &{}^{\phi|_{\bk^p_{\sf a}}} K^{(z|_{\bk^p_{\sf a}},c|_{\bk^p_{\sf a}})-n}_{G_{\bk^p_{\sf a}}}(\{\bk^p_{\sf a}\})\nonumber \\
    &\cong 
    {}^{\phi|_{\bk^p_{\sf a}}} \tilde K^{(z|_{\bk^p_{\sf a}},c|_{\bk^p_{\sf a}})}_{G_{\bk^p_{\sf a}}}(\{\bk^p_{\sf a}\} \times \tilde S^n)
    \label{eq:suspension_iso_k_K}
\end{align}
to represent the $K$-group by gapped Hamiltonians on the $n$-dimensional virtual sphere $\tilde S^n$ without changing the symmetry group $G_{\bk^p_{\sf a}}$, where $G_{\bk^p_{\sf a}}$ trivially acts on the sphere $\tilde S^n$. 

Now, we develop the detail of the calculation.
First, we note that for the calculation of the first differential $d_1$, it is sufficient to consider only the $p$-cells that intersect with the closure of the fundamental domain of the BZ. 
The first differential $d_1^{p,-n}$ is defined as the composition of homomorphisms 
\begin{align}
    &d_1^{p,-n}
    :{}^\phi K_G^{(z,c)+p-n}(X_p,X_{p-1}) 
    \to {}^\phi K_G^{(z,c)+p-n}(X_p) \nonumber \\
    &\to {}^\phi K_G^{(z,c)+p-n+1}(X_{p+1},X_{p}) = E_1^{p+1,-n}.
\end{align}
The first line is induced homomorphism of the inclusion $X_p \to (X_p,X_{p-1})$, and the second line is the connected homomorphism.
The relation 
\begin{align}
    d_1^{p+1,-n} \circ d_1^{p,-n} =0
\end{align}
holds. 
The $E_2$-page is defined as 
\begin{align}
    E_2^{p,-n} := \ker d_1^{p,-n}/\im d_1^{p-1,-n}. 
\end{align}
(For $p=0$, $E_2^{0.-n} = \ker d_1^{0,-n}$.)
The physical meaning of $d_1^{p,-n}$ is that the gap closes inside $p$-cells and creates gapless points in adjacent $(p+1)$-cells. 
Therefore, for a $(p+1)$-cell, only adjacent $p$-cells contribute to $d_1^{p,-n}$. 
Starting from the 3-cell, the fundamental domain, only the 2-cells on the boundary of the 3-cell contribute to $d_1^{2,-n}$, only the 1-cells on the boundary of these 2-cells contribute to $d_1^{1,-n}$, and finally, only the $0$-cells on the boundary of these 1-cells contribute to $d_1^{0,-n}$. 
This means that one can compute $d_1^{p,-n}$ with the $p$-cells adjacent to the fundamental domain. 

\begin{figure}[!]
\centering
\includegraphics[width=0.5\linewidth, trim=0cm 0cm 0cm 0cm]{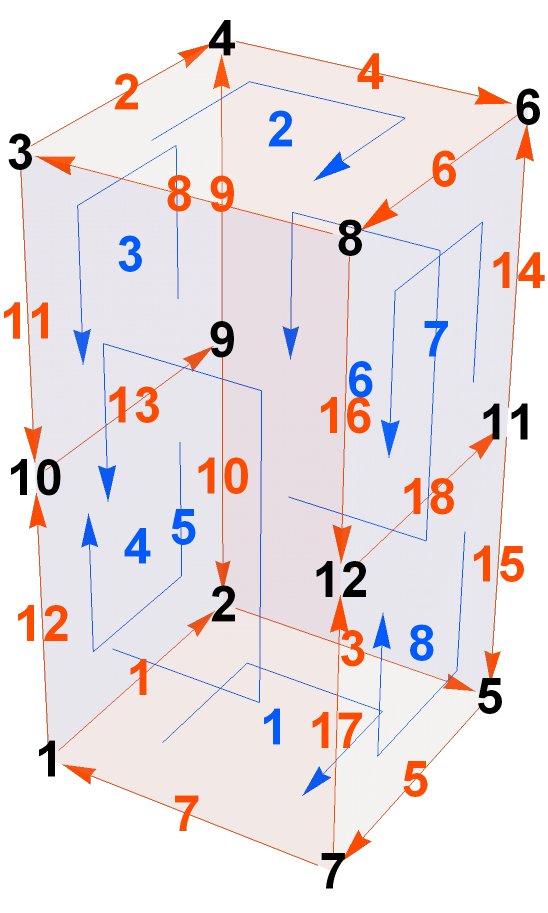}
\caption{
A cell decomposition of BZ for MSG $P2_11'$. 
The figure shows the fundamental domain (3-cell). 
The 0-, 1-, and 2-cells are shown with black, red, and blue Arabic numerals, respectively. 
The arrows of the 1- and 2-cells show orientations, which respect symmetry. 
The momenta of 0-cells are $(0,0,-\pi )$, $(0,\pi ,-\pi )$, $(0,0,\pi )$, $(0,\pi ,\pi )$, $(\pi ,\pi ,-\pi )$, $(\pi ,\pi ,\pi )$, $(\pi ,0,-\pi )$, $(\pi ,0,\pi )$, $(0,\pi ,0)$, $(0,0,0)$, $(\pi ,\pi ,0)$, $(\pi ,0,0)$ in order. }
\label{fig:kcell_P2_11'}
\end{figure}

Based on the above consideration, we introduce the integer lattices $E_0^p$ for $p=0,1,2,3$ as follows. 
Let $D^3$ be the fundamental domain. 
We define the set of relevant $p$-cells by 
\begin{align}
    &{\cal C}_p = \{ D^p_i | D^p_i \cap \overline{D^3} \neq \emptyset \}, \quad p=0,1,2, \\
    &{\cal C}_3 = \{D^3\}. 
\end{align}
Here, $\overline{D^p_i}$ is the closure of the open $p$-cell $D^p_i$. 
Note that ${\cal C}_p$ includes equivalent $p$-cells as independent ones. 
(For example, see Fig.~\ref{fig:kcell_P2_11'} for a choice of fundamental domain for the MSG $P2_11'$ together with the boundary 0-, 1-, and 2-cells.)
Also, we regard cells that are related to other cells by reciprocal lattice vectors as different cells.
For each $p$-cell, we pick a representative point $\bk^p_i \in D^p_i$ for each $p$-cell. 
For $\bk \in X$, introduce the left coset decomposition 
\begin{align}
    G_{\bk} = G_{\bk}^0 \coprod {\sf t} G_\bk^0 \coprod {\sf c} G_\bk^0 \coprod \gamma G_\bk^0, 
\end{align}
where $G_\bk^0 = \ker \phi \cap \ker c \cap G_\bk$, and 
\begin{align}
    &{\sf t} \in \{g \in G_\bk| -\phi_g=c_g=1\}, \\
    &{\sf c} \in \{g \in G_\bk| -\phi_g=-c_g=1\}, \\
    &\g \in \{g \in G_\bk| \phi_g=-c_g=1\}, 
\end{align}
are representatives.
In general, ${\sf t}, {\sf c}$, and $\g$ depend on the $\bk \in X$, but $\bk$ is omitted unless misunderstandings can arise.
At $\bk$, $\{\alpha_{j}(\bk)\}_{j=1}^{N_\bk}$ denotes the set of irreps of $G_\bk^0$ with the factor system $z_{g,h}(\bk)$. 
(See Appendix~\ref{app:Computing irreducible characters} for a numeral derivation of irreducible characters.)
The lattice $E_0^p$ is defined as the $\Z$-module generated by irreps 
\begin{align}
    E_0^p := \Braket{ \bigcup_{D^p_i \in {\cal C}_p} \bigcup_{r=1}^{N_{\bk^p_i}} \alpha_r(\bk^p_i)}, 
\end{align}
the $\Z$-module generated by the set of irreps $\bigcup_{D^p_i \in {\cal C}_p} \bigcup_{r=1}^{N_{\bk^p_i}} \alpha_r(\bk^p_i)$.

Introduce the integer $n_{i,i'} \in \{\pm 1,0\}$ such that $n_{i,i'} = 1 (-1)$ if $D^p_{i'} \in \p D^{p+1}_i$ and the orientation of $D^p_{i'}$ agrees (disagrees) with $\p D^{p+1}_i$, and $n_{i,i'} = 0$ if $D^p_{i'} \notin \p D^{p+1}_i$. 
We denote by $\chi^{\alpha_r(\bk^p_i)}_{g \in G_{\bk^p_i}}$ the irreducible character of the irrep $\alpha_r(\bk^p_i)$. 
We define the homomorphism $\delta_p: E_0^p \to E_0^{p+1}$ as 
how irreps at $p$-cells decompose into irreps at $(p+1)$-cells. 
Explicitly, $[\delta_p]_{ir,i'r'} =0$ when $D^p_{i'} \notin \p D^{p+1}_i$, and
\begin{align}\label{eq:diff_Lp}
    [\delta_p]_{ir,i'r'} 
    &:= n_{i,i'} \times 
    \frac{1}{|G^0_{\bk^{p+1}_i}|} \sum_{g \in G^0_{\bk^{p+1}_i}} (\chi^{\alpha_r(\bk^{p+1}_i)}_g)^* \nonumber\\
    & \times \chi^{\alpha_{r'}(\bk^p_{i'})}_g e^{-i \phi_g p_g (\bk^{p+1}_i-\bk^p_{i'}) \cdot \bm{a}_g}
\end{align}
when $D^p_{i'} \in \p D^{p+1}_i$. 
The last factor in (\ref{eq:diff_Lp}) is needed to match the factor systems between $\bk^p_{i'}$ and $\bk^{p+1}_i$.

\begin{table*}[]
    \centering
    $$
    \begin{array}{c|ccc|ccccccccccc}
&W^\alpha_T&W^\alpha_C&W^\alpha_\G&n=0&n=1&n=2&n=3&n=4&n=5&n=6&n=7\\
\hline
{\rm A}&0&0&0&(\Z,1)&0&(\Z,2)&0&(\Z,4)&0&(\Z,8)&0\\
{\rm AIII}&0&0&1&0&(\Z,2)&0&(\Z,4)&0&(\Z,8)&0&(\Z,16)\\
\hline
{\rm AI}&1&0&0&(\Z,1)&(\Z_2,2)&(\Z_2,4)&0&(\Z,8)&0&0&0\\
{\rm BDI}&1&1&1&(\Z_2,2)&(\Z_2,4)&0&(\Z,8)&0&0&0&(\Z,16)\\
{\rm D}&0&1&0&(\Z_2,2)&0&(\Z,4)&0&0&0&(\Z,8)&(\Z_2,16)\\
{\rm DIII}&-1&1&1&0&(\Z,4)&0&0&0&(\Z,8)&(\Z_2,16)&(\Z_2,32)\\
{\rm AII}&-1&0&0&(\Z,2)&0&0&0&(\Z,4)&(\Z_2,8)&(\Z_2,16)&0\\
{\rm CII}&-1&-1&1&0&0&0&(\Z,4)&(\Z_2,8)&(\Z_2,16)&0&(\Z,32)\\
{\rm C}&0&-1&0&0&0&(\Z,2)&(\Z_2,4)&(\Z_2,8)&0&(\Z,16)&0\\
{\rm CI}&1&-1&1&0&(\Z,2)&(\Z_2,4)&(\Z_2,8)&0&(\Z,16)&0&0\\
\hline
{\rm mapped\ rep.}&&&&c_h&s_h \phi_h&\phi_hc_h&s_h&c_h&s_h \phi_h&\phi_hc_h&s_h \\
\end{array}
$$
\caption{Wigner criteria, AZ classes, and classification of gapped Hamiltonian over the momentum-space $n$-sphere. 
In the table, $0, \Z$ and $\Z_2$ indicate the classification, and when the classification is nontrivial, the matrix size of the generator Dirac Hamiltonian is shown together.
Here, the ``matrix size" $m$ means that the minimal Dirac Hamiltonian realizing a unit topological invariant is represented by an $m \times m$ matrix.
The last row shows the sign change of the $\Z$ topological invariant under the action of the group element $h$.
}
    \label{tab:k_AZ}
\end{table*}

For each irrep $\alpha(\bk)$ at $\bk$, we identify the AZ class by the Wigner criteria~\cite{ShiozakiSatoGomi_Atiyah-Hirzebruch_2022}
\begin{align}
    &W^{\alpha(\bk)}_T:= \frac{1}{|G^0_\bk|} \sum_{g \in {\sf t}G_\bk^0} z_{g,g}(\bk) \chi^{\alpha(\bk)}_{g^2} \in \{\pm 1,0\}, \\
    &W^{\alpha(\bk)}_C:= \frac{1}{|G^0_\bk|} \sum_{g \in {\sf c}G_\bk^0} z_{g,g}(\bk) \chi^{\alpha(\bk)}_{g^2} \in \{\pm 1,0\}, \\
    &W^{\alpha(\bk)}_\G:= \frac{1}{|G^0_\bk|} \sum_{g \in G_\bk^0} (\chi^{\alpha(\bk)}_{g})^*\frac{z_{g,\g}(\bk)}{z_{\g,\g^{-1}g\g}(\bk)}\chi^{\alpha(\bk)}_{\g^{-1}g\g} \nonumber\\
    &\hspace{50pt}\in \{ 1,0\}.
\end{align}
We extend the notation as follows: 
$W^{\alpha(\bk)}_T=0$ also indicates the case where ${\sf t} \in G_\bk$ does not exists.
The same notation for $W^{\alpha(\bk)}_C$ and $W^{\alpha(\bk)}_\G$ is used. 
For each $(W^{\alpha(\bk)}_T,W^{\alpha(\bk)}_C,W^{\alpha(\bk)}_\G)$, the AZ class and the corresponding classification are listed in TABLE \ref{tab:k_AZ}. 

For each $\bk$, if $W^{\alpha(\bk)}_\G=1$, we pick an irrep $\alpha(\bk)+$ of $G^0_\bk \coprod \g G^0_\bk$ so that $\chi^{\alpha(\bk)+}_{g \in G^0_\bk}=\chi^{\alpha(\bk)}_g$.
We introduce another homomorphism $\delta^\G_p: E_0^p \to E_0^{p+1}$ as 
\begin{align}\label{eq:def_delta_gamma}
    [\delta^\G_p]_{ir,i'r'} 
    &:= n_{i,i'} \times 
    \frac{1}{|G^0_{\bk^{p+1}_i}|} \sum_{g \in \g G^0_{\bk^{p+1}_i}} (\chi^{\alpha_r(\bk^{p+1}_i)+}_g)^* \nonumber\\
    & \times \chi^{\alpha_{r'}(\bk^p_{i'})+}_g e^{-i \phi_g p_g (\bk^{p+1}_i-\bk^p_{i'}) \cdot \bm{a}_g}
\end{align}
for $D^p_{i'} \in \p D^{p+1}_i$ and $W^{\alpha(\bk^p_{i'})}_\G=W^{\alpha(\bk^{p+1}_{i})}_\G=1$, and $[\delta^\G_p]_{ir,i'r'}=0$ otherwise. 
Note that $\delta^\G_p$ depends on choices of irreps $\alpha(\bk^p_i)+$. 

For an irrep $\alpha(\bk)$ of $G^0_\bk$ at $\bk \in X$ and $h \in G$, we introduce the mapped irrep $h[\alpha(\bk)]$ by $h$, which is an irrep of $G^0_{\phi_h p_h \bk} = h G^0_\bk h^{-1}$ at $\phi_h p_h \bk \in X$ whose character is 
\begin{align}
    \chi^{h[\alpha(\bk)]}_{g \in h G^0_\bk h^{-1}} = \frac{z_{g,h}(\bk)}{z_{h,h^{-1}gh}(\bk)} \chi^{\alpha(\bk)}_{h^{-1}gh}.
\end{align}
When $W^{\alpha(\bk)}_\G=1$, the mapped irrep $h[\alpha(\bk)+]$ is defined in the same way. 
Note that the two irreps $h[\alpha(\bk)+]$ and $h[\alpha(\bk)]+$ of the group $h(G^0_{\bk} \coprod \g G^0_\bk) h^{-1}$ may not be unitary equivalent to each other. 
We define the sign $s_h \in \{ \pm 1\}$ as $s_h=1$ if $h[\alpha(\bk)+]$ is unitary equivalent to $h[\alpha(\bk)]+$ and $s_h=-1$ else. 
From (\ref{eq:chiral_sign}), $s_h$ can be computed from
\begin{align}
    s_h = \frac{1}{|G^0_\bk|} \sum_{g \in h (\g G^0_\bk)h^{-1}} (\chi^{h[\alpha(\bk)]+}_g)^* \chi^{h[\alpha(\bk)+]}_g.
\end{align}
The sign $s_h$ is the relative sign between the two chiral operators 
\begin{align}
    u_h(\bk) (\G_{\alpha(\bk)})^{\phi_h} u_h(\bk)^{-1} = s_h \G_{h[\alpha(\bk)]}.\label{eq:sign_chiral}
\end{align}

\subsection{$E_1$ page}
We will define two integer sublattices $\tilde E_1^{p,-n}$ and ${\cal P}_1^{p,-n}$ of $E_0^p$ such that $\tilde E_1^{p,-n}/{\cal P}_1^{p,-n} = E_1^{p,-n}$. 

Based on (\ref{eq:suspension_iso_k_K}), $E_1^{p,-n}$ is generated by an orbit of $n$-dimensional massive Dirac Hamiltonians
\begin{align}
H_{\bk^p_i}(\tilde \bk)=\sum_{\mu=1}^n \tilde k_\mu \g_\mu + m \g_0, \quad \tilde \bk \in {\bk^p_i} \times \tilde S^n
\end{align}
on $p$-cells $D^p_i$.
Here, $\bk^p_i \in D^p_i$ is a representative point within the $p$-cell $D^p_i$, and $\tilde S^n$ is the virtual $n$-sphere on which the little group $G_{\bk^p_i}$ acts trivially.
The symmetry implies that 
\begin{align}
    &H_{\phi_g p_g \bk^p_i}(\tilde \bk) = c_g u_g(\bk^p_i) H_{\bk^p_i}(\tilde \bk)^{\phi_g} u_g(\bk^p_i)^{-1},\nonumber \\
    &\tilde \bk \in \{\phi_g p_g \bk^p_i\} \times \tilde S^n, 
    \label{eq:sym_change}
\end{align}
for the $p$-cells in the same orbit. 
For a given irrep $\alpha(\bk^p_i)$ of $p$-cell $D^p_i$, the classification of the mass term $m \g_0$ is determined according to TABLE \ref{tab:k_AZ}.
In Table \ref{tab:k_AZ}, the matrix size of the generating Dirac Hamiltonian is shown on the right of the parentheses.
For example, $(\Z,4)$ indicates that the classification is $\Z$, and the generator $1 \in \Z$ is represented by a massive Dirac Hamiltonian whose matrix size is 4.
Note that the matrix size shown in the table does not include the dimension of the representation space of the irreducible representation $\alpha$.

The matrix sizes of the Dirac Hamiltonians in TABLE~\ref{tab:k_AZ} are derived as follows. 
First, the construction for zero dimensions ($n=0$), where only the mass term is present, can be easily carried out. 
Namely, for classes A and AI, the matrix size is 1 because there is no degeneracy; for class AII, the matrix size is 2 due to Kramers degeneracy; and for classes BDI and D, the matrix size is 2 because both occupied and unoccupied states must exist due to PHS. 
For general $n>0$, the sizes are obtained by the isomorphism of $K$-groups (shifting to the top right within the complex and real AZ classes, respectively, in TABLE~\ref{tab:k_AZ}) according to the following rules: 
For the isomorphism from a chiral class ($W^\alpha_\Gamma=1$) to a non-chiral class ($W^\alpha_\Gamma=0$), the dimension does not change because the chiral operator provides the additional gamma matrix for the added momentum component.  
Conversely, for the isomorphism from a non-chiral class to a chiral class, the dimension is doubled to introduce an additional chiral operator.

\subsubsection{$\Z$ classification}
For each irrep of the little group for each orbit, we construct a vector $\vec{a}^{p,-n} \in E_0^p$ consisting of the matrix dimensions of the generating Dirac Hamiltonians with the relative signs of $\Z$ numbers. 
Pick a representative $p$-cell $D^p_i$ of an orbit of equivalent $p$-cells in ${\cal C}_p$. 
For an irrep $\alpha_r(\bk^p_i)$, if the classification of degree $n$ is $\Z$, $a^{p,-n}_{ir}$ denotes the matrix size of the generator Dirac Hamiltonian listed in Table \ref{tab:k_AZ}. 
For other equivalent irreps $h[\alpha(\bk^p_i)]$ at $\phi_h p_h \bk^p_i \in X$, we should implement the relative sign of $\Z$ invariant from the relation (\ref{eq:sym_change}).

For even $n$, the $\Z$ invariant is the Chern number 
\begin{align}
    ch_{n/2} = \frac{1}{(n/2)!} \left(\frac{i}{2\pi}\right)^{n/2} \int_{\tilde S^n} \tr [{\cal F}^{n/2}], 
    \label{eq:def_Chern}
\end{align}
where ${\cal F}$ is the Berry connection of the occupied state of the Hamiltonian $H_{\bk^p_i}(\tilde{\bk})$, and integral is taken over the virtual $n$-sphere $\tilde S^n$.
For $n=0$, we define $ch_0$ as the number of occupied states minus the number of unoccupied states.
Since the Chern number of the unoccupied band is $-ch_{n/2}$, we have the factor $c_h$. 
Moreover, since the Berry curvature ${\cal F}$ changes its sign if $h$ is antiunitary, we have the factor $\phi_h$ when $n \in 4\Z+2$. 

For odd $n$, the $\Z$ invariant is the winding number 
\begin{align}
    w_{n} &= \frac{((n+1)/2)!}{(2\pi i)^{(n+1)/2}(n+1)!}\nonumber\\
    & \qquad \int_{\tilde S^n} \tr[(H_{\bk^p_i}^{-1} dH_{\bk^p_i})^{n}\G_{\alpha(\bk^p_i)}], \label{eq:winding_number}
\end{align}
where $\G_{\alpha(\bk^p_i)}$ is the chiral operator, whose sign depends on the choice of $\alpha(\bk^p_i)+$, and integral is taken over the virtual $n$-sphere $\tilde S^n$.
From the sign change between the chiral operators, we have the factor $s_h$. 
Moreover, since the winding number (\ref{eq:winding_number}) includes the imaginary unit, we have the factor $\phi_h$ when $n \in 4\Z+1$. 

Let us introduce $h(i)$ and $h(r)$ so that $D^p_{h(i)} = h(D^p_i)$ and $\alpha_{h(r)}((\phi_j p_h \bk)^p_{h(i)}) = h[\alpha_r(\bk^p_i)]$. 
Then, for other components, we set 
\begin{align}
    a^{p,-n}_{h(i)h(r)} = a^{p,-n}_{ir} \times \left\{\begin{array}{ll}
        c_h & (n=0),  \\
        s_h \phi_h & (n=1), \\
        \phi_h c_h & (n=2), \\
        s_h & (n=3), \\
        c_h & (n=4), \\
        s_h \phi_h & (n=5), \\
        \phi_h c_h & (n=6), \\
        s_h & (n=7). \\
    \end{array}\right.
    \label{eq:k_sign_map}
\end{align}
The last factor in (\ref{eq:k_sign_map}) is also shown in Table \ref{tab:k_AZ}. 
Constructing the vectors $\vec{a}^{p,-n}_1,\vec{a}^{p,-n}_2,\dots,$ for all inequivalent irreps and inequivalent orbits, we have the sublattice 
\begin{align}
    E_{1\Z}^{p,-n} := \braket{\vec{a}^{p,-n}_1,\vec{a}^{p,-n}_2,\dots } \subset E_0^p.
\end{align}

We note that since the $p$-cells in ${\cal C}_p$ are oriented symmetrically, there is no sign change due to the mismatch of orientations.

\subsubsection{$\Z_2$ classification}
No sign difference exists in the $\Z_2$ number. 
For each irrep of each orbit, we construct a vector $\vec{b}^{p,-n} \in E_0^p$ consisting of matrix sizes of generator Dirac Hamiltonians as follows. 
Pick a representative $p$-cell $D^p_i$ of an orbit of equivalent $p$-cells in ${\cal C}_p$. 
For an irrep $\alpha_r(\bk^p_i)$, if the classification of degree $n$ is $\Z_2$, we define $b^{p,-n}_{ir}$ by the matrix size of the generator Dirac Hamiltonian listed in Table~\ref{tab:k_AZ}, i.e., $b^{p,-n}_{ir}=\dim (H)$.
For other equivalent irreps $h[\alpha(\bk^p_i)]$ at $\phi_h p_h \bk \in X$, $b^{p,-n}_{h(i)h(r)}=b^{p,-n}_{ir}$. 
Constructing the vectors $\vec{b}^{p,-n}_1,\vec{b}^{p,-n}_2,\dots,$ for all inequivalent irreps and inequivalent orbits, we have 
\begin{align}
    E_{1\Z_2}^{p,-n} := \braket{\vec{b}^{p,-n}_1,\vec{b}^{p,-n}_2,\dots } \subset E_0^p
\end{align}
and ${\cal P}_1^{p,-n} := \{2x \in E_0^p|x \in E_{1\Z_2}^{p,-n}\}$. 

The group $E_1^{p,-n}$ is represented as the quotient group 
\begin{align}
    E_1^{p,-n} = (E_{1\Z}^{p,-n} \oplus E_{1\Z_2}^{p,-n})/{\cal P}^{p,-n}_1.
\end{align}
Note that by construction $E_{1\Z}^{p,-n} \cap E_{1\Z_2}^{p,-n} = \{\bm{0}\}$.
We write $\tilde E_1^{p,-n}:=E_{1\Z}^{p,-n} \oplus E_{1\Z_2}^{p,-n}$.

\subsection{The first differential $d_1$}
The homomorphism $d_1^{p,-n}$ is given by the expansion coefficients
\begin{align}
    &d_1^{p,-n}(\vec{a}^{p,-n}_\lambda) = \sum_\kappa \vec a^{p+1,-n}_{\kappa} [M^{p,-n}_{\Z\Z}]_{\kappa \lambda} \nonumber \\
    & \qquad \qquad + \sum_\kappa \vec b^{p+1,-n}_\kappa [M^{p,-n}_{\Z\Z_2}]_{\kappa \lambda},\\
    &d_1^{p,-n}(\vec{b}^{p,-n}_\lambda) = \sum_\kappa \vec b^{p+1,-n}_\kappa [M^{p,-n}_{\Z_2\Z_2}]_{\kappa \lambda},
\end{align}
where $[M^{p,-n}_{\Z\Z}]_{\kappa \lambda} \in \Z$ and $[M^{p,-n}_{\Z\Z_2}]_{\kappa \lambda},[M^{p,-n}_{\Z_2\Z_2}]_{\kappa \lambda} \in \{0,1\}$.
A vector $\vec v = (v_{ir}) \in E_1^{p,-n}$ represents a set of massive Dirac Hamiltonians over the $n$-spheres $\{\bk^p_i\} \times \tilde S^n$ with $\Z$ or $\Z_2$ invariants specified by $v_{ir}$. 
This set of the massive Dirac Hamiltonians over $\{\bk^p_i\} \times \tilde S^n$ may be incompatible on adjacent $(p+1)$-cells whose obstruction is given by the vector $d_1^{p,-n}(\vec v) \in E_1^{p+1,-n}$. 

The expansion coefficients are computed as follows. 
We denote the matrices consisting of generators of $E^{p,-n}_{1\Z}$ and $E_{1\Z_2}^{p,-n}$ by 
\begin{align}
    &A_1^{p,-n}:=(\vec{a}^{p,-n}_1,\vec a^{p,-n}_2,\dots), \\
    &B_1^{p,-n}:=(\vec{b}^{p,-n}_1,\vec b^{p,-n}_2,\dots).
\end{align}
We claim that $\delta_p(E_{1\Z}^{p,-n}) \subset E_{1\Z}^{p+1,-n}$ for even $n$: 
A $\Z$-valued vector $\vec{a} = (a_{ir}) \in E^{p,-n}_{1\Z}$ represents the set of massive Dirac Hamiltonians with $\Z$ numbers specified by $a_{ir}$ over the $p$-cells $D^p_i$. 
The vector $\delta_p \vec{a} \in E_0^{p+1}$ represents the $\Z$-valued obstruction to glue the massive Dirac Hamiltonians on adjacent $(p+1)$-cells. 
Suppose that an irrep $\alpha_{r'}(\bk^{p+1}_{i'})$ at a point of $\bk^{p+1}_{i'}$ of a $(p+1)$-cell is classified as $\Z_2$. 
The relation $d_1^{p,-n} \vec a \in E_1^{p+1,-n}$ implies that the total $\Z$ number of the irrep $\alpha_{r'}(\bk^{p+1}_{i'})$, the coefficient $[\delta_p \vec a]_{i'r'}$ must vanish. 
For odd $n$, the same relation $\delta_p^\G(E_{1\Z}^{p,-n}) \subset E_{1\Z}^{p+1,-n}$ holds true. 
Therefore, the expansion coefficient $M^{p,-n}_{\Z\Z}$ is given as 
\begin{align}
    &M^{p,-n}_{\Z\Z} \nonumber \\
    &= \left\{
    \begin{array}{ll}
        (A_1^{p+1,-n})^+\delta_p A_1^{p,-n} & ({\rm for\ even\ }n), \\
        (A_1^{p+1,-n})^+\delta^\G_p A_1^{p,-n} & ({\rm for\ odd\ }n).
    \end{array}\right.
\end{align}
Here, $X^+$ is the pseudoinverse of the matrix $X$. 
(Note that the pseudoinverse $X^+ = X^+ X X^+$ itself involves the projection $XX^+$ onto $\im X$.)
For the coefficient of $\Z_2$ groups, since the $\Z_2$ number is given by the matrix size of Dirac Hamiltonians, without considering either orientations or the sign of $\Z$ numbers, the expansion coefficients are given as 
\begin{align}
    &M^{p,-n}_{\Z\Z_2} = (B_1^{p+1,-n})^+|\delta_p| |A_1^{p,-n}|\quad \mod 2, \\
    &M^{p,-n}_{\Z_2\Z_2} = (B_1^{p+1,-n})^+|\delta_p| B_1^{p,-n}\quad \mod 2.
\end{align}
Here, $|X|$ is the matrix whose component is $|X_{ij}|$.

Introduce an integer lift 
\begin{align}
    \tilde d_1^{p,-n}: \tilde E_1^{p,-n} \to \tilde E_1^{p+1,-n}. 
\end{align}
By considering $M^{p,-n}_{\Z\Z_2}$ and $M^{p,-n}_{\Z_2\Z_2}$ as $\Z$-valued matrices, such a lift is given.
The group $E_2^{p,-n}$ is computed as the quotient of two integer sublattices of $E_0^p$: 
\begin{align}
    E_2^{p,-n} = \frac{\ker (\tilde d_1^{p,-n} \oplus {\rm Id}_{{\cal P}_1^{p+1,-n}})|_{\tilde E_1^{p,-n}}}{\im (\tilde d_1^{p-1,-n})+{\cal P}_1^{p,-n}}.
    \label{eq:k_e2_int}
\end{align}
Here, $\tilde d_1^{p,-n} \oplus {\rm Id}_{{\cal P}_1^{p+1,-n}}(\vec{u},\vec{v}) = \tilde d_1^{p,-n}(\vec{u})+\vec{v}$, and $\ker (\tilde d_1^{p,-n} \oplus {\rm Id}_{{\cal P}_1^{p+1,-n}})|_{\tilde E_1^{p,-n}}$ means the restriction to $\tilde E_1^{p,-n}$. 
See Appendix~\ref{app:Computation in Z-module} for a derivation of (\ref{eq:k_e2_int}).

\section{Real-space AHSS
\label{sec:Real-space AHSS}
}
We present a method for calculating the first differential $d^1$ of the real-space AHSS~\cite{Shiozaki_homology}.
In this section, $g, h, \dots$ refer to elements of the group ${\cal G}$. We denote the ${\cal G}$-action on the real space by $\bx \mapsto g(\bx) = p_g \bx + \bm{t}_g$ for $g \in {\cal G}$. 
This formulation shares similarities with the momentum-space AHSS, so our primary focus will be on the differences.

The real-space AHSS is based on the concept of ``crystalline topological liquid"~\cite{ThorngrenElse}: 
We assume the scale of lattice translations is much larger than the microscopic scale. 
A topological phase protected by an MSG is represented as a "patchwork" of the set of topological phases localized on $p$-cells. 
With this perspective, we denote a Hamiltonian near the position $\bx \in \mathbb{E}^3$ by a single variable as $H(\bx)$. The symmetry constraint on $H(\bx)$ is written as 
\begin{align}
    &u_g H(\bx)^{\phi_g} u_g^{-1} = c_g H(g(\bx)),\\
    &u_g u_h^{\phi_g} = z^{\rm int}_{g,h} u_{gh}, \quad g,h \in {\cal G}. 
\end{align}

We shall write the $K$-theory classification of Hamiltonians $H(\bx)$ with ${\cal G}$ symmetry by ${}^{\phi} K^{\cal G}_{(z^{\rm int},c)-n}(\R^3)$.
Here, $n \in \Z$ is the degree of the $K$-homology group. 
The physical meaning of the degree $n$ is opposite to the $K$-cohomology group ${}^{\phi} K_G^{(z,c)-n}(X)$: 
For $n=1,0,-1$, the group ${}^{\phi} K^{\cal G}_{(z^{\rm int},c)-n}(\R^3)$ represents the classification of gapless Hamiltonians, gapped Hamiltonians, and one-parameter families gapped Hamiltonians (adiabatic pumps, says), respectively. 
It is expected that 
\begin{align}
{}^{\phi} K^{\cal G}_{(z^{\rm int},c)-n}(\R^3) \cong {}^{\phi} K_G^{(z,c)+n}(X) \label{eq:iso_k_r}
\end{align}
since the two $K$-groups classify the physically same systems, and the isomorphism (\ref{eq:iso_k_r}) was proved in \cite{GomiKubotaThiang_crystallographic_T-duality2021}.

In the following, the superscript of the factor system $z^{\rm int}$ will be omitted and simply written as $z$.

\subsection{Preliminary}

For a cell decomposition $X_0 \subset X_1 \subset X_2 \subset X_3 = \R^3$ introduced in Sec.~\ref{sec:Cell decomposition}, the $E^1$-page $E^1_{p,-n}$ is defined as the relative $K$-homology group 
\begin{align}
    E^1_{p,-n}
    := {}^\phi K_{(z,c)+p-n}^{\cal G}(X_p,X_{p-1}). 
\end{align}
We denote the label set of orbits of $p$-cells by $I^p_{\rm orb}$. 
We have 
\begin{align}
    &E^1_{p,-n} \cong {}^\phi K^{\cal G}_{(z,c)+p-n} (\coprod_{{\sf a} \in I^p_{\rm orb}} {\cal G}/{\cal G}_{D^p_{\sf a}} \times D^p_{\sf a},\nonumber \\
    &\hspace{50pt} \coprod_{{\sf a} \in I^p_{\rm orb}} {\cal G}/{\cal G}_{D^p_{\sf a}} \times \p D^p_{\sf a}) \nonumber\\
    &\cong 
    \bigoplus_{{\sf a} \in I^p_{\rm orb}} {}^{\phi|_{D^p_{\sf a}}} K_{(z|_{D^p_{\sf a}},c|_{D^p_{\sf a}})+p-n}^{{\cal G}_{D^p_{\sf a}}}(D^p_{\sf a},\p D^p_{\sf a}). 
\end{align}
Here, $D^p_{\sf a}$ is a representative of the orbit ${\sf a} \in I^p_{\rm orb}$, ${\cal G}_{D^p_{\sf a}} = \{g \in {\cal G}| g(\bx) = \bx {\rm\ for\ } \bx \in D^p_{\sf a}\}$ is the little group of the $p$-cell $D^p_{\sf a}$, and $\phi|_{D^p_{\sf a}},c|_{D^p_{\sf a}},z|_{D^p_{\sf a}}$ are symmetry data restricted in the $p$-cell $D^p_{\sf a}$. 
Since the little group ${\cal G}_{D^p_{\sf a}}$ fixes the $p$-cell $D^p_{\sf a}$ pointwise, 
\begin{align}
    E^1_{p,-n} 
    &\cong \bigoplus_{{\sf a} \in I^p_{\rm orb}} {}^{\phi|_{D^p_{\sf a}}} \tilde K^{(z|_{D^p_{\sf a}},c|_{D^p_{\sf a}})+p-n}_{{\cal G}_{D^p_{\sf a}}}(D^p_{\sf a}/\p D^p_{\sf a}) \nonumber \\
    &\cong \bigoplus_{{\sf a} \in I^p_{\rm orb}} {}^{\phi|_{\bx^p_{\sf a}}} K^{(z|_{\bx^p_{\sf a}},c|_{\bx^p_{\sf a}})-n}_{{\cal G}_{\bx^p_{\sf a}}}(\{\bx^p_{\sf a}\}).
\end{align}
In the same way as the momentum-space AHSS, to represent the $K$-group with degree $-n$, we use the suspension isomorphism
\begin{align}
    &{}^{\phi|_{\bx^p_{\sf a}}} K_{(z|_{\bx^p_{\sf a}},c|_{\bx^p_{\sf a}})-n}^{{\cal G}_{\bx^p_{\sf a}}}(\{\bx^p_{\sf a}\})\nonumber \\
    &\cong 
    {}^{\phi|_{\bx^p_{\sf a}}} \tilde K_{(z|_{\bx^p_{\sf a}},c|_{\bx^p_{\sf a}})}^{{\cal G}_{\bx^p_{\sf a}}}(\{\bx^p_{\sf a}\}\times \tilde S^n)
    \label{eq:suspension_iso_r_K}, 
\end{align}
where the little group ${\cal G}_{\bx^p_{\sf a}}$ trivially acts on the sphere $\tilde S^n$. 
To distinguish it from real space, we use the symbol of the virtual sphere as $\tilde S^n$.

The first differential $d^1_{p,-n}$ is defined as the composition of the homomorphisms below 
\begin{align}
    &d^1_{p,-n}
    :{}^\phi K^{\cal G}_{(z,c)+p-n}(X_p,X_{p-1}) \nonumber \\
    &\to {}^\phi K^{\cal G}_{(z,c)+p-n-1}(X_{p-1}) \nonumber \\
    &\to {}^\phi K^{\cal G}_{(z,c)+p-n-1}(X_{p-1},X_{p-2}) = E^1_{p-1,-n}.
\end{align}
Here, the first homomorphism is the boundary map (physically, the bulk-boundary correspondence) and the second one is from the inclusion $X_{p-1} \to (X_{p-1},X_{p-2})$.
The relation 
\begin{align}
    d_1^{p-1,-n} \circ d_1^{p,-n} =0
\end{align}
holds and the the $E_2$-page is defined as 
\begin{align}
    E_2^{p,-n} := \ker d_1^{p,-n}/\im d_1^{p+1,-n}. 
\end{align}
(For $p=3$, $E_2^{3.-n} = \ker d_1^{3,-n}$.)
The physical meaning of $d^1_{p,-n}$ is to measure how much of the boundary gapless state remains in the adjacent $(p-1)$-cells for a gapped Hamiltonian inside $p$-cells.

\begin{figure}[!]
\centering
\includegraphics[width=0.7\linewidth, trim=0cm 0cm 0cm 0cm]{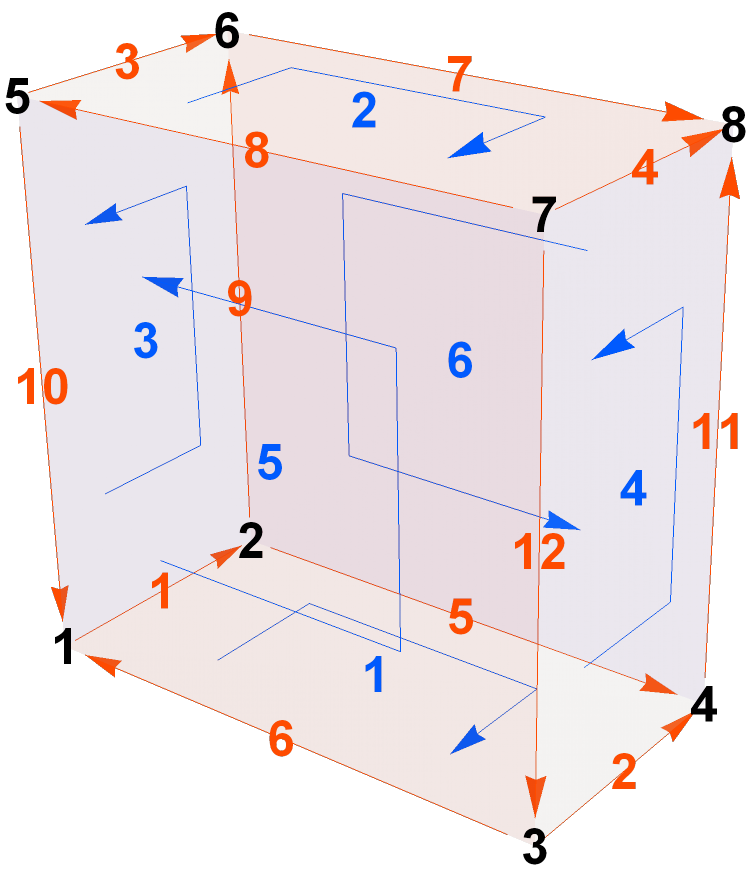}
\caption{A cell decomposition of real space for MSG $P2_11'$. 
The figure shows the fundamental domain (3-cell). 
The 0-, 1-, and 2-cells are shown with black, red, and blue Arabic numerals, respectively. 
The coordinates of 0-cells are $(-\frac{1}{2},-\frac{1}{4},-\frac{1}{2})$, $(-\frac{1}{2},\frac{1}{4},-\frac{1}{2})$, $(\frac{1}{2},-\frac{1}{4},-\frac{1}{2})$, $(\frac{1}{2},\frac{1}{4},-\frac{1}{2})$, $(-\frac{1}{2},-\frac{1}{4},\frac{1}{2})$, $(-\frac{1}{2},\frac{1}{4},\frac{1}{2})$, $(\frac{1}{2},-\frac{1}{4},\frac{1}{2})$, $(\frac{1}{2},\frac{1}{4},\frac{1}{2})$ in order. }
\label{fig:rcell_P2_11'}
\end{figure}

It turns out that we only need to consider the $p$-cells intersecting the closure of the fundamental domain (asymmetric unit). 
We introduce the integer lattices $E^0_p$ for $p=0,1,2,3$ as follows. 
Let $D^3$ be the fundamental domain. 
We define the set of relevant $p$-cells ${\cal C}_p$ as before.
(For example, see Fig.~\ref{fig:rcell_P2_11'} for a choice of fundamental domain for the MSG $P2_11'$ together with the boundary 0-, 1-, and 2-cells.)
For each $p$-cell, we pick a representative point $\bx^p_i \in D^p_i$ for each $p$-cell. 
For each $\bx$, introduce the left coset decomposition 
\begin{align}
    {\cal G}_{\bx} = {\cal G}_{\bx}^0 \coprod {\sf t} {\cal G}_\bx^0 \coprod {\sf c} {\cal G}_\bx^0 \coprod \gamma {\cal G}_\bx^0, 
\end{align}
where ${\cal G}_\bx^0 = \ker \phi \cap \ker c \cap {\cal G}_\bx$, and 
\begin{align}
    &{\sf t} \in \{g \in {\cal G}_\bx| -\phi_g=c_g=1\}, \\
    &{\sf c} \in \{g \in {\cal G}_\bx| -\phi_g=-c_g=1\}, \\
    &\g \in \{g \in {\cal G}_\bx| \phi_g=-c_g=1\}, 
\end{align}
are representatives.
In general, ${\sf t}, {\sf c}$, and $\g$ depend on the $\bx \in \R^3$, but $\bx$ is omitted unless misunderstanding arises.
We denote by $\alpha_1(\bx),\dots,\alpha_{N_\bx}(\bx)$ the set of irreps of ${\cal G}_\bx^0$ with the factor system $z_{g,h}$. 
The lattice $E^0_p$ is defined as the $\Z$-module generated by irreps inside the relevant $p$-cells 
\begin{align}
    E_0^p := \Braket{ \bigcup_{D^p_i \in {\cal C}_p} \bigcup_{r=1}^{N_{\bx^p_i}} \alpha_r(\bx^p_i)}. 
\end{align}
Denote by $\chi^{\alpha_r(\bx^p_i)}_{g \in {\cal G}_{\bx^p_i}}$ the irreducible character of the irrep $\alpha_r(\bx^p_i)$. 
For each irrep $\alpha(\bx)$ at $\bx$, we identify the AZ class by the Wigner criteria
\begin{align}
    &W^{\alpha(\bx)}_T:= \frac{1}{|{\cal G}^0_\bx|} \sum_{g \in {\sf t}{\cal G}_\bx^0} z_{g,g} \chi^{\alpha(\bx)}_{g^2} \in \{\pm 1,0\}, \\
    &W^{\alpha(\bx)}_C:= \frac{1}{|{\cal G}^0_\bx|} \sum_{g \in {\sf c}G_\bx^0} z_{g,g} \chi^{\alpha(\bx)}_{g^2} \in \{\pm 1,0\}, \\
    &W^{\alpha(\bx)}_\G:= \frac{1}{|{\cal G}^0_\bx|} \sum_{g \in G_\bx^0} (\chi^{\alpha(\bx)}_{g})^*\frac{z_{g,\g}}{z_{\g,\g^{-1}g\g}}\chi^{\alpha(\bx)}_{\g^{-1}g\g} \nonumber\\
    &\hspace{50pt}\in \{ 1,0\}.
\end{align}
For each pattern of $(W^{\alpha(\bx)}_T,W^{\alpha(\bx)}_C,W^{\alpha(\bx)}_\G)$, the AZ class and one of the $\Z$, $\Z_2$, or trivial classification are listed in Table~\ref{tab:r_AZ}. 
For each irrep $\alpha(\bx)$, if $W^{\alpha(\bx)}_\G=1$, we pick an irrep $\alpha(\bx)+$ of ${\cal G}^0_\bx \coprod \g {\cal G}^0_\bx$ so that $\chi^{\alpha(\bx)+}_{g \in {\cal G}^0_\bx}=\chi^{\alpha(\bx)}_g$.

We will define the homomorphisms $\delta_p, \delta^\G_p: E_0^p \to E_0^{p+1}$ in the same way as before. 
The integer $n_{i,i'} \in \{\pm 1,0\}$ is defined as $n_{i,i'}=1$ ($n_{i,i'}=-1$) if $D^p_{i'} \in \p D^{p+1}_i$ and the orientation of $D^p_{i'}$ agrees (disagrees, resp.), and $n_{i,i'}=0$ else. 
We define $[\delta_p]_{ir,i'r'} =0$ for $D^p_{i'} \notin \p D^{p+1}_i$ and 
\begin{align}\label{eq:rdiff_Lp}
    [\delta_p]_{ir,i'r'} 
    &:= n_{i,i'} \times 
    \frac{1}{|{\cal G}^0_{\bx^{p+1}_i}|} \nonumber \\
    & \sum_{g \in {\cal G}^0_{\bx^{p+1}_i}} (\chi^{\alpha_r(\bx^{p+1}_i)}_g)^*
    \chi^{\alpha_{r'}(\bx^p_{i'})}_g 
\end{align}
for $D^p_{i'} \in \p D^{p+1}_i$. 
We define $\delta_p^\G$ by 
\begin{align}
    [\delta^\G_p]_{ir,i'r'} 
    &:= n_{i,i'} \times 
    \frac{1}{|{\cal G}^0_{\bx^{p+1}_i}|} \nonumber\\
    &\sum_{g \in \g {\cal G}^0_{\bx^{p+1}_i}} (\chi^{\alpha_r(\bx^{p+1}_i)+}_g)^* \chi^{\alpha_{r'}(\bx^p_{i'})+}_g
\end{align}
when $D^p_{i'} \in \p D^{p+1}_i$ and $W^{\alpha(\bk^p_{i'})}_\G=W^{\alpha(\bk^{p+1}_{i})}_\G=1$ and $[\delta^\G_p]_{ir,i'r'}=0$ else.

For an irrep $\alpha(\bx)$ of ${\cal G}^0_\bx$ at $\bx$ and $h \in {\cal G}$, we introduce the mapped irrep $h[\alpha(\bx)]$ by $h$, which is an irrep of ${\cal G}^0_{h(\bx)} = h {\cal G}^0_\bx h^{-1}$ at $p_h \bx +\bm{t}_h \in X$ whose character is 
\begin{align}
    \chi^{h[\alpha(\bx)]}_{g \in h {\cal G}^0_\bx h^{-1}} = \frac{z_{g,h}}{z_{h,h^{-1}gh}} \chi^{\alpha(\bx)}_{h^{-1}gh}.
\end{align}
When $W^{\alpha(\bx)}_\G=1$, the mapped irrep $h[\alpha(\bx)+]$ is defined similarly. 
We define the sign $s_h \in \{ \pm 1\}$ as before.
\begin{align}
    s_h = \frac{1}{|{\cal G}^0_\bx|} \sum_{g \in h(\g {\cal G}^0_\bx)h^{-1}} (\chi^{h[\alpha(\bx)]+}_g)^* \chi^{h[\alpha(\bx)+]}_g, 
\end{align}
which measures the relative sign between the two chiral operators at $h(\bx)$ as in 
\begin{align}
    u_h (\G_{\alpha(\bx)})^{\phi_h} u_h = s_h \G_{h[\alpha(\bx)]}.
\end{align}

The new ingredient specific to the real-space AHSS is the homomorphisms $\delta_p^T$ and $\delta_p^{\G,T}$, which are needed in MSGs of types III and IV.
We define 
\begin{align}\label{eq:rdiffT_Lp}
    [\delta^T_p]_{ir,i'r'} 
    &:= n_{i,i'} \times 
    \frac{1}{|{\cal G}^0_{\bx^{p+1}_i}|} \nonumber \\
    & \sum_{g \in {\cal G}^0_{\bx^{p+1}_i}} (\chi^{\alpha_r(\bx^{p+1}_i)}_g)^*
    \chi^{{\sf t}[\alpha_{r'}(\bx^p_{i'})]}_g 
\end{align}
only if the following conditions are fulfilled, and $0$ otherwise: 
$D^p_{i'} \in \p D^{p+1}_i$, ${\sf t}{\cal G}^0_{\bx^{p+1}_{i}} = \emptyset$, and ${\sf t}{\cal G}^0_{\bx^{p}_{i'}} \neq \emptyset$. 
Similarly, we define 
\begin{align}
    [\delta^{\G,T}_p]_{ir,i'r'} 
    &:= n_{i,i'} \times 
    \frac{1}{|{\cal G}^0_{\bx^{p+1}_i}|} \nonumber \\
    & \sum_{g \in \g {\cal G}^0_{\bx^{p+1}_i}} (\chi^{\alpha_r(\bx^{p+1}_i)+}_g)^*
    \chi^{{\sf t}[\alpha_{r'}(\bx^p_{i'})+]}_g 
\end{align}
only if the following conditions are fulfilled, and $0$ otherwise: 
$D^p_{i'} \in \p D^{p+1}_i$, $\g {\cal G}^0_{\bx^p_{i'}} \supset \g {\cal G}^0_{\bx^{p+1}_i} \neq \emptyset$, ${\sf t}{\cal G}^0_{\bx^{p+1}_{i}} = \emptyset$, and ${\sf t}{\cal G}^0_{\bx^{p}_{i'}} \neq \emptyset$.

\subsection{$E^1$ page}
In the view of (\ref{eq:suspension_iso_r_K}), $E^1_{p,-n}$ is generated by an orbit of $n$-dimensional massive Dirac Hamiltonians over the virtual $n$-sphere labeled by $p$-cells as in 
\begin{align}
&H_{\bx^p_i}(\tilde \bx)=\sum_{\mu=1}^n -i \tilde \p_\mu \g_\mu + m \g_0,\nonumber\\
&\tilde \bx \in \{\bx^p_i\} \times \tilde S^n.
\end{align}
Here, $\tilde \p_\mu = \frac{\p}{\p \tilde x_\mu}$. 
The Hamiltonians for equivalent $p$-cells are related as 
\begin{align}
    &H_{g(\bx^p_i)}(\tilde \bx) = c_g u_g H_{\bx^p_i}(\tilde \bx)^{\phi_g} u_g^{-1},\nonumber\\
    &\tilde \bx \in \{g(\bx^p_i)\} \times \tilde S^n, 
    \label{eq:sym_change_r}
\end{align}
for $g \in {\cal G}$. 
For a given irrep $\alpha(\bx^p_i)$ of $p$-cell $D^p_i$, the classification of the mass term $m \g_0$ is determined according to Table\ref{tab:r_AZ}, which is the same as the periodic table of topological insulators and superconductors~\cite{Kitaev_periodic2009, RyuTenFold}.
In Table \ref{tab:r_AZ}, the matrix size of the generating Dirac Hamiltonian is shown on the right of the parentheses.
Note that the matrix size shown in TABLE~\ref{tab:r_AZ} does not include the dimension of the representation space of the irreducible representation $\alpha$. 
The matrix sizes of the Dirac Hamiltonians are derived in the same manner as in the momentum space (TABLE~\ref{tab:k_AZ}), except that the isomorphism of $K$-groups shifts to the bottom right within the complex and real AZ classes in TABLE~\ref{tab:r_AZ}.

\begin{table*}
$$
\begin{array}{c|ccc|cccccccccccc}
&W^\alpha_T&W^\alpha_C&W^\alpha_\G&n=0&n=1&n=2&n=3&n=4&n=5&n=6&n=7\\
\hline
{\rm A}&0&0&0&(\Z,1)&0&(\Z,2)&0&(\Z,4)&0&(\Z,8)&0\\
{\rm AIII}&0&0&1&0&(\Z,2)&0&(\Z,4)&0&(\Z,8)&0&(\Z,16)\\
\hline
{\rm AI}&1&0&0&(\Z,1)&0&0&0&(\Z,8)&0&(\Z_2,16)&(\Z_2,16)\\
{\rm BDI}&1&1&1&(\Z_2,2)&(\Z,2)&0&0&0&(\Z,16)&0&(\Z_2,32)\\
{\rm D}&0&1&0&(\Z_2,2)&(\Z_2,2)&(\Z,2)&0&0&0&(\Z,16)&0\\
{\rm DIII}&-1&1&1&0&(\Z_2,4)&(\Z_2,4)&(\Z,4)&0&0&0&(\Z,32)\\
{\rm AII}&-1&0&0&(\Z,2)&0&(\Z_2,4)&(\Z_2,4)&(\Z,4)&0&0&0\\
{\rm CII}&-1&-1&1&0&(\Z,4)&0&(\Z_2,8)&(\Z_2,8)&(\Z,8)&0&0\\
{\rm C}&0&-1&0&0&0&(\Z,4)&0&(\Z_2,8)&(\Z_2,8)&(\Z,8)&0\\
{\rm CI}&1&-1&1&0&0&0&(\Z,8)&0&(\Z_2,16)&(\Z_2,16)&(\Z,16)\\
\hline
{\rm mapped\ rep.}&&&&c_h&s_h&\phi_h c_h&s_h\phi_h&c_h&s_h&\phi_h c_h&s_h\phi_h\\
\end{array}    
$$
\caption{Wigner criteria, AZ classes, and classification of gapped Hamiltonian over the real-space $n$-sphere. 
In the table, $0, \Z$ and $\Z_2$ indicate the classification, and when the classification is nontrivial, the matrix size of the generator Dirac Hamiltonian is shown together.
Here, the ``matrix size" $m$ means that the minimal Dirac Hamiltonian realizing a unit topological invariant is represented by an $m \times m$ matrix.
The last row shows the sign change of the $\Z$ topological invariant under the action of the group element $h$.}
\label{tab:r_AZ}
\end{table*}

\subsubsection{$\Z$ classification}
For each irrep of each orbit, we construct a vector $\vec{a}^{p,-n} \in E^0_p$.
Pick a representative $p$-cell $D^p_i$ of an orbit of equivalent $p$-cells in ${\cal C}_p$. 
For an irrep $\alpha_r(\bx^p_i)$, if the classification of degree $-n$ is $\Z$, we set $a^{p,-n}_{ir}$ to be the matrix size of generator Dirac Hamiltonian listed in Table~\ref{tab:r_AZ}. 
The components for other equivalent irreps $h[\alpha(\bx^p_i)]$ are determined by the relation (\ref{eq:sym_change_r}). 

For even $n$, the $\Z$ invariant is the Chern number (\ref{eq:def_Chern}). 
Since the Chern number of the unoccupied band is $-ch_{n/2}$, we have the factor $c_h$. 
Moreover, since the Berry curvature ${\cal F}$ changes its sign if $h$ is antiunitary, we have the factor $\phi_h$ when $n \in 4\Z+2$. 
No sign change arises from the flip of momenta $\bk \to -\bk$. 

For odd $n$, the $\Z$ invariant is the winding number (\ref{eq:winding_number}) with the chiral operator $\G_{\alpha(\bx^p_i)}$.
We have the factor $s_h$ for all $n \in 2\Z +1$ and $\phi_h$ for $n \in 4\Z+1$ as before. 
Moreover, from the flip of momenta $\bk \to -\bk$, we have the additional factor $\phi_h$ for all $n \in 2\Z+1$. 

Let us introduce $h(i)$ and $h(r)$ as before, i.e., $D^p_{h(i)} = h(D^p_i)$ and $\alpha_{h(r)}(h(\bx^p_{h(i)})) = h[\alpha_r(\bx^p_i)]$. 
In sum, we have 
\begin{align}
    a^{p,-n}_{h(i)h(r)} = a^{p,-n}_{ir} \times \left\{\begin{array}{ll}
        c_h & (n=0),  \\
        s_h & (n=1), \\
        \phi_h c_h & (n=2), \\
        s_h \phi_h & (n=3), \\
        c_h & (n=4), \\
        s_h & (n=5), \\
        \phi_h c_h & (n=6), \\
        s_h \phi_h & (n=7). \\
    \end{array}\right.
    \label{eq:r_sign_map}
\end{align}
The last factor in (\ref{eq:r_sign_map}) is also shown in Table \ref{tab:r_AZ}. 
Constructing the vectors $\vec{a}^{p,-n}_1,\vec{a}^{p,-n}_2,\dots,$ for all inequivalent irreps and inequivalent orbits, we have the sublattice 
\begin{align}\label{eq:rE1Z}
    E^{1\Z}_{p,-n} := \braket{\vec{a}^{p,-n}_1,\vec{a}^{p,-n}_2,\dots } \subset E^0_p.
\end{align}

\subsubsection{$\Z_2$ classification}
For an irrep $\alpha_r(\bx^p_i)$, if the classification of degree $-n$ is $\Z_2$, set $b^{p,-n}_{ir}=\dim (H)$, the matrix dimension listed in Table \ref{tab:r_AZ}. 
For other equivalent irreps $h[\alpha(\bx^p_i)]$ at $h(\bx) \in X$, set $b^{p,-n}_{h(i)h(r)}=b^{p,-n}_{ir}$. 
Constructing the vectors $\vec{b}^{p,-n}_1,\vec{b}^{p,-n}_2,\dots,$ for all inequivalent irreps and inequivalent orbits, we have 
\begin{align}
    E^{1\Z_2}_{p,-n} := \braket{\vec{b}^{p,-n}_1,\vec{b}^{p,-n}_2,\dots } \subset E^0_p
\end{align}
and ${\cal P}^1_{p,-n} = \{2x \in E^0_p|x \in E^{1\Z_2}_{p,-n}\}$. 

The group $E^1_{p,-n}$ is represented as the quotient group 
\begin{align}
    E^1_{p,-n} = (E^{1\Z}_{p,-n} \oplus E^{1\Z_2}_{p,-n})/{\cal P}_{p,-n}^1.
\end{align}
By construction, $E^{1\Z}_{p,-n} \cap E^{1\Z_2}_{p,-n} = \{\bm{0}\}$.
We write $\tilde E^1_{p,-n}=E^{1\Z}_{p,-n} \oplus E^{1\Z_2}_{p,-n}$.

\subsection{The first differential $d^1$}

\begin{figure}[]
\centering
\includegraphics[width=0.6\linewidth, trim=0cm 0cm 0cm 0cm]{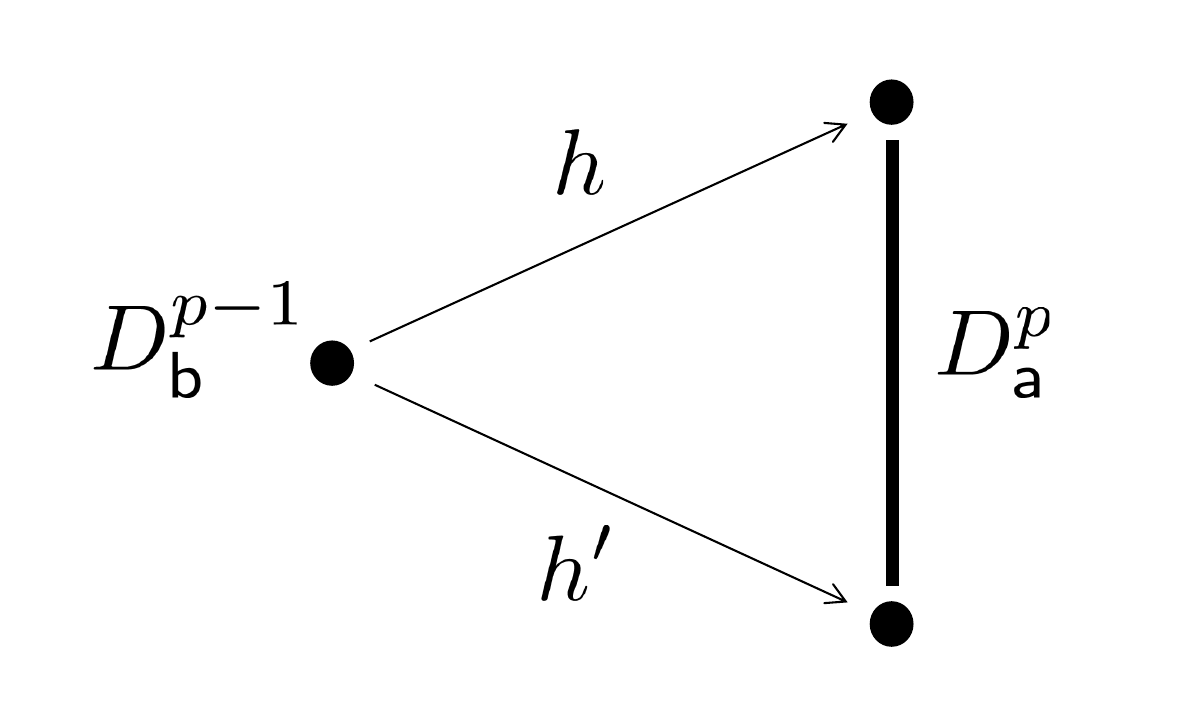}
\caption{The relationship between the boundary $(p-1)$-cells of the representative $p$-cell $D^p_{\sf a}$ and the ${\sf b}$th orbit of $(p-1)$-cells.}
\label{fig:cell_map}
\end{figure}

Let us focus on the ${\sf a}$-th and ${\sf b}$-th orbits of $p$- and $(p-1)$-cells, respectively.
Let $D^p_{\sf a}$ and $D^{p-1}_{\sf b}$ be representative of each orbit. 
The small block of the first differential is 
\begin{widetext}
\begin{align}
    &[d^1_{p,-n}]_{\sf ab}: 
    {}^\phi K^{{\cal G}}_{(z,c)+p-n} ({\cal G}/{\cal G}_{D^p_{\sf a}} \times D^p_{\sf a},{\cal G}/{\cal G}_{D^{p}_{\sf a}} \times \p D^p_{\sf a}) \nonumber \\
    &\to 
    {}^\phi K^{{\cal G}}_{(z,c)+(p-1)-n} ({\cal G}/{\cal G}_{D^{p-1}_{\sf b}} \times D^{p-1}_{\sf b},{\cal G}/{\cal G}_{D^{p-1}_{\sf b}} \times \p D^{p-1}_{\sf b}).
\end{align}
\end{widetext}
This is done by ``expanding" the orbit of Hamiltonians over the $p$-cells ${\cal G}({\cal G}/{\cal G}_{D^p_{\sf a}} \times D^p_{\sf a})$ by the orbit of Hamiltonians over the $(p-1)$-cells ${\cal G}({\cal G}/{\cal G}_{D^{p-1}_{\sf b}} \times D^{p-1}_{\sf b})$.~\footnote{
Note that, as a matter of mathematical fact, for a subgroup $H \subset G$, the group $G$ naturally acts on the set of left cosets $G/H$ of $H$ in $G$:
Let $G/H = \prod_{\sigma} g_\sigma H$ be a complete set of left coset representatives.
For $k \in G$, we have a unique representative $g_{k(\sigma)}$ such that $kg_{\sigma} = g_{k(\sigma)}h$, with $h \in H$.
The map $\sigma \mapsto k(\sigma)$ is well-defined, as for $h' \in H$, $kg_{\sigma}h = g_{k(\sigma)}hh' \in g_{k(\sigma)} H$.
The orbit of the action $G$ on $G/H$ in this sense is denoted by $G(G/H)$.}
There, each boundary of the $p$-cell $D^p_{\sf a}$, $D^{p-1}_i \in \p D^p_{\sf a}$ say, contribute to $[d^1_{p,-n}]_{\sf ab}$ if the $(p-1)$-cell $D^{p-1}_i$ is in the $b$-th orbit ${\cal G}({\cal G}/{\cal G}_{D^{p-1}_{\sf b}} \times D^{p-1}_{\sf b})$. 
To implement this, we write the boundary of $D^p_{\sf a}$ as the sum of orbit of $(p-1)$-cells as in 
\begin{align}
&\p D^p_{\sf a} \nonumber\\
&= \sum_{{\sf b} \in I^{p-1}_{\rm orb}} \sum_{h \in {\cal G}, h(D^{p-1}_{\sf b}) \in \p D^p_{\sf a}} n_{{\sf a},h({\sf b})} h(D^{p-1}_{\sf b}).
\label{eq:r_deco_ab}
\end{align}
See Fig.~\ref{fig:cell_map}. 
It is important to note that since PHS is an internal symmetry the sum of $h$ can be limited to $h \in {\cal G}^0 \coprod {\sf t}{\cal G}^0$ for the symmetry classes discussed in this paper. 

Fix $h \in {\cal G}^0 \coprod {\sf t}{\cal G}^0$ such that $h(D^{p-1}_{\sf b}) \in \p D^p_{\sf a}$. 
For an irrep $\alpha_r(\bx^p_{\sf a})$ of ${\cal G}^0_{\bx^p_{\sf a}}$, the mapped irrep $h^{-1}[\alpha_r(\bx^p_{\sf a})]=\alpha_{h^{-1}(r)}(h^{-1}(\bx^p_{\sf a}))$ is the irrep at $h^{-1}(\bx^p_{\sf a})$. 
Let $\{g_\s\}_{\s=1,\dots,|{\cal G}_{\bx^{p-1}_{\sf b}}|/|{\cal G}_{\bx^p_{\sf a}}|}$ be a complete set of the left coset representatives of ${\cal G}_{h^{-1}(\bx^p_{\sf a})}$ in ${\cal G}_{\bx^{p-1}_{\sf b}}$, say 
\begin{align}
    &{\cal G}_{\bx^{p-1}_{\sf b}}
    =
    \prod_{\s=1}^{|{\cal G}_{\bx^{p-1}_{\sf b}}|/|{\cal G}_{\bx^p_{\sf a}}|} g_\s {\cal G}_{h^{-1}(\bx^p_{\sf a})}.
\end{align}
Here, $g_\s$ takes an element in the group according to the following cases: 
$g_\s \in {\cal G}^0 \coprod {\sf t}{\cal G}^0$ if ${\sf t}{\cal G}_{\bx^{p-1}_{\sf b}}\neq \emptyset$ and ${\sf t}{\cal G}_{\bx^p_{\sf a}} = \emptyset$, and $s_\s \in {\cal G}^0$ otherwise.
The $p$-cells 
\begin{align}
(g_\s h^{-1})(D^p_{\sf a}),\quad \s=1,\dots,|{\cal G}_{\bx^{p-1}_{\sf b}}|/|{\cal G}_{\bx^p_{\sf a}}|,
\end{align}
adjacent to the representative $(p-1)$-cell $D^{p-1}_{\sf b}$.
We introduce the ``induced representation" of $h^{-1}[\alpha_r(\bx^p_{\sf a})]$ by $g_\s$s, which is a representation of ${\cal G}_{\bx^{p-1}_{\sf b}}$, as in 
\begin{align}
&{\rm Ind}_{\{g_\s\}}(h^{-1}[\alpha_r(\bx^p_{\sf a})])\nonumber\\
&:= \bigoplus_{\s=1}^{|{\cal G}_{\bx^{p-1}_{\sf b}}|/|{\cal G}_{\bx^p_{\sf a}}|} (g_\s h^{-1})[\alpha_r(\bx^p_{\sf a})],
\end{align}
whose character is 
\begin{align}
&\chi^{{\rm Ind}_{\{g_\s\}}(h^{-1}[\alpha_r(\bx^p_{\sf a})])}_{g \in {\cal G}^0_{\bx^{p-1}_{\sf b}}}\nonumber\\
&=\sum_{\s=1}^{|{\cal G}_{\bx^{p-1}_{\sf b}}|/|{\cal G}_{\bx^p_{\sf a}}|}
\delta_{g \in {\cal G}^0_{(g_\s h^{-1})(\bx^p_{\sf a})}} \chi^{(g_\s h^{-1})[\alpha_r(\bx^p_{\sf a})]}_g.
\end{align}
Note that when $\{g_\s\}_{\s=1,\dots,|{\cal G}_{\bx^{p-1}_{\sf b}}|/|{\cal G}_{\bx^p_{\sf a}}|} \subset {\cal G}^0$, this is the standard induced representation ${\rm Ind}_{{\cal G}^0_{h^{-1}(\bx^p_{\sf a})}}^{{\cal G}^0_{\bx^{p-1}_{\sf b}}} (h^{-1}[\alpha_r(\bx^p_{\sf a})])$.
We expand the representation ${\rm Ind}_{\{g_\s\}}(h^{-1}[\alpha_r(\bx^p_{\sf a})])$ by the irreps $\beta_{r'}(\bx^{p-1}_{\sf b})$ at $\bx^{p-1}_{\sf b}$.
Let us denote the inner product between two reps $\alpha,\beta$ of $G_0$ by $\braket{\alpha,\beta} = \frac{1}{|G_0|}\sum_{g \in G_0} (\chi^\alpha_g)^*\chi^\beta_g \in \Z_{\geq 0}$. 
Then, the expansion coefficient is given by 
\begin{widetext}
\begin{align}
&\Braket{\beta_{r'}(\bx^{p-1}_{\sf b}),{\rm Ind}_{\{g_\s\}}(h^{-1}[\alpha_r(\bx^p_{\sf a})])}_{{\cal G}^0_{\bx^{p-1}_{\sf b}}} \nonumber\\
&= 
\sum_{\s=1}^{|{\cal G}_{\bx^{p-1}_{\sf b}}|/|{\cal G}_{\bx^p_{\sf a}}|}
\frac{1}{|{\cal G}^0_{\bx^{p-1}_{\sf b}}|} \sum_{g \in {\cal G}^0_{\bx^{p-1}_{\sf b}}} (\chi^{\beta_{r'}(\bx^{p-1}_{\sf b})}_g)^* 
\delta_{g \in {\cal G}^0_{(g_\s h^{-1})(\bx^p_{\sf a})}} 
\chi^{(g_\s h^{-1})[\alpha_r(\bx^p_{\sf a})]}_g \nonumber\\
&= 
\sum_{\s=1}^{|{\cal G}_{\bx^{p-1}_{\sf b}}|/|{\cal G}_{\bx^p_{\sf a}}|}
\frac{1}{|{\cal G}^0_{\bx^{p-1}_{\sf b}}|} 
\sum_{g \in {\cal G}^0_{(g_\s h^{-1})(\bx^p_{\sf a})}} (\chi^{\beta_{r'}(\bx^{p-1}_{\sf b})}_g)^* \chi^{(g_\s h^{-1})[\alpha_r(\bx^p_{\sf a})]}_g \nonumber\\
&= 
\frac{|{\cal G}^0_{\bx^p_{\sf a}}|}{|{\cal G}^0_{\bx^{p-1}_{\sf b}}|}
\sum_{\s=1}^{|{\cal G}_{\bx^{p-1}_{\sf b}}|/|{\cal G}_{\bx^p_{\sf a}}|}
\Braket{\beta_{r'}(\bx^{p-1}_{\sf b})|_{{\cal G}^0_{(g_\s h^{-1})(\bx^p_{\sf a})}} , (g_\s h^{-1})[\alpha_r(\bx^p_{\sf a})]}_{{\cal G}^0_{(g_\s h^{-1})(\bx^p_{\sf a})}}\nonumber\\
&= 
\frac{|{\cal G}^0_{\bx^p_{\sf a}}|}{|{\cal G}^0_{\bx^{p-1}_{\sf b}}|}
\sum_{\s=1}^{|{\cal G}_{\bx^{p-1}_{\sf b}}|/|{\cal G}_{\bx^p_{\sf a}}|}
\Braket{(h g_\s^{-1})[\beta_{r'}(\bx^{p-1}_{\sf b})]|_{{\cal G}^0_{\bx^p_{\sf a}}} , \alpha_{r}(\bx^p_{\sf a})}_{{\cal G}^0_{\bx^p_{\sf a}}}.
\end{align}
Because $g_\s \in {\cal G}^0_{\bx^{p-1}_{\sf b}} \coprod {\sf t}{\cal G}^0_{\bx^{p-1}_{\sf b}}$, the equivalent class of the mapped irrep $g_\s^{-1}[\beta_{r'}(\bx^{p-1}_{\sf b})]$ depends only on which $g_\s$ belongs to ${\cal G}^0_{\bx^{p-1}_{\sf b}}$ or ${\sf t}{\cal G}^0_{\bx^{p-1}_{\sf b}}$. 
Namely, $g_\s^{-1}[\beta_{r'}(\bx^{p-1}_{\sf b})] \sim \beta_{r'}(\bx^{p-1}_{\sf b})$ for $g_\s \in {\cal G}^0_{\bx^{p-1}_{\sf b}}$ and $g_\s^{-1}[\beta_{r'}(\bx^{p-1}_{\sf b})] \sim {\sf t}[\beta_{r'}(\bx^{p-1}_{\sf b})]$ for $g_\s \in {\sf t}{\cal G}^0_{\bx^{p-1}_{\sf b}}$. 
Therefore, 
\begin{align}
    &\Braket{\beta_{r'}(\bx^{p-1}_{\sf b}),{\rm Ind}_{\{g_\s\}}(h^{-1}[\alpha_r(\bx^p_{\sf a})])}_{{\cal G}^0_{\bx^{p-1}_{\sf b}}} \nonumber\\
    &=\left\{\begin{array}{ll}
    \Braket{\left\{h [\beta_{r'}(\bx^{p-1}_{\sf b})]\oplus (h {\sf t})[\beta_{r'}(\bx^{p-1}_{\sf b})]\right\}|_{{\cal G}^0_{\bx^p_{\sf a}}},\alpha_{r}(\bx^p_{\sf a})}_{{\cal G}^0_{\bx^p_{\sf a}}}    
    & ({\sf t}{\cal G}_{\bx^{p-1}_{\sf b}}\neq \emptyset, {\sf t}{\cal G}_{\bx^p_{\sf a}} = \emptyset), \\
    \Braket{h[\beta_{r'}(\bx^{p-1}_{\sf b})]|_{{\cal G}^0_{\bx^p_{\sf a}}} , \alpha_{r}(\bx^p_{\sf a})}_{{\cal G}^0_{\bx^p_{\sf a}}} & ({\rm else}). \\
    \end{array}\right. 
    \label{eq:T_Frobenius}
\end{align}
Since (\ref{eq:T_Frobenius}) is a slight extension of the Frobenius reciprocity, we will refer to it as the Frobenius reciprocity as well.
Using the formula (\ref{eq:T_Frobenius}) and the connectivity (\ref{eq:r_deco_ab}) of representative $p$-cells, we find that the irrep $\alpha_r(\bx^p_{\sf a})$ at the $p$-cell $D^p_{\sf a}$ contributes to the irrep $\beta_{r'}(\bx^{p-1}_{\sf b})$ at the $(p-1)$-cell $D^{p-1}_{\sf b}$ by the integer 
\begin{align}
    \sum_{h \in {\cal G}, h(D^{p-1}_{\sf b}) \in \p D^p_{\sf a}} [\delta_{p-1}+\delta^T_{p-1}]_{{\sf a}r,h({\sf b})r'}.
    \label{eq:r_d1_n=0}
\end{align}
(Note that (\ref{eq:rdiff_Lp}) and (\ref{eq:rdiffT_Lp}) involve the connectivity $n_{i,i'}$.)
Here, we have used $h{\sf t}[\beta_{r'}(\bx^{p-1}_{\sf b})] \sim ({\sf t}' h)[\beta_{r'}(\bx^{p-1}_{\sf b})]$ with ${\sf t}' \in {\sf t}{\cal G}^0_{h(\bx^{p-1}_{\sf b})}$.
Since the $(p-1)$-cell $h(D^{p-1}_{\sf b}) = (h{\sf t})(D^{p-1}_{\sf b})$ is a boundary of $D^p_{\sf a}$, all the $p$- and $(p-1)$-cells in the expression (\ref{eq:T_Frobenius}) are in the relevant $p$-cells ${\cal C}_p$, meaning that the first differential $d^1_{p,-n}$ can be computed in the lattice $E^p_0$. 

In the same way, for a pair of irreps $\alpha_r(\bx^p_{\sf a})$ and $\beta_{r'}(\bx^p_{\sf a})$ with the values $W_\G^{\alpha_r(\bx^p_{\sf a})}=W_\G^{\beta_{r'}(\bx^{p-1}_{\sf b})}=1$ so that the winding number is well-defined for odd $n$, we have 
\begin{align}
    &\Braket{\beta_{r'}(\bx^{p-1}_{\sf b})+,{\rm Ind}_{\{g_\s\}}(h^{-1}[\alpha_r(\bx^p_{\sf a})+])}_{{\cal G}^0_{\bx^{p-1}_{\sf b}}\coprod \g {\cal G}^0_{\bx^{p-1}_{\sf b}}} \nonumber\\
    &=\left\{\begin{array}{ll}
    \Braket{\left\{ h [\beta_{r'}(\bx^{p-1}_{\sf b})+]\oplus (h {\sf t})[\beta_{r'}(\bx^{p-1}_{\sf b})+]\right\}|_{{\cal G}^0_{\bx^{p-1}_{\sf b}}\coprod \g {\cal G}^0_{\bx^{p-1}_{\sf b}}},\alpha_{r}(\bx^p_{\sf a})+}_{{\cal G}^0_{\bx^{p-1}_{\sf b}}\coprod \g {\cal G}^0_{\bx^{p-1}_{\sf b}}}
    & ({\sf t}{\cal G}_{\bx^{p-1}_{\sf b}}\neq \emptyset, {\sf t}{\cal G}_{\bx^p_{\sf a}} = \emptyset), \\
    \Braket{h[\beta_{r'}(\bx^{p-1}_{\sf b})+]|_{{\cal G}^0_{\bx^{p-1}_{\sf b}}\coprod \g {\cal G}^0_{\bx^{p-1}_{\sf b}}} , \alpha_{r}(\bx^p_{\sf a})+}_{{\cal G}^0_{\bx^{p-1}_{\sf b}}\coprod \g {\cal G}^0_{\bx^{p-1}_{\sf b}}} & ({\rm else}),  \\
    \end{array}\right. 
\end{align}
and restricting the summation to elements of chiral type and subtracting (\ref{eq:T_Frobenius}), we obtain
\begin{align}
    &\Braket{\beta_{r'}(\bx^{p-1}_{\sf b})+,{\rm Ind}_{\{g_\s\}}(h^{-1}[\alpha_r(\bx^p_{\sf a})+])}_{\g {\cal G}^0_{\bx^{p-1}_{\sf b}}} \nonumber\\
    &=\left\{\begin{array}{ll}
    \Braket{\left\{ h [\beta_{r'}(\bx^{p-1}_{\sf b})+]\oplus (h {\sf t})[\beta_{r'}(\bx^{p-1}_{\sf b})+]\right\}|_{{\cal G}^0_{\bx^{p-1}_{\sf b}}\coprod \g {\cal G}^0_{\bx^{p-1}_{\sf b}}},\alpha_{r}(\bx^p_{\sf a})+}_{\g {\cal G}^0_{\bx^{p-1}_{\sf b}}}
    & ({\sf t}{\cal G}_{\bx^{p-1}_{\sf b}}\neq \emptyset, {\sf t}{\cal G}_{\bx^p_{\sf a}} = \emptyset), \\
    \Braket{h[\beta_{r'}(\bx^{p-1}_{\sf b})+]|_{{\cal G}^0_{\bx^{p-1}_{\sf b}}\coprod \g {\cal G}^0_{\bx^{p-1}_{\sf b}}} , \alpha_{r}(\bx^p_{\sf a})+}_{\g {\cal G}^0_{\bx^{p-1}_{\sf b}}} & ({\rm else}).  \\
    \end{array}\right. 
\end{align}
Here, we have introduced the notation $\braket{\alpha+,\beta+}_{\g G_0} := \frac{1}{|G_0|} \sum_{g \in \g G_0} (\chi^{\alpha+}_g)^*\chi^{\beta+}_g \in \Z$.

With the preparation above, we develop the formula to compute $d^1_{p,-n}$.
We want to compute the expansion coefficients
\begin{align}
    &d^1_{p,-n}(\vec{a}^{p,-n}_\lambda) 
    = \sum_\kappa \vec a^{p-1,-n}_{\kappa} [M^{p,-n}_{\Z\Z}]_{\kappa \lambda}
    + \sum_\kappa \vec b^{p-1,-n}_\kappa [M^{p,-n}_{\Z\Z_2}]_{\kappa \lambda},\\
    &d^1_{p,-n}(\vec{b}^{p,-n}_\lambda) = \sum_\kappa \vec b^{p-1,-n}_\kappa [M^{p,-n}_{\Z_2\Z_2}]_{\kappa \lambda},
\end{align}
where $[M^{p,-n}_{\Z\Z}]_{\kappa \lambda} \in \Z$ and $[M^{p,-n}_{\Z\Z_2}]_{\kappa \lambda},[M^{p,-n}_{\Z_2\Z_2}]_{\kappa \lambda} \in \{0,1\}$.
Let $\pi^p_j: E^p_0 \to E^p_0$ the projection onto irreps at $j$-th $p$-cell. 
Namely, for $\vec{v} = (v_{ir}) \in E^p_0$, 
\begin{align}
    \pi^p_j(\vec{v}) = (0,\dots,0,\underbrace{v_{j1},v_{j2},\dots}_{i=j},0,\dots,0)^{tr}.
\end{align}

The $\lambda$-th basis vector $\vec{a}_\lambda^{p,-n} = (a^{p,-n}_{\lambda,ir})$ of $E^{1\Z}_{p,-n}$ represents the set of the massive Dirac Hamiltonians $H_{\bx^p_i}^{\alpha_r(\bx^p_i)}(\tilde \bx)$ of the irrep $\alpha_r(\bx^p_i)$ over the virtual $n$-sphere $\tilde \bx \in \{\bx^p_i\} \times \tilde S^n$ with $\Z$ invariant of $a^{p,-n}_{\lambda,ir}$. 
We pick a representative $p$-cell $D^p_{\sf a}$ from $\vec{a}_\lambda^{p,-n}$ such that $a^{p,-n}_{\lambda,{\sf a}r} \neq 0$ and consider the projection onto the ${\sf a}$-th $p$-cell 
\begin{align}
    \pi^p_{\sf a}(\vec a_\lambda^{p,-n})
    &= (0,\dots,0,\underbrace{a^{p,-n}_{\lambda,{\sf a} 1},a^{p,-n}_{\lambda,{\sf a} 2},\dots}_{i={\sf a}},0,\dots,0)^{tr}.
\end{align}
For the moment, $n = 0$.
From the Frobenius reciprocity (\ref{eq:T_Frobenius}), the vector  
\begin{align}
    \vec{w} = (\delta_{p-1}+\delta_{p-1}^T)^{tr} \pi_{\sf a}(\vec a_\lambda^{p,0}) \in E^{p-1}_0
\end{align}
represents how irreps $\alpha_1(\bx^p_{\sf a})^{\oplus a^{p,0}_{\lambda,{\sf a} 1}} \oplus \alpha_2(\bx^p_{\sf a})^{\oplus a^{p,0}_{\lambda,{\sf a} 2}} \oplus \cdots $ at the representative $p$-cell $D^p_{\sf a}$ form the ``induced representations" at adjacent $(p-1)$-cells $D^{p-1}_j \in \p D^p_{\sf a}$ and are expanded by the irreps at $(p-1)$-cells $D^{p-1}_j$.
Here, $\delta_{p-1}^{tr}, (\delta_{p-1}^T)^{tr}: E^p_0 \to E^{p-1}_0$ are the transpose of $\delta_{p-1},\delta_{p-1}^T$. 
Since $\vec{w}$ contains only the irreps at $(p-1)$-cells adjacent to $D^p_{\sf a}$, 
\begin{align}
    \vec{w} = \sum_{j \in {\cal C}_{p-1}, D^{p-1}_j \in \p D^p_{\sf a}} \pi^{p-1}_j (\vec w).
\end{align}
We compare the vector $\vec{w}$ with the basis $\{\vec a^{p-1,0}_{\kappa}\}_{\kappa}$ of $E^{1\Z}_{p-1,0}$.
In the view of (\ref{eq:r_d1_n=0}), each $(p-1)$-cells $D^{p-1}_j \in \p D^p_{\sf a}$ contributes to the expansion coefficient $[M^{\Z\Z}_{p,0}]_{\kappa\lambda}$ independently. 
We have 
\begin{align}
    [M^{\Z\Z}_{p,0}]_{\kappa\lambda}=\sum_{j \in {\cal C}_{p-1}, D^{p-1}_j \in \p D^p_{\sf a}} [\pi^{p-1}_j(\vec a^{p-1,0}_{\kappa})]^+ \pi^{p-1}_j(\vec w).
    \label{eq:MZZ_tmp}
\end{align}
Here, $[\pi^{p-1}_j(\vec a^{p-1,0}_{\kappa})]^+$ is the pseudoinverse of the vector $\pi^{p-1}_j(\vec a^{p-1,0}_{\kappa})$ as a $\dim(E^0_{p-1}) \times 1$ matrix.
(Note that the pseudoinverse of the null matrix is also the null matrix.)
Eq.(\ref{eq:MZZ_tmp}) can also be written as 
$[M^{\Z\Z}_{p,0}]_{\kappa\lambda}=\sum_{j \in {\cal C}_{p-1}} [\pi^{p-1}_j(\vec a^{p-1,0}_{\kappa})]^+ \pi^{p-1}_j(\vec w)$, and $[M^{\Z\Z}_{p,0}]_{\kappa\lambda}$ should not depend on a choice of representative $p$-cell $D^p_{\sf a}$, we get 
\begin{align}
     [M^{\Z\Z}_{p,0}]_{\kappa\lambda}
     =\frac{1}{{\cal N}(\vec a^{p,0}_\lambda)} \sum_{j \in {\cal C}_{p-1}} [\pi^{p-1}_j(\vec a^{p-1,0}_{\kappa})]^+ \pi^{p-1}_j\left((\delta_{p-1}+\delta_{p-1}^T)^{tr} \vec a_\lambda^{p,0}\right).
\end{align}
Here, ${\cal N}(\vec v) := |\{i \in {\cal C}_p|v_{ir} \neq 0\}|\in \N$ is the number of $p$-cells contained in $\vec{v}$.
For other degrees $n$, we have to implement the sign change (\ref{eq:r_sign_map}) if the orbit $\{ h(D^p_{\sf a})\}_{h \in {\cal G}}$ includes a TRS-type element $h \in {\sf t}{\cal G}^0$. 
We eventually get the formula
\begin{align}
     [M^{\Z\Z}_{p,-n}]_{\kappa\lambda}
     =\frac{1}{{\cal N}(\vec a^{p,-n}_\lambda)} \sum_{j \in {\cal C}_{p-1}} [\pi^{p-1}_j(\vec a^{p-1,-n}_{\kappa})]^+ \cdot \left\{
     \begin{array}{ll}
     \pi^{p-1}_j\left((\delta_{p-1}+\delta_{p-1}^T)^{tr} \vec a_\lambda^{p,-n}\right) & (n \in 4\Z),  \\
     \pi^{p-1}_j\left((\delta^\G_{p-1}+\delta_{p-1}^{\G,T})^{tr} \vec a_\lambda^{p,-n}\right) & (n \in 4\Z+1),  \\
     \pi^{p-1}_j\left((\delta_{p-1}-\delta_{p-1}^T)^{tr} \vec a_\lambda^{p,-n}\right) & (n \in 4\Z+2),  \\
     \pi^{p-1}_j\left((\delta^\G_{p-1}-\delta_{p-1}^{\G,T})^{tr} \vec a_\lambda^{p,-n}\right) & (n \in 4\Z+3).
     \end{array}\right.
\end{align}

For the coefficients $[M^{p,-n}_{\Z\Z_2}]_{\kappa \lambda}, [M^{p,-n}_{\Z_2\Z_2}]_{\kappa \lambda}$, the derivation is similar. 
We neglect the orientation of $p$-cells and the sign of components of vector. 
We have 
\begin{align}
     [M^{\Z\Z_2}_{p,-n}]_{\kappa\lambda}
     =\frac{1}{{\cal N}(\vec a^{p,-n}_\lambda)} \sum_{j \in {\cal C}_{p-1}} [\pi^{p-1}_j(\vec b^{p-1,-n}_{\kappa})]^+ \pi^{p-1}_j\left((|\delta_{p-1}|+|\delta_{p-1}^T|)^{tr} |\vec a_\lambda^{p,-n}|\right) \quad \mod 2, 
\end{align}
and 
\begin{align}
     [M^{\Z_2\Z_2}_{p,-n}]_{\kappa\lambda}
     =\frac{1}{{\cal N}(\vec b^{p,-n}_\lambda)} \sum_{j \in {\cal C}_{p-1}} [\pi^{p-1}_j(\vec b^{p-1,-n}_{\kappa})]^+ \pi^{p-1}_j\left((|\delta_{p-1}|+|\delta_{p-1}^T|)^{tr} \vec b_\lambda^{p,-n}\right) \quad \mod 2.
\end{align}

In this way, we get the first differential $d^1_{p,-n}$. 
The computation of $E^2$-page is the same as for the momentum-space AHSS. 
Introduce an integer lift 
\begin{align}
    \tilde d^1_{p,-n}: \tilde E^1_{p,-n} \to \tilde E^1_{p-1,-n}
\end{align}
of $d^1_{p,-n}$. 
Considering $M_{p,-n}^{\Z\Z_2}$ and $M_{p,-n}^{\Z_2\Z_2}$ as $\Z$-valued matrices does work for such a lift. 
Then, the group $E^2_{p,-n}$ is computed as the quotient of two integer sublattices of $E^0_p$: 
\begin{align}
    E^2_{p,-n} = \frac{\ker (\tilde d^1_{p,-n} \oplus {\rm Id}_{{\cal P}^1_{p-1,-n}})|_{\tilde E^1_{p,-n}}}{\im (\tilde d^1_{p-1,-n})+{\cal P}^1_{p,-n}}.
    \label{eq:r_e2_int}
\end{align}

\end{widetext}

\section{Constraint on $K$-group by $E_2$-pages
\label{sec:Analysis of E2 pages}
}
The $E_2$- or $E^2$-page is sometimes sufficient to determine the $K$-group as a $\Z$-module uniquely.
However, in general, to obtain the $K$-group from an $E_2$- or $E^2$-page, one must compute higher-order derivatives and then solve an extension problem.
Nevertheless, one can constrain the possible $K$-groups by comparing the momentum-space $E_2$-page and real-space $E^2$-page, which we discuss in this section.

To illustrate our method, Tables~\ref{tab:E_2} and \ref{tab:E^2} show the $E_2$- and $E^2$-pages for spinful class D superconductors with MSG $P2_11'$ and $A$-representation of pairing symmetry, respectively.
This kind of pair of $E_2$- and $E^2$-pages is the input for the following analysis.

\begin{table}[]
    \centering
    $$
    \begin{array}{l|cccc}
     n=0&0&0&0&0\\
     n=1&\Z&\Z_2\oplus\Z_4^{\oplus 3}&\Z&0\\
     n=2&0&\Z_2^{\oplus 3}&\Z_2&0\\
     n=3&0&\Z\oplus\Z_2^{\oplus 3}&\Z_2&\Z\\
     n=4&0&0&0&0\\
     n=5&\Z&0&\Z&0\\
     n=6&\Z_2^{\oplus 5}&0&\Z_2&0\\
     n=7&\Z_2^{\oplus 5}&\Z&\Z_2&\Z\\
     \hline
     E_2^{p,-n}&p=0&p=1&p=2&p=3\\
    \end{array}
    $$
    \caption{The momentum space $E_2$-page of spinful class D SC with MSG $P2_11'$ and $A$ representation of pairing symmetry.}
    \label{tab:E_2}
\end{table}

\begin{table}[]
    \centering
    $$
    \begin{array}{l|cccc}
     n=0&0&0&0&0\\
     n=1&\Z_2&\Z_2^{\oplus 3}&\Z_2^{\oplus 3}&\Z_2\\
     n=2&\Z_2&\Z_2^{\oplus 3}&\Z_2^{\oplus 3}&\Z_2\\
     n=3&\Z&\Z\oplus\Z_2^{\oplus 2}&\Z&\Z\\
     n=4&0&0&0&0\\
     n=5&0&0&0&0\\
     n=6&0&0&0&0\\
     n=7&\Z&\Z\oplus\Z_2^{\oplus 2}&\Z&\Z\\
     \hline
     E^2_{p,-n}&p=0&p=1&p=2&p=3\\
    \end{array}
    $$
    \caption{The real space $E^2$-page of spinful class D SC with MSG $P2_11'$ and $A$ representation of pairing symmetry}
    \label{tab:E^2}
\end{table}

\subsection{Higher differential}
\label{sec:Higher_differential}
In the momentum-space AHSS, the differential $d_r$ and $E_r$-page for $r \geq 1$ are defined successively as
\begin{align}
&d_r^{p,-n}: E_r^{p,-n} \to E_r^{p+r,-n-r+1}, \\
&E_{r+1}^{p,-n} := \ker d_r^{p,-n}/\im d_r^{p-r,-n+r-1}.
\end{align}
Here, $d_r \circ d_r=0$ holds.
Although the existence of higher-order differentials $d_r$ for $r\geq 2$ is mathematically guaranteed, and there is a physical picture of $d_r$ such as the creation of gapless points to generic momenta associated with band inversion at a high-symmetry point, there is no known formula to calculate $d_r$ automatically at this time.
See Appendix~\ref{app:HigherDifferentials} for the formal mathematical definition of the higher differentials via connecting homomorphisms.
Recall that we are interested in $K$-groups over a space in three dimensions.
The differentials $d_r$ for $r\geq 4$ are trivial, and the $E_4$-page is the limit.
In such cases, one of the composite $d_r^{p+r,-n-r+1} \circ d_r^{p,-n}$ for $r=2,3$ is always trivial, and the composite $d_r^{p+r,-n-r+1} \circ d_r^{p,-n}$ vanishes automatically, meaning that no constraints on $d_r^{p+r,-n-r+1}$ and $d_r^{p,-n}$ exist. 
In other words, any homomorphism $f \in {\rm Hom}_\Z(E_r^{p,-n},E_r^{p+r,-n-r+1})$ is a candidate for the true $d_r^{p,-n}$.
Listing all homomorphisms $f \in {\rm Hom}_\Z(E_r^{p,-n},E_r^{p+r,-n-r+1})$ for $r=2$ and $r=3$, we obtain all candidates for $E_4$-pages.

The same is the case for the real-space AHSS. 
The differential $d^r$ and $E^r$-page for $r \geq 1$ are defined successively as 
\begin{align}
    &d^r_{p,-n}: E^r_{p,-n} \to E^r_{p-r,-n+r-1}, \\
    &E^{r+1}_{p,-n} := \ker d^r_{p,-n}/\im d^r_{p+r,-n-r+1}, 
\end{align}
where $d^r \circ d^r=0$ holds. 
There is no known formula to compute the higher-differential $d^r$ for $r\geq 2$. 
(Nevertheless, we should note that Ref.~\cite{OnoShiozakiWatanabe_classification_SC_2022} successfully computed $d^2$ for the time-reversal symmetric SCs for spinful electrons for conventional pairing symmetry, based on the physical picture that $d^2$ is equivalent to the inevitable vortex zero modes enforced by crystalline symmetries.)
The $E^4$-page is the limit, one of the composite $d^r_{p-r,-n+r-1} \circ d^r_{p,-n}$ for $r=2,3$ is always trivial, and the composite $d^r_{p-r,-n+r-1} \circ d^r_{p,-n}$ vanishes automatically. 
Listing all homomorphisms $f \in {\rm Hom}_\Z(E^r_{p,-n},E^r_{p-r,-n+r-1})$ for $r=2$ and $r=3$, we get all candidates for $E^4$-pages. 

Let us consider how to list all possible homomorphisms $f \in {\rm Hom}_\Z(A,B)$ for a given pair of $\Z$-modules $A$ and $B$. Precisely, we are only interested in the $\Z$-modules $\ker f$ and $\coker f$ per isomorphisms. (Different homomorphisms $f_1,f_2 \in {\rm Hom}_\Z(A,B)$ may give the same kernel and cokernel as $\Z$-modules.) 
To simplify the problem and maintain physical relevance, we adopt the following working assumptions:

\begin{assumption}
\label{ass:1}
The rank of the $K$-group ${}^\phi K_G^{(z,c)-n}(X)$ is the same as the rank of the $E_2$-page $\bigoplus_{p=0}^3 E_2^{p,-n-p}$.
\end{assumption}

\begin{figure}[!]
\centering
\includegraphics[width=\linewidth, trim=0cm 0cm 0cm 0cm]{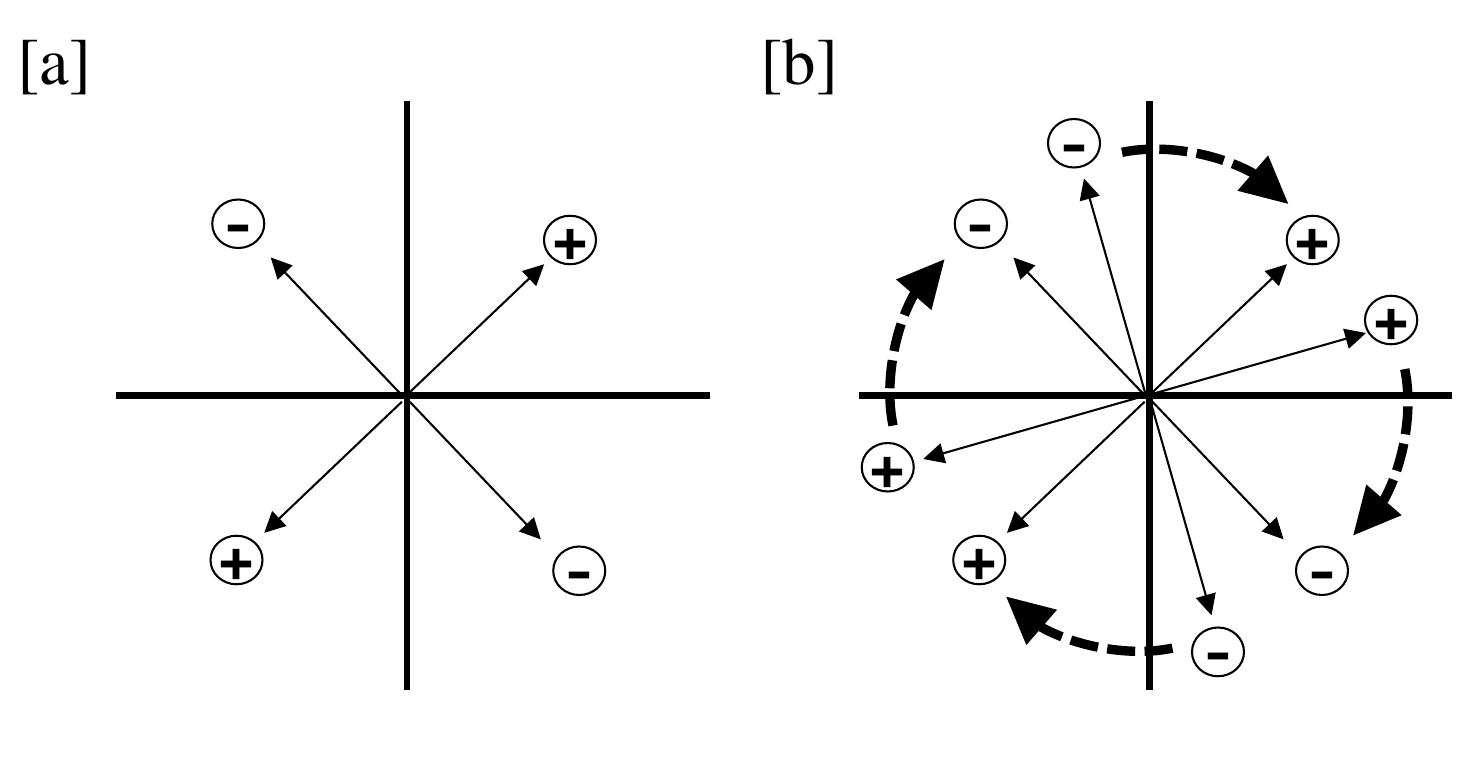}
\caption{
[a] The higher differentials $d_{r \geq 2}$ can be understood as band inversion followed by the creation of gapless points.
[b] When the same gapless point creation occurs twice, they can cancel each other out.
}
\label{fig:dr}
\end{figure}

Here, the rank of the $K$-group refers to the number of $\mathbb{Z}$ factors in its decomposition, i.e., the dimension of the free part, and does not include torsion components such as $\mathbb{Z}_2$. 
This property is known to hold for equivariant $K$-theory over any finite proper $G$-complex~\cite{LuckOliver_Chern_characters2001}. Although no proof exists for twisted and equivariant $K$-theory, we provide some reasons below for why this conjecture might be correct.
The higher differentials $d^{p,-n}_{r \geq 2}$ can be understood as the band inversion at a high-symmetry $p$-cell followed by the creation of gapless point inside adjacent $(p+r)$-cells~\cite{ShiozakiSatoGomi_Atiyah-Hirzebruch_2022}. 
For example, consider a $C_4$-symmetric $0$-cell and a band inversion at the $0$-cell followed by the creation of four gapless points in adjacent $2$-cells, as shown in Fig.~\ref{fig:dr} [a]. This is an example of nontrivial $d_2^{0,-n}(x)$ for $x \in E_2^{0,-n}$. The gapless points in the 2-cells cannot be removed on 1-cells by the $C_4$ symmetry.
Let us consider the same band inversion twice as shown in Fig.~\ref{fig:dr} [b], namely, the element $2 d_2^{0,-n}(x)$. We have two quartets of gapless points. Since each quartet can pass through 1-cells while preserving $C_4$ symmetry, the two quartets can annihilate each other. This implies that $2 d_2^{0,-n}(x) = 0$, that is, the image of $d_2^{0,-n}$ is a torsion. We expect a similar picture to be true for any higher-differentials $d_{r \geq 2}^{p,-n}$.

For the real-space AHSS, we assume the same one:
\begin{assumption}\label{ass:2}
The rank of the $K$-group ${}^\phi K^{\cal G}_{(z^{\rm int},c)-n}(\mathbb{E}^3)$ is the same as the rank of the $E^2$-page $\bigoplus_{p=0}^3 E^2_{p,-n-p}$.
\end{assumption}
We leave a discussion of the validity of this assumption for SCs. 
As discussed in \cite{Shiozaki_homology}, the origin of the second differential $d^2$ is a vortex zero mode of topological SC with a unit Chern number. 
The two-layered SC with Chern number 1 for each is equivalent to a single SC with Chern number 2 under the stable equivalence, and there are no vortex zeros. 
This means that the image of $d^2$ should be a 2-torsion. 

These assumptions imply that higher-order differentials do not reduce the rank, i.e., there is no $\Z$ in the image of $d_r$ and $d^r$ for $r=2,3$. 
Thus, the homomorphisms to be considered have the following form
\begin{align}
    f: \Z^{\oplus q} \oplus A \to B 
\end{align}
with $A$ and $B$ torsion $\Z$-modules. 
Let us denote $f_1=f|_{\Z^{\oplus q}}$ and $f_2=f|_A$ so that $f = f_1 \oplus f_2$. 
$\ker f$ and $\coker f$ are computed as follows. 
The rank of $\ker f$ is the same as $\Z^{\oplus q}$, and the torsion part of $\ker f$ is $\ker f_2$. 
(If $f_1(x)+f_2(a)=0$ and there exists $k \in \N$ such that $k(x,a)=0$, then $x=0$ and $f_2(a)=0$.)
Therefore, $\ker f$ is in a form $\Z^{\oplus q} \oplus \ker f_2$ and is determined by $f_2$ alone.
Using the third-isomorphism theorem, $\coker f = B/(\im f_1+\im f_2) \cong (B/\im f_2)/((\im f_1+\im f_2)/\im f_2)$. 
The corresponding theorem says there is a bijection between the set of sub-$\Z$-modules of $B$ including $\im f_2$ and the set of sub-$\Z$-modules of $B/\im f_2$. 
Here, the bijection is given by $\im f_1 + \im f_2 \mapsto \im [f_1': \Z^{\oplus q} \to B/\im f_2]$ where $f_1'$ is defined by $f_1'(n):= f_1(n) \mod \im f_2$. 
Therefore, 
\begin{align}
    \coker f 
    &\cong \frac{B/\im f_2}{\im [f_1': \Z^{\oplus q} \to B/\im f_2]} \nonumber \\
    &=\coker f_1'.
\end{align}

From the observation above, to list all the possible pairs of $\ker f$ and $\coker f$, we do the following.
First, we tabulate all possible homomorphisms $f_2:A \to B$. 
Next, we tabulate all possible homomorphisms $f_1':\Z^{\oplus q} \to B/\im f_2$ for each $f_2$.  

Computing the $E_2$ and $E^2$-pages explicitly for the symmetry classes summarized in Sec.~\ref{sec:symmetry and factor system}, we find that the torsion sub-$\Z$-modules appearing in the homomorphism $f_2: A\to B$ is in the following form
\begin{align}
f_2: \Z_2^{\oplus k} \oplus \Z_4^{\oplus l} \to \Z_2^{\oplus m} \oplus \Z_4^{\oplus n}.
\end{align}
Let us denote the matrix representation of $f_2$ as 
\begin{align}
    M_{f_2} = \left[ \begin{array}{cc}
        M^{\Z_2\to\Z_2}_{m \times k} & M^{\Z_4\to\Z_2}_{l \times m} \\
        M^{\Z_2\to\Z_4}_{n \times k} & M^{\Z_4\to\Z_4}_{l \times n} \\
    \end{array}\right]. 
\end{align}
By the basis transformation, the diagonal blocks can be Smith normal forms 
\begin{align}
    &M^{\Z_2\to\Z_2}_{m \times k} = \begin{pmatrix}
        I_{r_1\times r_1}&O\\
        O&O \\
        \end{pmatrix}, \nonumber\\
    &0 \leq r_1 \leq \min(m,k), 
        \label{eq;SNF_Z2}
\end{align}
\begin{align}
    &M^{\Z_4\to\Z_4}_{l \times n}=\begin{pmatrix}
        I_{r_2 \times r_2}&O&O\\
        O&2\times I_{r_3 \times r_3}&O\\
        O&O&O\\
    \end{pmatrix}, \nonumber\\
    &0\leq r_2+r_3 \leq \min(l,n).
\end{align}
For the off-diagonal ones $M^{\Z_4\to\Z_2}_{l \times m}$ and $M^{\Z_2\to\Z_4}_{n \times k}$, we consider all possible $\Z_2$- and $\Z_4$-valued matrices. 
For a given matrix $M_{f_2}$, applying the method in Appendix~\ref{app:Computation in Z-module}, we get $\ker f_2$ and $\coker f_2$.

For a given $f_2$, the quotient group $B/\im f_2$ is also in a form $\Z_2^{\oplus s} \oplus \Z_4^{\oplus t}$. 
We tabulate all homomorphisms $f_1': \Z^{\oplus q} \to \Z_2^{\oplus s} \oplus \Z_4^{\oplus t}$, of which matrix expression is 
\begin{align}
    M_{f_1'} = \left[ 
    \begin{array}{cc}
        M^{\Z \to \Z_2}_{q \times s} \\
        M^{\Z \to \Z_4}_{q \times t}
    \end{array}
    \right].
\end{align}
By the basis transformation, $M^{\Z \to \Z_2}_{q \times s}$ can be the Smith normal form, the same as (\ref{eq;SNF_Z2}). 
For $M^{\Z \to \Z_4}_{q \times t}$, we consider all possible $\Z_4$-valued matrices. 
Using the method in Appendix~\ref{app:Computation in Z-module}, we get $\coker f_1'$.

Following the procedure above, from a $E_2$-page of the momentum space , we first compute possible $d_2^{p,-n}$ to get the $E_3$-page as $E_3^{p,-n}=\ker d_2^{p,-n}$ and $E_3^{p+2,-n-1}=\coker d_2^{p,-n}$ for $p=0,1$. 
For each candidate $E_3$-page, we compute possible $d_3^{p,-n}$ to get $E_4^{0,-n}=\ker d_3^{0,-n}$ and $E_4^{3,-n-2}=\coker d_3^{0,-n}$. 
Note that $E_4^{p,-n} = E_3^{p,-n}$ for $p=1,2$. 
Eventually, we get the set of candidates of $E_4$-pages, which we denote by ${\cal S}_k = \{cE_{4,i}=\{cE_{4,i}^{p,-n}\}_{p=0,\dots,3, n=0,\dots,7}\}_{i=1,\dots,|{\cal S}_k|}$.
Similarly, from a $E^2$-page, we compute possible $d^3_{p,-n}$ to get the candidates of $E^3$-pages by $E^3_{p,-n} = \ker d^2_{p,-n}$ and $E^3_{p-2,-n+1}=\coker d^2_{p,-n}$ for $p=2,3$, and for each $E^3$-page we compute possible $d^3_{3,-n}$ to get $E^4_{3,-n}=\ker d^3_{3,-n}$ and $E^4_{0,-n+2} = \coker d^3_{3,-n}$ with $E^4_{p,-n} =E^3_{p,-n}$ for $p=1,2$. 
We denote the set of candidate $E^4$-pages by ${\cal S}_r = \{cE^{4,i}=\{cE^{4,i}_{p,-n}\}_{p=0,\dots,3, n=0,\dots,7}\}_{i=1,\dots,|{\cal S}_r|}$.

\subsection{Extension}
\label{sec:extension}
The $E_4$-page approximates the momentum space $ K$-group in the following sense. 
We have a filtration
\begin{align}
    &0 = F_4K^{-n}\subset F_3K^{-n} \subset F_2K^{-n} \nonumber\\
    &\subset F_1K^{-n} \subset F_0K^{-n} ={}^\phi K^{(z,c)-n}_G(X), 
\end{align}
where each quotient group is given as 
\begin{align}
    F_pK^{-n}/F_{p+1}K^{-n} \cong E_4^{p,-p-n}.\label{eq:ext_kK}
\end{align}
The $K$-group is obtained by solving the extension problems as $\Z$-module sequentially.
$F_3K^{-n}=E_4^{3,-3-n}$. 
Then, $F_pK^{-n}$ is given by an extension of $E_4^{p,-p-n}$ by $F_{p+1}K^{-n}$ for $p=2,1,0$ in order.

Similarly, the $E^4$-page approximates the real space $K$-group as 
\begin{align}
    &0 = F_{-1}K_{-n} \subset F_0K_{-n} \subset F_1K_{-n} \nonumber\\
    &\subset F_2K_{-n} \subset F_3K_{-n}={}^\phi K_{(z^{\rm int},c)-n}^{{\cal G}}(\R^3)
\end{align}
with 
\begin{align}
    F_pK_{-n}/F_{p-1}K_{-n} \cong E^4_{p,-p-n}.\label{eq:ext_rK}
\end{align}
We have $F_0K_{-n}=E^4_{0,-n}$, and $F_pK_{-n}$ is an extension of $E^4_{p,-p-n}$ by $F_{p-1}K_{-n}$ for $p=1,2,3$ in order.

Given a pair of $\Z$-modules $A$ and $B$, equivalence classes of $\Z$-module extensions of $A$ by $B$ are classified by ${\rm Ext}_\Z^1(A,B)$~\cite{Weibel_book_homological_algebra}. 
The group ${\rm Ext}^1_\Z(A,B)$ has the following properties 
\begin{align}
{\rm Ext}_\Z^1\Big(\bigoplus_i A_i ,\bigoplus_j B_j\Big) = \bigoplus_{i,j}
{\rm Ext}_\Z^1(A_i,B_j), 
\end{align}
and 
\begin{align}
&{\rm Ext}_\Z^1(\Z_n,\Z) = \Z_n,\\
&{\rm Ext}_\Z^1(\Z_n,\Z_m) = \Z_{{\rm gcd}(n,m)},\\
&{\rm Ext}_\Z^1(\Z,\Z_n) ={\rm Ext}_\Z^1(\Z,\Z) = 0.
\end{align}
The free $\Z$-module $\Z$ is not extended. 
Hereafter we assume $A$ is a torsion $\Z$-module. 
The abelian group structure of ${\rm Ext}^1_\Z(A,B)$ is represented by the Baer sum of two extensions $0 \to B \to E_i \to A \to 0$ for $i=0,1$. 
(See Appendix~\ref{app:BaerSum} for the construction of the Baer sum.)
Thus, all extensions can be constructed from the extensions corresponding to the generators of the $\Z$-module ${\rm Ext}_\Z^1(A,B)$.

However, it is not feasible to compute extensions for all values of ${\rm Ext}_\Z^1(A,B)$.
For example, $|{\rm Ext}_\Z^1(\Z_2^{\oplus n},\Z^{\oplus m})| = 2^{nm}$.
Changing the basis of $A$ and $B$ induces an automorphism of ${\rm Ext}_\Z^1(A,B)$, but the $\Z$-module obtained as an extension via this basis transformation does not change.
For example, ${\rm Ext}^1_\Z(\Z_4,\Z)= \Z_4$, and each value of $\Z_4$ gives $\Z\oplus \Z_4,\Z,\Z\oplus \Z_2,\Z$, respectively.
Here, the second and fourth extensions are related via the basis transformation of $\Z_4$ (or $\Z$) multiplying the basis by $-1$.  
Thus, if we are only interested in the $\Z$-module obtained as extensions, we can contract the elements of ${\rm Ext}_\Z^1(A,B)$ via basis transformations.
We will now formalize this contraction procedure.

We express an element of the group ${\rm Ext}_\Z^1(A,B)$ as a matrix $M$, which we call an extension matrix, using the following procedure.
First, we decompose $A$ into its invariant factors as
\begin{align}
A = \Z_{p_1}^{\oplus n_1} \oplus \cdots \oplus \Z_{p_k}^{\oplus n_k},
\quad
p_k|p_{k-1}|\cdots |p_1.
\end{align}
Here, $p|q$ denotes that $p$ divides $q$.
We denote the basis of $A$ as
\begin{align}
a^{(p_1)}_1,\dots,a^{(p_1)}_{n_1},\dots, a^{(p_k)}_1,\dots,a^{(p_k)}_{n_k}.
\end{align}
Similarly, we decompose $B$ into its invariant factors as
\begin{align}
&B =\Z^{\oplus m_0}\oplus \Z_{q_1}^{\oplus m_1} \oplus \cdots \oplus \Z_{q_l}^{\oplus q_l},\nonumber \\
&q_l|q_{l-1}|\cdots |q_1.
\end{align}
We denote the basis of $B$ as
\begin{align}
{\cal B} 
&= (b^{(q_0)}_1,\dots,b^{(q_0)}_{m_0},\nonumber \\
&\quad \quad b^{(q_1)}_1,\dots,b^{(q_1)}_{m_1},
\dots, b^{(q_l)}_1,\dots,b^{(q_l)}_{m_l}),
\end{align}
where we formally wrote $q_0 = \infty$.
An element of ${\rm Ext}^1_\Z(A,B) \supset {\rm Ext}^1_\Z(\Z_p,B)$ represents how $``p" \in \Z_p$ is ``carried-up" by an element of $B$. 
We denote this correspondence as an extension matrix $M$ by
\begin{align}
&f_M ({\cal A}) = {\cal B} M, 
\end{align}
with the basis 
\begin{align}
    {\cal A} &= (p_1 a^{(p_1)}_1,\dots,p_1 a^{(p_1)}_{n_1},\nonumber \\
    &\quad \quad \dots, p_k a^{(p_k)}_1,\dots,p_k a^{(p_k)}_{n_k}).
\end{align}

Next, let's consider possible basis transformations for ${\cal A}$ and ${\cal B}$.
For ${\cal B}$, any basis transformations that do not change the structure of the $\Z$-module $B$ are allowed.
Since it is allowed to add linear combinations of basis elements $b^{(q_j)}_t$ with $i\leq j$ to a basis $b^{(q_i)}_s$, we can use the following block lower triangular unimodular matrix in the form
\begin{align}
V =
\left[ \begin{array}{ccccc}
v_{00} & O & O &\cdots \\
v_{10} & v_{11} & O \\
v_{20} & v_{21} & v_{33} \\
\vdots \\
\end{array} \right]
\end{align}
as a general basis transformation for ${\cal B}$.
On the other hand, the basis transformation of ${\cal A}$ to be considered is not a basis transformation that does not change the structure as $\Z$-module $A$, but a basis transformation that keeps the integer lattice spanned by ${\cal A}$ invariant.
For example, if we add $r$ times $p_j a^{(p_j)}_t$ to $p_i a^{(p_i)}_s$, we have
\begin{align}
p_i a^{(p_i)}_s + r p_j a^{(p_j)}_t
= p_i (a^{(p_i)}_s + r \frac{p_j}{p_i} a^{(p_j)}_t)
\end{align}
which is well-defined only if $p_i|p_j$, i.e., in the case where $i \geq j$. 
Therefore, the basis transformations for ${\cal A}$ are given by block upper triangular unimodular matrices of the form
\begin{align}
U =
\left[ \begin{array}{ccccc}
u_{11} & u_{12} & u_{13} &\cdots \\
O & u_{22} & u_{23} \\
O & O & u_{33} \\
\vdots \\
\end{array} \right].
\end{align}
Through these basis transformations, the entension matrix $M$ changes to
\begin{align}
M \mapsto V^{-1} M U. 
\label{eq:extM_tr}
\end{align}
Since the inverse of a lower-triangular block unimodular matrix is also a lower-triangular block unimodular, the above transformation is nothing but a ``directional" Gauss-Jordan elimination from the upper left to the lower right.
We write the extension matrix $M$ as a block matrix
\begin{align}
&M = (M^{(q_i,p_j)})_{0 \leq i \leq l, 1 \leq j \leq k},\\
&M^{(q_i,p_j)} = (M^{(q_i,p_j)}_{st})_{1 \leq s \leq m_i, 1\leq t \leq n_j},
\end{align}
where $M^{(q_i,p_j)}_{st}$ takes values in $\Z_{{\rm gcd}(p_j,q_i)}$.
(Here, we set ${\rm gcd}(p_j,\infty) = \Z_{p_j}$.)
The operations allowed in the ``directional" Gauss-Jordan elimination are as follows:

- Multiply a row vector by $(-1)$.

- Multiply a column vector by $(-1)$.

- (Row reduction)
Add an integer multiple $r v^{(q_i)}_s, r \in \Z,$ of a $(i,s)$th row vector $v^{(q_i)}_s = (M^{(q_i,p_j)}_{st})_{1 \leq j \leq k, 1 \leq t \leq n_j}$ to other row vector $v^{(q_{i'})}_{s'}$ with $(q_i,s) \neq (q_{i'},s')$ and $i \geq i'$.
Here, when adding the row vector $r v^{(q_i)}_s$ to the row vector $v^{(q{i'})}_{s'}$ with $i>i'$, take the modulo ${\rm gcd}(q_{i'},p_j)$ of the entries $r M^{(q_i,p_j)}_{st}$.

- (Column reduction) Add an integer multiple $r v^{(p_j)}_t, r \in \Z,$ of a $(j,t)$th column vector $v^{(p_j)}_t = (M^{(q_i,p_j)}_{st})_{1 \leq i \leq l, 1 \leq s \leq m_i}$ to other column vector $v^{(p_{j'})}_{t'}$ with $(p_j,t) \neq (p_{j'},t')$ and $j \geq j'$.
Here, when adding the column vector $r v^{(p_j)}_t$ to the column vector $v^{(p_{j'})}_{t'}$ with $j>j'$, take the modulo ${\rm gcd}(q_{i},p_{j'})$ of the entries $r M^{(q_i,p_j)}_{st}$.

The extension matrix $M$ can be transformed to a standard form to some extent by using the transformation (\ref{eq:extM_tr}), which performs the Smith decomposition in order, starting from the top-left block.
\begin{align}
M = \begin{array}{c|c|c}
M^{(q_0,p_1)}&M^{(q_0,p_2)} &\cdots \\
\hline
M^{(q_1,p_1)}&M^{(q_1,p_2)} \\
\hline 
\vdots & & \\
\end{array}.
\end{align}
First, take $M^{(q_0,p_1)}$ in Smith normal form.
\begin{align}
\to 
\begin{array}{c|c|c}
\begin{array}{c|c}
\Lambda^{(q_0,p_1)}&O\\
\hline
O&O\\
\end{array}
&M^{(q_0,p_2)} &\cdots \\
\hline
M^{(q_1,p_1)}&M^{(q_1,p_2)} \\
\hline 
\vdots & & \\
\end{array}, 
\end{align}
where $\Lambda^{(q_0,p_1)}$ is a diagonal matrix.
In the $(q_0,p_1)$ block, the components of the $(q_0,p_1)$ block remain unchanged at 0 even after performing basis transformations on the rows and columns that have component 0.
So, the lower half of $M^{(q_0,p_2)}$ can be taken in Smith normal standard form.
\begin{align}
\to 
\begin{tabular}{c|c|c|c|c}
$\Lambda^{(q_0,p_1)}$&$O$&\multicolumn{2}{c|}{$*$}&\multirow{3}{*}{$M^{(q_0,p_3)}$}\\
\cline{1-4}
\multirow{2}{*}{$O$}&\multirow{2}{*}{$O$}&$\Lambda^{(q_0,p_2)}$&$O$&\\
\cline{3-4}
&&$O$&$O$&\\
\cline{1-5}
\multicolumn{2}{c|}{$M^{(q_1,p_1)}$}& \multicolumn{2}{c|}{$M^{(q_1p_2)}$}&$M^{(q_1,p_3)}$\\
\end{tabular}, 
\end{align}
where $*$ is some integer matrix.
Similarly, $M^{(q_1,p_1)}$ can also be taken in Smith normal form on the right side.
\begin{align}
\to 
\begin{tabular}{c|cc|cc|c}
$\Lambda^{(q_0,p_1)}$&\multicolumn{2}{c|}{$O$}&\multicolumn{2}{c|}{$*$}&\multirow{3}{*}{$M^{(q_0,p_3)}$}\\
\cline{1-5}
\multirow{2}{*}{$O$}&\multicolumn{2}{c|}{\multirow{2}{*}{$O$}}&\multicolumn{1}{c|}{$\Lambda^{(q_0,p_2)}$}&$O$&\\
\cline{4-5}
&&&\multicolumn{1}{c|}{$O$}&$O$&\\
\cline{1-6}
\multirow{2}{*}{$*$}&\multicolumn{1}{c|}{$\Lambda^{(q_1,p_1)}$}&$O$& \multicolumn{2}{c|}{\multirow{2}{*}{$M^{(q_1,p_2)}$}}&\multirow{2}{*}{$M^{(q_1,p_3)}$}\\
\cline{2-3}
&\multicolumn{1}{c|}{$O$}&$O$&&& \\
\end{tabular}.
\end{align}
In this way, taking the Smith normal form of the extension matrix $M$ in order from the upper left, we can partially reduce it. 
The matrix $\Lambda^{(q_i,p_j)}$ is a $\Z_{{\rm gcd}(p_j,q_i)}$-valued diagonal matrix, but its entries can be limited from 1 to the integer no larger than ${\rm gcd}(p_j,q_i)/2$ due to the sign change of the basis.
Furthermore, if an entry of the diagonal matrix $\Lambda^{(q_i,p_j)}$ is $\lambda$, then all entries in the blocks to the right and below it (denoted by $*$ in the above matrix) can be replaced by their remainders when divided by $\lambda$. 
In particular, if $\lambda=1$, the entry can be replaced by 0.

The standard form of the extension matrix $M$ in the sense described above can be enumerated by the following procedure. 
First, determine the non-zero entries of the Smith normal form $\Lambda^{(q_i,p_j)}$. 
This procedure simply constructs a set of pairs between the direct summands of $A$ and $B$: $A = \bigoplus_{i=1}^k \Z_{p_i}^{\oplus n_i}, B = \bigoplus_{j=0}^{l} \Z_{q_j}^{\oplus m_j}$. 
To enumerate all such pairs where there are $d$ pairs and ${\tilde p_1,\dots,\tilde p_d}$ has no more than $n_i$ occurrences of $p_i$ and ${\tilde q_1,\dots,\tilde q_d}$ has no more than $m_j$ occurrences of $q_j$, we need to consider all possible sets of $d$ pairs $\left\{(\tilde p_r,\tilde q_r) \in \{p_1,\dots,p_k\}\times \{q_0,\dots,q_l\}\right\}_{r=1}^d$. 
Moreover, the upper bound for the number of pairs $d$ is given by ${\rm min}(\sum_{i=1}^k n_k, \sum_{j=0}^l m_j)$. Therefore, the set of possible sets of pairs is given by the following expression:
\begin{widetext}
\begin{align}
    &\Big\{\big\{(\tilde p_r,\tilde q_r) \in \{p_1,\dots,p_k\}\times \{q_0,\dots,q_l\}\big\}_{r=1}^d \Big| \sum_{r=1}^d \delta_{p_i,\tilde p_r} \leq n_i {\rm\ for\ } i = 1,\dots,k, \nonumber \\
    &\quad  {\rm \ and\ } \sum_{r=1}^d \delta_{q_j,\tilde q_r} \leq m_j {\rm \ for\ } j=1,\dots,l, 
    {\rm \ and\ } d = 0,\dots, {\rm min}(\sum_{i=1}^k n_i, \sum_{j=0}^l m_j) \Big\}.
\end{align}
\end{widetext}
Next, for each set of pairs ${(\tilde p_r,\tilde q_r)}_{r=1}^d$, we assign possible integer values to the diagonal entries of the diagonal matrix $\Lambda^{(\tilde q_r,\tilde p_r)}$. 
Based on this assignment, we enumerate all possible assignments of entries for the right and bottom blocks of the diagonal matrix $\Lambda^{(q_j,p_i)}$ to obtain a set of reduced extension matrices $M$.

As an example, consider the case where $A=\Z_4^{\oplus 2} \oplus \Z_2^{\oplus 3}$ and $B=\Z^{\oplus 3} \oplus \Z_4^{\oplus 2} \oplus \Z_2^{\oplus 2}$, and the set of pairs is chosen to be ${(4,\infty),(2,\infty),(4,4),(2,2)}$. 
Using this, we obtain the following initial extension matrix $M_0$. 
\begin{align}
    M_0 = \begin{array}{c|cc|ccc}
         & \Z_4&\Z_4&\Z_2&\Z_2&\Z_2 \\
         \hline
        \Z & \lambda_1 &&&&\\
        \Z &&&\lambda_2&&\\
        \Z&&&&&\\
        \hline 
        \Z_4&&\lambda_3&&&\\
        \Z_4&&&\\
        \hline
        \Z_2&&&&\lambda_4&\\
        \Z_2&&&&&\\
    \end{array}. \label{eq:ext_m0}
\end{align}
Here, $\lambda_1,\lambda_3 \in {1,2}$ and $\lambda_2,\lambda_4 \in {1}$. Blanks represent $0$. 
For each of the four possible combinations of $(\lambda_1,\lambda_2,\lambda_3,\lambda_4)$, namely $(1,1,1,1)$, $(1,1,2,1)$, $(2,1,1,1)$, and $(2,1,2,1)$, we obtain the extension matrices $M$ of the following forms.
\begin{align}
    \begin{array}{c|cc|ccc}
         & \Z_4&\Z_4&\Z_2&\Z_2&\Z_2 \\
         \hline
        \Z & 1 &&&&\\
        \Z &&&1&&\\
        \Z&&&&&\\
        \hline 
        \Z_4&&1&&&\\
        \Z_4&&&\\
        \hline
        \Z_2&&&&1&\\
        \Z_2&&&&&\\
    \end{array},
\end{align}
\begin{align}
    \begin{array}{c|cc|ccc}
         & \Z_4&\Z_4&\Z_2&\Z_2&\Z_2 \\
         \hline
        \Z & 1 &&&&\\
        \Z &&&1&&\\
        \Z&&&&&\\
        \hline 
        \Z_4&&2&*&*&*\\
        \Z_4&&&\\
        \hline
        \Z_2&&*&&1&\\
        \Z_2&&*&&&\\
    \end{array},
\end{align}
\begin{align}
    \begin{array}{c|cc|ccc}
         & \Z_4&\Z_4&\Z_2&\Z_2&\Z_2 \\
         \hline
        \Z &2&&*&*&*\\
        \Z &&&1&&\\
        \Z&&&&&\\
        \hline 
        \Z_4&*&1&&&\\
        \Z_4&*&&\\
        \hline
        \Z_2&*&&&1&\\
        \Z_2&*&&&&\\
    \end{array},
\end{align}
\begin{align}
    \begin{array}{c|cc|ccc}
         & \Z_4&\Z_4&\Z_2&\Z_2&\Z_2 \\
         \hline
        \Z &2&&*&*&*\\
        \Z &&&1&&\\
        \Z&&&&&\\
        \hline 
        \Z_4&*&2&*&*&*\\
        \Z_4&*&&\\
        \hline
        \Z_2&*&*&&1&\\
        \Z_2&*&*&&&\\
    \end{array}.\label{eq:extmat_ex}
\end{align}
The symbol $* \in {0,1}$ is independent, and all combinations are considered.

The set of extension matrices $M$ obtained by the above procedure is still large, so we further perform the following (I) and (II) reductions:

(I)
The row and column reduction. 
For example, the set of extension matrices in the form (\ref{eq:extmat_ex}) contains the extension matrix in the form 
\begin{align}
    \begin{array}{c|cc|ccc}
         & \Z_4&\Z_4&\Z_2&\Z_2&\Z_2 \\
         \hline
        \Z &2&&*&1&\\
        \Z &&&1&&\\
        \Z&&&&&\\
        \hline 
        \Z_4&*&2&*&*&1\\
        \Z_4&1&&\\
        \hline
        \Z_2&*&*&&1&\\
        \Z_2&*&*&&&\\
    \end{array}, 
\end{align}
(where $*$s are arbitrary), but it can be further reduced to
\begin{align}
    \to \begin{array}{c|cc|ccc}
         & \Z_4&\Z_4&\Z_2&\Z_2&\Z_2 \\
         \hline
        \Z &2&&&1&\\
        \Z &&&1&&\\
        \Z&&&&&\\
        \hline 
        \Z_4&&2&&&1\\
        \Z_4&1&&\\
        \hline
        \Z_2&&*&&&\\
        \Z_2&&*&&&\\
    \end{array}.
\end{align}

(II)
Furthermore, for the direct summands of $\Z_2$, we transform the off-diagonal blocks to Hermite normal form by a basis transformation: 
we decompose the extension matrix $M$ into $\Z_2$ and non-$\Z_2$ components as
\begin{align}
M = \begin{pmatrix}
M^{(*,*)} & M^{(*,2)} \\
M^{(2,*)} & M^{(2,2)}
\end{pmatrix}. 
\end{align}
We use unimodular transformations $U$ and $V$ to bring $H^{(2,*)} = UM^{(2,*)}$ into the row-style Hermite normal form, and $H^{(*,2)}= M^{(*,2)}V$ into the column-style Hermite normal form to get 
\begin{align}
M &\mapsto 
\begin{pmatrix}
1\\
&U \\ 
\end{pmatrix} 
M 
\begin{pmatrix}
1 \\
&V\\
\end{pmatrix}\nonumber \\
&= 
\begin{pmatrix}
M^{(*,*)} & H^{(*,2)} \\
H^{(2,*)} & UM^{(2,2)}V\\
\end{pmatrix}.
\end{align}

Repeat steps (I) and (II) in this order twice to further compress the set of extension matrices $\{M\}$.
For a given extension matrix $M$, the extension corresponding to the matrix $M$ is computed by the Baer sum. 
Thus, we can collect all the possible $\Z$-modules obtained by extensions. 
Supplemental Material~\cite{Supplemental_Material} lists the set of $\Z$-modules obtained as extensions of $A$ by $B$ for a given $A$ and $B$ that appeared in the calculations of this paper. 

\subsection{Candidates of $K$-groups}
For the $i$-th candidate $cE_{4,i} \in {\cal S}_k$ on the $E_4$-page, we write the set of $K$-groups $^{\phi}K^{(z,c)-n}_G(X)$ that can be obtained as a result of the extension (\ref{eq:ext_kK}) as
\begin{align}
cK^{-n}_i= \{cK^{-n}_{i,a}\}_{a=1,\dots,|cK^{-n}_i|}, \quad n=0,\dots,7.
\end{align}
Similarly, for the $j$-th candidate $cE^{4,j} \in {\cal S}_r$ on the $E^4$ page, we write the set of $K$-groups $^{\phi} K^{\cal G}_{(z^{\rm int},c)-n}(\mathbb{E}^3)$ that can be obtained as a result of the extension (\ref{eq:ext_rK}) as
\begin{align}
cK_{-n}^j= \{cK_{-n}^{j,a}\}_{a=1,\dots,|cK_{-n}^j|}, \quad n=0,\dots,7.
\end{align}
Now, for all degrees $n=0,\dots,7$, if the intersection of the $K$-group candidates $cK^{n}_i$ and $cK_{-n}^j$ is not an empty set, there is no contradiction with the isomorphism (\ref{eq:iso_k_r}). 
The set of compatible pairs $(i,j)$ is defined as
\begin{align}
&{\cal S}_{\rm pair}=\{(i,j)\nonumber\\
&|cK^{n}_i \cap cK^j_{-n} \neq \emptyset {\rm\ for\ all\ } n=0,\dots,7\}.
\end{align}
The set of $\mathbb{Z}$-modules that can be candidates for the $K$-groups of degree $(-n)$ in real space (i.e., the $K$-groups of degree $n$ in momentum space) is obtained by the intersection
\begin{align}
SK_{-n}:= \bigcap_{(i,j) \in {\cal S}_{\rm pair}} cK^{n}_i \cap cK^j_{-n}. 
\end{align}
The set of $K$-group candidates $SK_{-n}$ is shown in the Supplemental material~\cite{Supplemental_Material} and is also available at http \href{https://www2.yukawa.kyoto-u.ac.jp/~ken.shiozaki/ahss/e2.html}{URL}.

For the example of spinful class D SC with MSG $P2_11'$ and $A$-representation of pairing symmetry, of which $E_2$- and $E^2$-pages are shown in Tables~\ref{tab:E_2} and \ref{tab:E^2}, respectively, the set of candidates for the $K$-groups is shown in Table~\ref{tab:SK}.
In cases like this example, despite the existence of multiple possibilities for higher-order derivatives and extensions, comparing the $E_2$-page in momentum space and the $E^2$-page in real space allows us to impose significant constraints on the $K$-groups.

\begin{table}[]
    \centering
    $$
    \begin{array}{ll}
    SK_{-n} & \mbox{The set of candidate $K$-groups}\\
    \hline 
    n=0&\Z \oplus \Z_4^{\oplus 3}, \Z\oplus \Z_4^{\oplus 2}\oplus \Z_2^{\oplus 2}\\
    n=1&\Z \oplus \Z_2^{\oplus 4}\\
    n=2&\Z \oplus \Z_2^{\oplus 3}\\
    n=3&\Z\\
    n=4&\Z\\
    n=5&\Z\\
    n=6&\Z \oplus \Z_2^{\oplus 3}\\
    n=7&\Z \oplus \Z_4 \oplus \Z_2^{\oplus 2}, \Z \oplus \Z_2^{\oplus 4}
    \end{array}
    $$
    \caption{The set of candidates $K$-groups of spinful class D SC with MSG $P2_11'$ and $A$-representaion of pairing symmetry.}
    \label{tab:SK}
\end{table}

In three dimensions with MSG and one-dimensional representations of pairing symmetries, three are 31050 symmetry settings in total. 
For the degree $n=0$ $K$-groups, which are the groups of classification of TIs and SCs, approximately 59\% of the $K$-groups are uniquely determined by comparing the $E_2$ and $E^2$ pages. 
For other degrees, see Table~\ref{tab:fixed_K}.

Through this calculation, we can also obtain the set of pairs of $E_4, E^4$ pages $(cE_{4,i}, cE^{4,j}), (i,j)\in {\cal S}_{\rm pair}$ that have no contradiction with the isomorphism (\ref{eq:iso_k_r}). 
However, since the number of elements in the set ${\cal S}_{\rm pair}$ is generally large, we do not show the results in this paper.

\section{Other symmetry settings
\label{sec:Other symmetry settings}
}
Up until the previous section, we have discussed the calculation methods for $E_2$ and $E^2$ pages in three-dimensional space, as well as the enumeration of possible higher-order derivatives and extensions. 
Extending these methods to magnetic layer groups in two dimensions and magnetic rod groups in one dimension is straightforward. 
We will not go into details here.
For magnetic layer and magnetic rod groups, we constructed the symmetry operations using the BNS symbols listed in \cite{Litvin_Magnetic}. 

It is also possible to calculate the $E_2$ and $E^2$ pages and enumerate the candidate $K$-groups for TIs and SCs protected solely by magnetic point group symmetries, with spatial translations removed from the magnetic space groups. 
The corresponding real-space $K$-groups are the relative $K$-groups ${}^\phi K^{\cal G}_{(z^{\rm int},c)-n}(\R^3,\p \R^3)$.
For some magnetic point groups, the $E^2$ pages for TIs were calculated in \cite{OkumaSatoShiozaki2019}. 
See also \cite{ZhangRenQiFang_intrinsicTSC_2022} for SCs. 
The momentum space $K$-groups are also the relative $K$-groups ${}^\phi K_G^{(z,c)-n}(\R^3,\p \R^3)$, where $\R^3$ here represents infinite momentum space. 
In both real and momentum space, the cell decomposition can be obtained by retaining only the $p$-cells touching the origin $O$ from the cell decomposition of the corresponding magnetic space group. 
As the other formulations are exactly the same as for the magnetic space group case, we will not go into details here. 
The same applies to magnetic point groups in one and two-dimensional spaces. 


Similar to the magnetic space groups, Table~\ref{tab:fixed_K} shows the percentage of symmetry classes for which there is exactly one candidate $K$-group in the set $SK_{-n}$ for each degree $(-n)$.

\begin{table*}[]
    \centering
    $$
\begin{array}{l|c|lllllllll}
\mbox{Symmetry type} & \mbox{Number of symmetry settings} & n=0 & n=1 & n=2 & n=3 & n=4 & n=5 & n=6 & n=7\\
\hline
 \mbox{3D, magnetic space groups} &31050& 0.594 & 0.779 & 0.886 & 0.859 & 0.749 & 0.663 & 0.539 & 0.519 \\
 \mbox{3D, magnetic point groups} &2346& 0.788 & 0.869 & 0.931 & 0.939 & 0.941 & 0.817 & 0.649 & 0.682 \\
 \mbox{2D, magnetic layer groups} &9264& 0.88 & 0.966 & 0.975 & 0.964 & 0.941 & 0.914 & 0.801 & 0.751 \\
 \mbox{2D, magnetic point groups} &2364& 0.951 & 0.978 & 0.976 & 0.974 & 0.984 & 0.936 & 0.769 & 0.795 \\
 \mbox{1D, magnetic rod groups} &6488& 0.943 & 0.996 & 0.999 & 0.996 & 0.981 & 0.975 & 0.945 & 0.897 \\
 \mbox{1D, magnetic point groups} &1946& 0.958 & 0.997 & 0.998 & 0.997 & 0.995 & 0.984 & 0.93 & 0.904 \\
\end{array}
$$
    \caption{
    The proportion of symmetry classes for which the candidate $K$-group $SK_{-n}$ is uniquely determined. The first column indicates the spatial dimension and the presence or absence of lattice translational symmetry.
    }
    \label{tab:fixed_K}
\end{table*}

\section{Conclusion
\label{sec:Conclusion}
}
In this paper, we have developed the calculation methods for the $E_2$-page and $E^2$-page of AHSS, and presented the results for the classification problem of topological phases in free fermion systems, namely TIs and SCs. 
For a given symmetry class, the $K$-groups in momentum space and real space are defined independently, and both give the same $K$-group.
The physical interpretation of AHSS has already been detailed in momentum space in \cite{ShiozakiSatoGomi_Atiyah-Hirzebruch_2022} and in real space in \cite{Shiozaki_homology}. 
In this paper, we have focused on the technical aspects of the calculations. 
The $E_2$ and $E^2$ pages serve as a first approximation in the classification problem, and in general, it is necessary to solve higher differentials and extension problems as $\Z$-modules. 
Although the physical interpretations of higher differentials and extension problems are known, systematic calculation methods are still unknown except for a few cases. 
In this paper, assuming that the rank of the $K$-group does not change by higher differentials, we have enumerated possible combinations of higher differentials and extensions for the obtained $E_2$ and $E^2$ pages, and calculated to what extent the classification results can be restricted by the isomorphism of the $K$-groups between momentum space and real space. 
As a result, about 59\% of the classifications were determined for TIs and SCs protected by MSGs. 
The calculation results can be obtained from Supplemental Material~\cite{Supplemental_Material} and \href{https://www2.yukawa.kyoto-u.ac.jp/~ken.shiozaki/ahss/e2.html}{URL}.

Finally, we summarize the calculation methods for $K$-groups that were not mentioned in this paper.

-- For simple MSG symmetries, the $K$-group may be determined by the Mayer-Vietoris sequence. 
For example, in the case of Table~\ref{tab:SK}, the $n=0$ degree is known to be $\Z \oplus \Z_4^{\oplus 2} \oplus \Z_2^{\oplus 2}$ in Ref.~\cite{ShiozakiSatoGomi_nonsymmorphic2016}. 

-- The choice of filtration for introducing AHSS is not unique. 
In Sec.~\ref{sec:Cell decomposition}, we set the condition that if the action of the group element $g \in G$ on the $p$-cell is closed in $D^p$, then the action of $g$ preserves $D^p$ pointwise, but this condition can be relaxed. 
For example, one can take the dual cell decomposition of the cell decomposition introduced in Sec.~\ref{sec:Cell decomposition}.

-- In the momentum-space AHSS, higher differentials $d_2^{0,0}$ and $d_3^{0,0}$ correspond to symmetry indicators that detect gapless points at generic momenta on 2-cells and 3-cells, respectively~\cite{ShiozakiSatoGomi_Atiyah-Hirzebruch_2022}.
In normal states, the correspondence between symmetry indicators and (higher-order) topological insulators~\cite{Song_quantitative_mappings_2018, Khalaf_PRX_2018, Fang_diagnosis_magnetic_2022}, or gapless semimetals~\cite{SongZhangFang_Diagnosis_AI_2018, Fang_diagnosis_magnetic_2022}, is well-established. 
As a result, some calculations in Sec.~\ref{sec:Higher_differential} can be replaced with known higher differentials to get more correct $E_\infty$ page. 
Although symmetry indicators for superconductors have been proposed and classified~\cite{SI_Luka, Ono-Po-Watanabe2020, Ono-Po-Shiozaki2021}, a comprehensive correspondence between (higher-order) topological superconductors and gapless superconductors has yet to be achieved.

-- In the real-space AHSS, some higher differentials have been systematically calculated for some symmetry classes. 
In Ref.~\cite{OnoShiozakiWatanabe_classification_SC_2022}, for time-reversal symmetric SCs with trivial pairing symmetry, a part of $d^2$ is calculated based on the equivalence between the second differential $d^2_{2,-2}$ of the real-space AHSS and the superconducting vortex zero modes.

-- $K$-groups are not just $\Z$-modules but are modules over some $K$-group. 
For example, in the problem setting of this paper, the $K$-group is a module over the representation ring $R(G_0)$ of the point group $G_0$~\cite{ShiozakiSatoGomi_crystalline_2017}. 
The differentials are homomorphisms as an $R(G_0)$-module, and the extension is that as an $R(G_0)$-module. 
Although we did not discuss the implementation of the module structure in this paper, it is expected to impose strong constraints on the construction of higher differentials and the solution of extension problems.

-- While the $K$-groups obtained in this work classify the topological phases, they do not directly provide explicit formulas for the topological invariants. 
However, a systematic construction of topological invariants is possible based on the intermediate data of the AHSS computation, as recently demonstrated in Ref.~\cite{Ono-Shiozaki2023}.

-- Generalizing our comparative approach between real and momentum spaces to bosonic and interacting spin systems remains an open problem. 
While the real-space AHSS is applicable to such many-body systems~\cite{Shiozaki_homology, FuHermele_prx_point_group_2017, PhysRevB.96.205106, SongFangQi_Real-space_recipes2020}, the momentum-space AHSS inherently relies on the single-particle picture. 
It remains unclear whether a dual description exists for many-body systems to enable a systematic computation as performed in Sec.~\ref{sec:Analysis of E2 pages}.


\begin{acknowledgments}
We thank Hoi Chun Po for collaborations on earlier works.
KS thanks Yosuke Kubota for helpful discussions.
KS was supported by The Kyoto University Foundation, JST CREST Grant No.~JPMJCR19T2, and JSPS KAKENHI Grant No.~22H05118. 
SO was supported by KAKENHI Grant No.~JP20J21692 from the Japan Society for the Promotion of Science.
We thank the YITP workshop YITP-T22-02 on ``Novel Quantum States in Condensed Matter 2022", which was useful in completing this work.
\end{acknowledgments}

\appendix

\section{An algorithm computing cell decomposition
\label{app:An algorithm computing cell decomposition}
}
This section presents an example of how to construct the cell decomposition introduced in Sec.~\ref{sec:Cell decomposition}.
We describe the cell decomposition of real space in three dimensions.
The cell decomposition in momentum spaces in lower dimensions can also be obtained in the same way.
The input is an MSG ${\cal G}$, which is equipped with the set of symmetry actions $x \to g(x)$ for $g \in {\cal G}$ on points $x \in \R^3$.
We first describe how a stereohedron, a convex polyhedron that fills space isohedrally, is given.
Such a stereohedron is also called a fundamental domain or an asymmetric unit.
Next, we explain a method to compute 2-, 1-, and 0-cells in order, starting from the boundary of the stereohedron.

\subsection{Fundamental domain}
Let ${\cal G}$ be an MSG.
Pick a reference point $a \in \R^3$ such that the little group ${\cal G}_a = \{g\in {\cal G}|g(a)=a\}$ is the same as that for generic points in $\R^3$.
From such a reference point, a stereohedron is given by (\cite{Schmitt_thesis2016}, Theorem 1.2.1.)
\begin{align}
DV(a) 
&=\{x \in \R^3 | \nonumber \\
&|x-a| < |x-g(a)| \mbox{ for all } g \in {\cal G} \backslash {\cal G}_a\}.
\end{align}
($DV$ means the Dirichlet-Voronoi stereohedron in \cite{Schmitt_thesis2016}.)
Note that $DV(a)$ is convex.

The set of vertices of the stereohedron $DV(a)$ is computed as follows.
Only points in the orbit ${\cal G}(a)=\{g(a) | g\in {\cal G}\}$ near the reference point $a$ are needed.
We represent the lattice translation group elements $\tau \in \Pi$ by $\tau = \{1|t\}$, where $t \in \mathbb{R}^3$ is the lattice translation vector.
For $\tau = \{1|t\} \in \Pi$, we introduce the open slab
\begin{align}
&S_a(t) \nonumber\\
&:=\{x \in \R^3 | (a-t,t)<(x,t)<(a+t,t)\}.
\end{align}
Let $t_1,t_2,t_3$ be a basis of the group of lattice translations.
The following holds true (\cite{Schmitt_thesis2016}, Lemma B.2.).
The subset ${\cal O}$ of the orbit ${\cal G}(a)$ defined by
\begin{align}
{\cal O}
&= \left( ({\cal G}(a)\backslash \{a\}) \cap \bigcap_{i=1}^3 S_a(t_i) \right) \nonumber\\
&\cup \{ \pm t_1(a), \pm t_2(a), \pm t_3(a) \}
\end{align}
suffices for computing $DV(a)$.
For a set of points ${\cal S}$, let us introduce the Dirichlet-Voronoi stereohedron
\begin{align}
    DV(a;{\cal S})
    &:=\{x \in \R^3 | \nonumber \\
    &|x-a| < |x-p| \mbox{ for all } p \in {\cal S}\}.
\end{align}
The above implies that $DV(a)=DV(a;{\cal O})$.
We call ${\cal O}$ the set of relevant points of $a$.

To efficiently compute $DV(a)$, we can do the following.
Introduce two subsets of ${\cal O}$ by ${\cal O}_0:=\{ \pm t_1(a), \pm t_2(a), \pm t_3(a) \}$ and ${\cal O}' := ({\cal G}(a)\backslash \{a\})\cap \bigcap_{i=1}^3 S_a(t_i)$ such that ${\cal O}_0 \cup {\cal O}'={\cal O}$.
As a starting big open polyhedron, define $DV(a)_0=DV(a;{\cal O}_0)$.
We sort the set of points in order of distance from $a$ and do the following for all the points $p_1,\dots,p_{|{\cal O}'|} \in{\cal O}'$ step by step.
For $p_i \in {\cal O}'$, if there is a vertex of $DV(a)_{i-1}$ in the region $\{x \in \R^3||x-a|>|x-p_i|\}$, set ${\cal O}_i={\cal O}_{i-1} \cup \{p_i\}$ and update $DV(a)_{i-1}$ by $DV(a;{\cal O}_i)$, otherwise ${\cal O}_i={\cal O}_{i-1}$ and $DV(a)_i=DV(a)_{i-1}$.
In the end, we get $DV(a)=DV(a)_{|{\cal O}'|}$.

Note that $DV(a)$ depends on the reference points $a$.

\subsection{Cell decomposition of the boundary of $DV(a)$}
In this section, we denote the interior of the convex hull of a convex set $e$ by $\tilde e$.
To specify the 3-cell $DV(a)$, we use the set of corner vertices $e^3 = \{x_1,x_2,\dots\}$ of $DV(a)$.
Similarly, let $e^2_i \subset e^3, i=1,\dots,$ be the sets of corner vertices of boundary convex polygons of $DV(a)$.

The convex polygon $\tilde e^2_i$ may not be a 2-cell in the definition in Sec.~\ref{sec:Cell decomposition}, since there may exist $x \in \tilde e^2_i$ and $g \in {\cal G}$ such that $g(x)\neq x$.
If this is the case, we properly divide $e^2_i$ into smaller polygons.
There are only two cases:
(i) There is an inversion center in $\tilde e^2_i$.
(ii) There is a two-fold axis in $\tilde e^2_i$.
Otherwise, there exists a reflection plane intersecting $\tilde e^2_i$.
The case (i) can be detected by comparing the little group $G_{\bar x}$ of the midpoint $\bar x=\frac{1}{|e^2_i|}\sum_{x \in e^2_i} x$ with the little group $G_x$ of a generic point in $\tilde e^2_i$.
The case (ii) can be detected by introducing the barycentric subdivision of a convex polygon $e^2_i$ and comparing the little groups $G_{x'}$ of a generic point $x'$ inside the line segments of the barycentric subdivision and the little group $G_x$ of a generic point of $\tilde e^2_i$.
In this way, we get the set of 2-cells.

For the line segments $e^1_i,i=1,\dots,$ of the boundary of each 2-cell, we divide $e^1_i$ into a set of line segments, if necessary.
The little group of the midpoint $\bar x$ of the line segment $e^1_i$ may be strictly larger than the little group of a generic point of the line segment $e^1_i$.
If this is the case, we divide $e^1_i$ into the two line segments.

\section{Computing irreducible characters
\label{app:Computing irreducible characters}
}
In this section we give a method to derive all the irreducible characters for a given finite group $G$ with a factor system $z_{g,h}$~\cite{Watanabe_kotaibutsuri}. 
For the basis $\{\ket{g}\}_{g \in G}$ labeled by the group elements, the (left) regular representation is defined by $\hat g \ket{k} = z_{g,k} \ket{gk}$.
The representation matrix $u^R_g$, which is defined by $\hat g \ket{k}=\sum_{h \in G} [u^R_g]_{hk}$, is 
\begin{align}
[u^R_g]_{hk} = z_{g,k} \delta_{h,gk}.     
\end{align}
As a matter of fact, the regular representation decomposes as a direct sum of the irreps, and each irrep appears with the number of its dimension of the irrep.
Namely, 
\begin{align}
R = \bigoplus_{\alpha \in \{{\rm irreps}\}} \alpha^{\oplus \dim \alpha}.
\end{align}
The key point here is that the regular representation includes all the irreps. 

Given a Hermitian random matrix $H$, symmetrizing it by 
\begin{align}
    H^G = \frac{1}{|G|}\sum_{g \in G} u_g H u_g^{-1} 
\end{align}
and diagonalizing $H^G$, we get the eigenvector $\ket{\phi_\alpha}$ for irreps $\alpha$. 
If the matrix $H$ is taken at random, there is no accidental degeneracy of $\dim \alpha$ numerically.
The irreducible character is given by $\chi^\alpha_g = \tr[\braket{\phi_\alpha|\hat g|\phi_\alpha}]$

\section{Computation in $\Z$-module
\label{app:Computation in Z-module}
}
Let $A$ and $B$ $\Z$-modules. 
For a homomorphism 
\begin{align}
f: A \to B, 
\end{align}
we summarize how to compute $\im f$ and $\ker f$. 
Firstly, we express the group $A$ as the quotient groups between torsion free $\Z$-modules as $A=\tilde A/P_A$ and $B=\tilde B/P_B$. 
For example, if $A = \Z^n \oplus \Z/k_1\Z \oplus \cdots \oplus \Z/k_u\Z$ be the invariant factor decomposition of $A$, then $\tilde A$ and $P_A$ can be $\tilde A = \Z^n \oplus \underbrace{\Z \oplus \cdots \oplus \Z}_{u}$ and $P_A = k_1\Z\oplus \cdots \oplus k_u \Z$. 
Set a lifted homomorphism $\tilde f:\tilde A \to \tilde B$ such that the following diagram commutes.
\begin{align}
\begin{CD}
0@>>>P_A@>>>\tilde A@>>>A@>>>0\\
@. @V\tilde f VV @V\tilde f VV @Vf VV\\
0@>>>P_B@>>>\tilde B@>>>B@>>>0\\
\end{CD}
\end{align}
Assuming commutativity on the right side of the diagram, 
\begin{align}
\tilde f(\tilde a \in \tilde A) \mod P_B = f(\tilde a \mod P_A), \label{ap_1}
\end{align}
commutativity on the left side, $\tilde f(P_A) \subset P_B$, follows from $\tilde f(P_A) \mod P_B = f(P_A \mod P_A) = f(0) = 0$.
Given a matrix expression $M_f$ of $f$ for bases of $A$ and $B$, a matrix expression $M_{\tilde f}$ of such $\tilde f$ is obtained by considering the matrix $M_f$ as $\Z$-valued.

\subsection{$\im f$}
From (\ref{ap_1}), we have 
\begin{align}
\im f
= \tilde f(\tilde A) \mod P_B 
= \tilde f(\tilde A)/(\tilde f(\tilde A) \cap P_B). 
\end{align}
Using the second isomorphism theorem, we get 
\begin{align}
\im f = (\tilde f(\tilde A)+P_B)/P_B.
\end{align}
Here, $\tilde f (\tilde A) + P_B$ is the union 
\begin{align}
\tilde f (\tilde A) + P_B = \{x+y \in \tilde B | x \in \tilde f(\tilde A), y \in P_B\}. 
\end{align}

\subsection{$\ker f$}
Note that 
\begin{align}
\ker f 
&= \{a \in A | f(a)=0\}\nonumber \\
&= \{\tilde a \in \tilde A | \tilde f(\tilde a) \in P_B\} \mod P_A.
\end{align}
Using $\tilde f(P_A) \subset P_B$, 
\begin{align}
\ker f =\{\tilde a \in \tilde A | \tilde f(\tilde a) \in P_B\}/P_A.
\end{align}
Here, the numerator can be expressed as 
\begin{align}
&\{\tilde a \in \tilde A | \tilde f(\tilde a) \in P_B\}\nonumber \\
&=\{\tilde a \in \tilde A | {}^\exists \tilde b \in P_B, {\rm\ s.t.\ }\tilde f(\tilde a)+\tilde b =0 \}. 
\end{align}
Thus, introducing the homomorphism 
\begin{align}
\tilde f \oplus {\rm Id}_{P_B}
&: \tilde A \oplus P_B \to \tilde B,\nonumber \\
&(\tilde a,\tilde b)\mapsto \tilde f(\tilde a)+\tilde b, 
\end{align}
we have 
\begin{align}
\{\tilde a \in \tilde A | \tilde f(\tilde a) \subset P_B\}
=\ker (\tilde f \oplus {\rm Id}_{P_B})|_{\tilde A}.
\end{align}
Here, $\ker (\tilde f \oplus {\rm Id}_{P_B})|_{\tilde A}$ is the projection of $\ker (\tilde f \oplus {\rm Id}_{P_B})$ onto $\tilde A$. 
We get 
\begin{align}
\ker f = \ker (\tilde f \oplus {\rm Id}_{P_B})|_{\tilde A}/P_A.
\end{align}

\subsection{$\ker g/\im f$}
Now let us consider a sequence of homomorphisms 
\begin{align}
\begin{CD}
A @>f>>B@>g>>C
\end{CD}
\end{align}
satisfying $\im f \subset \ker g$. 
By use of the formulas above, the quotient group $\ker g/\im f$ is given by 
\begin{align}
\ker g/\im f
=\frac{\ker (\tilde g \oplus {\rm Id}_{P_C})|_{\tilde B}/P_B}{(\tilde f(\tilde A)+P_B)/P_B}. 
\end{align}
Applying the third isomorphism theorem to it, this is a quotient group between torsion-free abelian subgroups of $\tilde B$, 
\begin{align}
\ker g/\im f
=\frac{\ker (\tilde g \oplus {\rm Id}_{P_C})|_{\tilde B}}{\tilde f(\tilde A)+P_B}. \label{eq:kerg/imf}
\end{align}
Practically, this quotient group can be computed by first expanding the basis of $\tilde f(\tilde A)+P_B$ in the basis of $\ker (\tilde g \oplus {\rm Id}_{P_C})|_{\tilde B}$ (the pseudo-inverse matrix can be used) and then computing the Smith normal form of the matrix consisting of the expansion coefficients.

\section{Higher differentials and connecting homomorphisms
\label{app:HigherDifferentials}
}
In this appendix, we briefly review the mathematical formulation of the higher differentials in the Atiyah-Hirzebruch spectral sequence. 
As mentioned in the main text, while explicit calculational formulas for higher differentials are generally difficult to obtain, they can be formally expressed using connecting homomorphisms between pairs of subspaces of different dimensions.

Let us explain this for the homology case following the standard formulation~\cite{CartanEilenberg1999}. 
The higher differential $d^r_{p,q}: E^r_{p,q} \to E^r_{p-r,q+r-1}$ can be understood from the following commutative diagram:
\begin{equation}
\xymatrix{
K_{p+q}(X_p,X_{p-r}) \ar[r]^{\varphi^r_{p,q}}\ar[d]^\partial & K_{p+q}(X_{p+r-1},X_{p-1})\ar[d]^\partial \\
K_{p+q-1}(X_{p-r}) \ar[d] & K_{p+q-1}(X_{p-1}) \ar[d] \\
K_{p+q-1}(X_{p-r},X_{p-2r}) \ar[r]^{\varphi^r_{p-r,q-1}} & K_{p+q-1}(X_{p-1},X_{p-r-1})
}
\end{equation}
Here, $\partial$ is the boundary map (connecting homomorphism), and $\varphi^r_{p,q}$ is the map induced by the inclusion of pairs $(X_p,X_{p-r}) \subset (X_{p+r-1},X_{p-1})$. 
In this formulation, the $E^r$-page is given by the image of $\varphi^r_{p,q}$, i.e., $E^r_{p,q} = {\rm im}\, \varphi^r_{p,q}$. 
The higher differential $d^r_{p,q}$ is then defined as 
\begin{align}
d^r_{p,q}: {\rm im}\, \varphi^r_{p,q} \to {\rm im}\, \varphi^r_{p-r,q-1}. 
\end{align}
Thus, the higher differential $d^r_{p,q}$ can be explicitly understood as the map induced by the vertical composition (which essentially consists of the connecting homomorphism) on the right-hand side of the commutative diagram, restricted to the image of the horizontal maps. 
The same logical structure applies to the cohomology case for the momentum-space AHSS with appropriate reversals of arrows and indices.

\section{Baer Sum}
\label{app:BaerSum}

Consider two extensions of $\mathbb{Z}$-modules as follows:
\begin{align}
&0\to B \xrightarrow{f_0}E_0 \xrightarrow{g_0}A \to 0, \\
&0\to B \xrightarrow{f_1}E_1 \xrightarrow{g_1}A \to 0
\end{align}
We want to construct an extension corresponding to the sum in ${\rm Ext}_\Z^1(A,B)$.
\begin{align}
&f: B\to E_0 \oplus E_1,\quad
b \mapsto (f_0(b),-f_1(b)), \\
&g: E_0 \oplus E_1 \to A,\quad
(e_0,e_1) \mapsto g_0(e_0)-g_1(e_1),
\end{align}
Define the maps as above. 
The inclusion $\im f \subset \ker g$ follows from $b \mapsto (f_0(b),-f_1(b)) \mapsto g_0\circ f_0(b) + g_1 \circ f_1 (b) = 0+0$.
The Baer sum is given by~\cite{Weibel_book_homological_algebra}
\begin{align}
&0\to B \xrightarrow{f'}\ker g/\im f \xrightarrow{g'}A \to 0.
\end{align}
Here,
\begin{align}
&f': B\to \ker g/ \im f,\nonumber\\
&b \mapsto [(f_0(b),0)] = [(0,f_1(b))], \label{eq:baer_fp}\\
&g': \ker g/ \im f \to A,\nonumber\\
&(e_0,e_1) \mapsto g_0(e_0)=g_1(e_1).\label{eq:baer_gp}
\end{align}
Suppose $(f_0(b),0) =(f_0(b'),-f_1(b'))$. 
Then $(f_0(b-b'),f_1(b'))=(0,0)$, and due to the injectivity of $f_0$ and $f_1$, we have $b=b'=0$. 
Notice that there is no intersection between $\im[ B \to \ker g, b \mapsto (f_0(b),0)]$ and $\im f$. 
The well-definedness of $g'$ follows from $(f_0(b),-f_1(b)) \mapsto g_0(f_0(b))=0=g_1(f_1(b))$.
The exactness is as follows. 
For the injectivity of $f'$, suppose $[(f_0(b),0)]=0$. 
Then, there exists $b' \in B$ such that $(f_0(b),0)=(f_0(b'),-f_1(b'))$, and similarly, we have $b=0$.
The surjectivity of $g'$ follows from that $g_0$ and $g_1$ are surjective. 
The inclusion $\im f' \subset \ker g'$ follows from $[f_0(b),0] \mapsto g_0(f_0(b))=0$.
We will show $\ker g' \subset \im f'$. 
Suppose $[(e_0,e_1)] \mapsto g_0(e)=g_1(e')=0$. 
Due to the surjectivity of $g_0$ and $g_1$, there exist $b, b' \in B$ such that $e_0=f_0(b), e_1=f_1(b')$. 
Then, $[(e_0,e_1)]=[(f_0(b),f_1(b'))]=[(f_0(b-b'),0)] \in \im f'$.

If we are only interested in the $\mathbb{Z}$-module resulting from the Baer sum of extensions, it is sufficient to calculate $\ker f/\im g$. 
If we want to take the Baer sum of the extensions obtained from the Baer sum itself, we also need to compute the homomorphism $f'$ and $g'$. 
We summarize the calculation method of the Baer sum, including the homomorphisms $f'$ and $g'$, below.

We can assume that $A$ is a torsion $\mathbb{Z}$-module. 
We start with the following commutative diagram:
\begin{align}
\begin{CD}
P_B @. P_{E_0} \oplus P_{E_1} @.P_A\\
@VVV @VVV @VVV\\
\tilde B@>\tilde f>> \tilde E_0 \oplus \tilde E_1 @>\tilde g>>\tilde A\\
@VVV @VVV@VVV\\
B@>f>>E_0 \oplus E_1 @>g>>A\\
\end{CD}
\end{align}
Here, for a $\mathbb{Z}$-module $X$, $\tilde X$ and $P_X$ are free $\mathbb{Z}$-modules such that $X = \tilde X / P_X$.
Applying the formula (\ref{eq:kerg/imf}), we have
\begin{align}
\frac{\ker g}{\im f} \cong \frac{\ker (\tilde g \oplus {\rm Id}_{P_A})|{\tilde E_0 \oplus \tilde E_1}}{\im \tilde f +(P_{E_0}\oplus P_{E_1})}
\end{align}
Let $b_1, \dots, b_n$ be basis vectors for the sublattice $\ker (\tilde g \oplus {\rm Id}_{P_A})|{\tilde E_0 \oplus \tilde E_1} \subset \tilde E_0 \oplus \tilde E_1$, and introduce the matrix $M_g = (b_1, \dots, b_n)^T$. 
Let the generators of the sublattice $\im \tilde f +(P_{E_0}\oplus P_{E_1}) \subset \tilde E_0 \oplus \tilde E_1$ be given by $\im \tilde f +(P_{E_0}\oplus P_{E_1}) = {\rm Span}(a_1, \dots, a_m)$, and introduce the matrix $M_f = (a_1, \dots, a_m)^T$. 
(There is no need to take $a_1, \dots, a_M$ to be linearly independent.)
Since $\im \tilde f +(P_{E_0}\oplus P_{E_1}) \subset \ker (\tilde g \oplus {\rm Id}_{P_A})|{\tilde E_0 \oplus \tilde E_1}$, $a_1, \dots, a_m$ can be expanded in terms of the basis $b_1, \dots, b_n$, and the expansion coefficients are given by the matrix $M_f M_g^+$, where $M_g^+$ is the generalized inverse matrix of $M_g$. Perform the Smith decomposition of the matrix $M_f M_g^+$:
\begin{align}
&U (M_f M_g^+) V =
\left[\begin{array}{cc}
\Lambda &O \\
O & O\\
\end{array}\right],\nonumber \\
&\Lambda = {\rm diag}(\la_1,\dots,\la_k), \nonumber \\
&\la_i|\la_{i+1} {\rm \ for\ }i=1,\dots,k-1.
\end{align}
We obtain:
\begin{align}
\frac{\ker g}{\im f}
\cong \Z^{\oplus (n-k)} \oplus \bigoplus_{i=1}^k \Z_{\la_i}. 
\end{align}

Now, consider the transformed basis of the sublattice $\ker (\tilde g \oplus {\rm Id}{P_A})|{\tilde E_0 \oplus \tilde E_1}$:
\begin{align}
V^{-1} M_g = (\tilde b_1,\dots,\tilde b_n)^T
\end{align}
By expressing the homomorphisms $f'$ and $g'$ defined in (\ref{eq:baer_fp}) and (\ref{eq:baer_gp}) in terms of the basis $\tilde b_1,\dots,\tilde b_n$, we can obtain their matrix representations. 
(Note that for $i$ such that $\la_i=1$, the basis $\tilde b_i$ is for a trivial $\Z$-module $\Z_1$.)

The matrix representation of the homomorphism $f': b \mapsto [(f_0(b),0)]$ is given as follows. 
Let the matrix representation of $f_0: B \to E_0$ be $M_{f_0}^T$ (where $M_{f_0}$ is a ${\rm dim}\tilde B \times {\rm dim}\tilde E_0$ matrix). 
Add ${\rm dim} \tilde E_1$ zeros to it and then expand it with the basis $V^{-1} M_g$. 
That is,
\begin{align}
(M_{f_0},O_{{\rm dim}\tilde B \times {\rm dim} \tilde E_1}) (V^{-1}M_g)^+
\end{align}
is the matrix representation of $f'$.

To find the matrix representation of the homomorphism $g':(e_0,e_1) \mapsto g_0(e_0)$, first project the components of the basis vectors $\tilde b_1,\dots,\tilde b_n$ onto $\tilde E_0$:
\begin{align}
(v^{-1}M_g)|_{E_0}.
\end{align}
This is an $n \times {\rm dim} \tilde E_0$ matrix. 
Let the matrix representation of $g_0: E_0 \to A$ be $M_{g_0}^T$. 
$M_{g_0}$ is a ${\rm dim} \tilde E_0 \times {\rm dim} \tilde A$ matrix. 
The matrix representation of $g'$ is given by:
\begin{align}
(V^{-1}M_g)|_{E_0} M_{g_0}.
\end{align}

Note that any extension can be obtained by a finite number of Baer sums of the following two types of extensions. 
${\rm Ext}(\Z_n,\Z)=\Z_n \ni 1:$
\begin{align}
\quad &0\to\Z\xrightarrow{1 \mapsto n}\Z\xrightarrow{1\mapsto 1}\Z_n\to 0,
\end{align}
and 
${\rm Ext}(\Z_n,\Z_m)=\Z_{{\rm gcd}(n,m)} \ni 1:$
\begin{align}
\quad &0\to\Z_m\xrightarrow{1 \mapsto n}\Z_{mn}\xrightarrow{1\mapsto 1}\Z_n\to 0.
\end{align}
Therefore, by the formulation presented in this section, extensions corresponding to any element of ${\rm Ext}_\Z^1(A,B)$ can be constructed.

We give an example of a nontrivial Baer sum.
Set $A=\Z_4$ and $B=\Z\oplus \Z_2$.
In the sense described in Sec.~\ref{sec:extension}, there exist five reduced extension matrices $M$:
\begin{align}
M = \begin{pmatrix}
0\\0\\
\end{pmatrix},
\begin{pmatrix}
1\\0\\
\end{pmatrix},
\begin{pmatrix}
2\\0\\
\end{pmatrix},
\begin{pmatrix}
0\\1\\
\end{pmatrix},
\begin{pmatrix}
2\\1\\
\end{pmatrix}
\end{align}
The matrix $M=(0,0)^T$ corresponds to the trivial extension to give $\Z \oplus \Z_2 \oplus \Z_4$. 
For both matrices $M=(1,0)^T$ and $M=(2,0)^T$, $\Z_2 \subset B$ do not contribute to the extension, resulting in $\Z\oplus \Z_2$ and $\Z \oplus \Z_2^{\oplus 2}$, respectively.
For $M=(0,1)^T$, $\Z \subset B$ does not contribute to the extension, yielding $\Z \oplus \Z_8$.
Finally, for $M=(2,1)^T$, both $\Z$ and $\Z_2 \subset B$ contribute to the extension.
Computing the Baer sum of $(2,0)^T$ and $(0,1)^T$, we have $\Z \oplus \Z_4$.

\bibliography{refs}

\end{document}